\documentclass[a4paper,fleqn,usenatbib]{mnras}

\usepackage{newtxtext,newtxmath}

\usepackage[T1]{fontenc}
\usepackage{ae,aecompl}

\usepackage{graphicx}	
\usepackage{amsmath}	
\usepackage{amssymb}	

\newcommand{\rkl}[1]{\left(#1\right)}
\newcommand{\ekl}[1]{\left[#1\right]}

\newcommand{\skl}[1]{\left\langle#1\right\rangle}

\title[SILCC V: The magnetised ISM]{The SILCC project - V. The impact of magnetic fields on the chemistry and the formation of molecular clouds}

\author[Philipp Girichidis et al.]{Philipp~Girichidis$^{1,2}$\thanks{E-mail: philipp@girichidis.com}, Daniel Seifried$^4$, Thorsten Naab$^3$, Thomas Peters$^3$,\newauthor Stefanie Walch$^4$, Richard~W\"{u}nsch$^{5}$, Simon~C.~O.~Glover$^{6}$, Ralf~S.~Klessen$^{6,7}$
\\
$^1$Leibniz-Institut f\"{u}r Astrophysik Potsdam (AIP), An der Sternwarte 16, 14482 Potsdam, Germany\\
$^2$Heidelberger Institut f\"{u}r theoretische Studien, Schloss-Wolfsbrunnenweg 35, 69118 Heidelberg, Germany\\
$^3$Max-Planck-Institut f\"{u}r Astrophysik, Karl-Schwarzschild-Str. 1, 85741 Garching, Germany\\
$^4$Physikalisches Institut, Universit\"{a}t zu K\"{o}ln, Z\"{u}lpicher Str. 77, 50937 K\"{o}ln, Germany\\
$^5$Astronomical Institue, Czech Academy of Sciences, Bo\v{c}n\'{i} II 1401, 141 00 Prague, Czech Republic\\
$^6$Universit\"{a}t Heidelberg, Zentrum f\"{u}r Astronomie, Institut f\"{u}r Theoretische Astrophysik, Albert-Ueberle-Str. 2, 69120 Heidelberg, Germany\\
$^7$Universit\"{a}t Heidelberg, Interdisziplin\"{a}res Zentrum f\"{u}r wissenschaftliches Rechnen, Im Neuenheimer Feld 2015, 69120 Heidelberg, Germany
}

\date{Accepted XXX. Received YYY; in original form ZZZ}

\pubyear{2017}

\begin{document}
\begin{NoHyper}
\label{firstpage}
\pagerange{\pageref{firstpage}--\pageref{lastpage}}
\maketitle

\begin{abstract}
Magnetic fields are ubiquitously observed in the interstellar medium (ISM) of present-day star-forming galaxies with dynamically relevant energy densities. Using three-dimensional magneto-hydrodynamic (MHD) simulations of the supernova (SN) driven ISM in the flux-freezing approximation (ideal MHD) we investigate the impact of the magnetic field on the chemical and dynamical evolution of the gas, fragmentation and the formation of molecular clouds. We follow the chemistry with a network of six species (H$^{+}$, H, H$_2$, C$^+$, CO, free electrons) including local shielding effects. We find that magnetic fields thicken the disc by a factor of a few to a scale height of $\sim100\,\mathrm{pc}$, delay the formation of dense (and molecular) gas by $\sim25\,\mathrm{Myr}$ and result in differently shaped gas structures. The magnetised gas fragments into fewer clumps, which are initially at subcritical mass-to-flux ratios, $M/\Phi\approx0.3(M/\Phi)_\mathrm{crit}$, and accrete gas preferentially parallel to the magnetic field lines until supercritial mass-to-flux ratios of up to order 10 are reached. The accretion rates onto molecular clouds scale with $\dot{M}\propto M^{1.5}$. The median of the inter-cloud velocity dispersion is $\sim2-5\,\mathrm{km\,s}^{-1}$ and lower than the internal velocity dispersion in the clouds ($\sim3-7\,\mathrm{km\,s}^{-1}$). However, individual cloud-cloud collisions occur at speeds of a few $10\,\mathrm{km\,s}^{-1}$.
\end{abstract}




\section{Introduction}

Magnetic fields are ubiquitously observed in the interstellar medium (ISM, \citealt{Beck2009, Crutcher2012, Haverkorn2015}) and might play a vital role for the evolution of galaxies \citep[e.g.][]{NaabOstriker2017}. With mean field strengths in the spiral structures of galaxies of a few $\mu\mathrm{G}$ the magnetic energy density is typically lower than the kinetic but overall comparable to the mean thermal energy density \citep[e.g.][]{BoularesCox1990, Cox2005}. The high conductivity and the partial ionisation in the ISM result in an efficient coupling of the magnetic field with the gas and the assumption of ideal MHD is valid for a large fraction of the ISM. This allows for a dynamical impact of the magnetic field on the motions of the gas from scales of the entire galaxy \citep{Beck2001, Hanasz2009, KotarbaEtAl2009, PakmorSpringel2013, RiederTeyssier2017}, representative parts of the ISM \citep{deAvillezBreitschwerdt2005, KoyamaOstriker2009a, KoyamaOstriker2009b, HillEtAl2012, KimOstriker2015b, SILCC1, GirichidisSILCC2, PardiEtAl2017}, molecular clouds \citep{ShuAdamsLizano1987, PadoanNordlund1999, Heitsch02, MacLowKlessen2004, BanerjeeEtAl2009} down to star-forming regions \citep{HennebelleFromang2008, HennebelleTeyssier2008, HennebelleCiardi2009, PetersEtAl2011, SeifriedEtAl2010, SeifriedEtAl2011, SeifriedEtAl2012, KlassenPudritzKirk2017, KoertgenBanerjee2015}.

Turbulent or turbulence-like motions, that are trans- or supersonic \citep{GoldreichKwan1974, ZuckermanEvans1974, ElmegreenScalo2004, ScaloElmegreen2004} stir the ISM. Supernovae (SNe, e.g. \citealt{GattoEtAl2015, WalchNaab2015}), stellar feedback processes like radiation \citep{WalchEtAl2013, KimKimOstriker2016, PetersEtAl2017a, HaidEtAl2018} and winds \citep{GattoEtAl2017}, gravitational instability, and differential rotation in the galactic disc form a multitude of complex dynamical interactions that are expected to shape the ISM on different scales and with different efficiencies \citep{KlessenGlover2016}. SNe and gravitational contraction are regarded as the most energetic dynamical drivers.

Theoretical considerations describe the interstellar medium in present-day star-forming disc galaxies as a multi-component fluid consisting of two (cool and warm) stable \citep{Field1965,FieldGoldsmithHabing1969, WolfireEtAl1995, WolfireEtAl2003} and one hot meta-stable phases \citep{CoxSmith1974, McKeeOstriker1977, Ferriere2001}. However, due to the strong dynamics a significant fraction of the gas is in the unstable regime. In addition, there is a constant exchange between the phases. The hot gas with temperatures of the order of $10^6\,\mathrm{K}$ originates from supernovae and fills most of the volume. The warm and cold component coexist in approximate pressure equilibrium \citep[e.g.][]{Cox2005}. The hot medium is expected to be dominated by thermal and kinetic rather than magnetic energy. Contrary, the cold regions can be permeated by strong magnetic fields that can retard or theoretically completely prevent gravitational contraction and the formation of dense clouds and stars \citep{CrutcherEtAl2010,Crutcher2012,PardiEtAl2017}. The transition from magnetically sub- to supercritical regions, which is needed for gravitational forces to overcome pressure balance is still a matter of active discussion \citep[e.g.][]{KoertgenBanerjee2015, ValdiviaEtAl2016}. In the case of strong fields the contraction perpendicular to the field lines is prohibited due to the magnetic pressure. However, condensations can form along the field lines \citep{Field1965, HeitschHartmann2014, SILCC1, GirichidisSILCC2, IffrigHennebelle2017}.

This paper is part of the SILCC project\footnote{\url{https://hera.ph1.uni-koeln.de/~silcc/}} (SImulating the Life Cycle of molecular Clouds) that investigates the processes in the interstellar medium using hydrodynamical simulations. The first two studies \citep{SILCC1, GirichidisSILCC2} focus on different modes of SN driving and varying SN rates. Both find that SNe partially need to explode in low-density environments to create a realistic chemical composition and launch outflows from the disc. \citet{GattoEtAl2017} uses active star clusters that self-consistently form in dense regions and accrete gas. The study also includes stellar winds in addition to SN feedback. \citet{PetersEtAl2017a} further added radiation from star clusters using the same formalism. All of these studies have a maximum grid resolution of $4\,\mathrm{pc}$, which is too coarse to study molecular clouds. The SILCC-zoom study \citep{SeifriedEtAl2017} specifically investigates the condensation of gas and the formation of clouds with a maximum resolution of $0.1\,\mathrm{pc}$ in the non-magnetised ISM and without the formation of star or star cluster particles. The high resolution accounts for a more detailed chemical evolution but restricts the study to only a few clouds. With this study we specifically aim at probing the impact of magnetic fields on the chemical evolution along with the formation of molecular clouds while keeping the dynamical connection to the large-scale ISM flows \citep[see dicussion in][]{SeifriedEtAl2017}. We increase the resolution to $1\,\mathrm{pc}$, which allows us to more accurately follow the evolution in dense regions and at the same time resolve many clouds with higher resolution to have a statistically relevant sample of molecular clouds.

Following the numerics and the simulation setup (Sections~\ref{sec:methods} and \ref{sec:setup-parameters}) we first give an overview of the morphological evolution of the simulations (Section~\ref{sec:morphology-global-dynamics}), followed by an analysis of the degree of magnetisation of the gas (Section~\ref{sec:magnetisation}). We then focus on the formation of shielded and molecular gas (Section~\ref{sec:optthick}) and the fragmentation into clouds and their properties (Section~\ref{sec:clouds}). Discussions and conclusions are presented in Sections~\ref{sec:discussion} and \ref{sec:conclusions}.

\section{Numerical Methods}
\label{sec:methods}

A detailed description of the numerical setup, the equations to solve as well as the numerical methods used is given in \citet{SILCC1} and \citet{GirichidisSILCC2}. Here we only present a brief summary.
We use the hydrodynamical code \textsc{Flash} in version 4 (\citealt{FLASH00, DubeyEtAl2008}, \url{http://flash.uchicago.edu/site/}) which is parallelised using the Message Passing Interface (MPI). The Eulerian grid divides the computational domain in blocks of $8^3$ cells which can be adaptively refined (Adaptive Mesh Refinement, AMR).

The magneto-hydrodynamic (MHD) equations are solved using the HLLR5 finite-volume scheme for ideal MHD \citep{Bouchut2007, Bouchut2010, Waagan2009, Waagan2011}. The directionally split solver ensures positivity of the density and pressure by construction and is suitable for high-Mach-number flows.

Gravitational effects are included in two ways. We solve the Poisson equation for the effects of self-gravity using the tree-based method by \citet{WuenschEtAl2018}. In addition we include an external potential that accounts for the acceleration due to the stellar component of the disc. We use tabulated values based on an isothermal sheet \citep{Spitzer1942} with a stellar surface density of $30\,\mathrm{M}_\odot\,\mathrm{pc}^{-2}$ and a scale height of $100\,\mathrm{pc}$. The acclerations are comparable to the Milky-Way values by \citet{KuijkenGilmore1989} for low heights above the midplane.

In order to both follow the chemical state of the gas and compute radiative cooling accurately, we use a chemical network that includes ionised (H$^+$), atomic (H) and molecular hydrogen (H$_2$) as well as singly ionized carbon (C$^+$) and carbon monoxide (CO). This allows us to include non-equilibrium abundances as in \citet{GloverMacLow2007a} and \citet{MicicEtAl2012} using the carbon chemistry by \citet{NelsonLanger1997}. We assume a temporally constant interstellar radiation field of $G_0=1.7$ \citep{Habing1968, Draine1978}, which is locally attenuated based on how strongly shielded the computational cell is. The column densities and the derived optical depth in every cell are computed using the \textsc{Flash} implementation \citep{WuenschEtAl2018} of the \textsc{TreeCol} algorithm \citep{ClarkGloverKlessen2012}. For every cell we integrate the column density out to a radial distance of $50\,\mathrm{pc}$. The column density dependent attenuation factor follows \citet{GloverClark2012b}.

For radiative cooling we follow the atomic and molecular cooling functions of \citet{GloverEtAl2010} and \citet{GloverClark2012b}. High-temperature cooling above $10^4\,\mathrm{K}$ is based on the rates described in \citet{GnatFerland2012}. Heating includes a constant cosmic ray ionisation and corresponding heating rate \citep{GoldsmithLanger1978}, X-ray heating \citep{WolfireEtAl1995}, and photoelectric heating that is coupled to the optical depth \citep{BakesTielens1994, Bergin2004, WolfireEtAl2003}. We assume a constant dust-to-gas mass ratio of 0.01 using dust opacities of \citet{MathisMezgerPanagia1983} and \citet{OssenkopfHenning1994}.

\section{Simulation setup and parameters}
\label{sec:setup-parameters}

\begin{table}
    \caption{List of performed simulations}
    \label{tab:simulations}
    \begin{tabular}{lcccccc}
      Name   & $B_0$ & eff. res. & $\Delta x$ & ref. & deref.\\
      & ($\mu$G) &      & (pc)       & $f_{\mathrm{H}_2}$ & $f_{\mathrm{H}_2}$\\
      \hline
      \texttt{B0-2pc} & $0$   & $256^3$ & $1.95$ & $0.01$ & $0.005$\\
      \texttt{B3-2pc} & $3$   & $256^3$ & $1.95$ & $0.01$ & $0.005$\\
      \texttt{B6-2pc} & $6$   & $256^3$ & $1.95$ & $0.01$ & $0.005$\\
      \texttt{B0-1pc} & $0$   & $512^3$ & $0.98$ & $0.01$ & $0.005$\\
      \texttt{B3-1pc} & $3$   & $512^3$ & $0.98$ & $0.01$ & $0.005$\\
      \texttt{B6-1pc} & $6$   & $512^3$ & $0.98$ & $0.01$ & $0.005$\\
      \hline
      \texttt{B0-1pc-ht} & $0$   & $512^3$ & $0.98$ & $0.1$ & $0.05$\\
      \texttt{B3-1pc-ht} & $3$   & $512^3$ & $0.98$ & $0.1$ & $0.05$\\
      \hline
    \end{tabular}

    \medskip
    Shown are the name of the simulation, the initial magnetic field strength in the midplane, the effective numerical resolution, the corresponding spatial resolution of the smallest cell as well as the threshold values for refinement and derefinement.
\end{table}

We set up a stratified box with a size of $(0.5\,\mathrm{kpc})^3$. The boundary conditions in $x$ and $y$ are periodic. For the boundary in $z$ we use diode boundary conditions, which allow gas to leave but not enter the box.  We emphasise that recent numerical studies of the SN-driven ISM with different mean densities and SN positioning in periodic boxes discuss in detail the problem of an over-pressurised ISM (\emph{thermal runaway}, \citealt{GattoEtAl2015, LiEtAl2015}), which is aggravated by a closed box. The gas density follows a Gaussian distribution in $z$ with a scale height of $30\,\mathrm{pc}$. The temperature is set such that the gas is in pressure equilibrium. In the densest regions of the disc at $z=0$ the composition of the gas is purely atomic with a minimum temperature of $4600\,\mathrm{K}$. At the lowest density ($\rho_\mathrm{min}=10^{-28}\,\mathrm{g\,cm}^{-3}$) at large $|z|$ we set $T=4\times10^8\,\mathrm{K}$ and assume that all the gas is ionised. The gas surface density is $10\,\mathrm{M}_\odot\,\mathrm{pc}^{-2}$, which yields a total mass of $2.5\times10^6\,\mathrm{M}_\odot$. Initially the gas is at rest.

We distinguish between a non-magnetic setup and two magnetic setups with different initial field strengths. The magnetic field is oriented along the $x$ direction an with initial field strengths in the midplane of $B_{x,0}=3\,\mu\mathrm{G}$ and $B_{x,0}=6\,\mu\mathrm{G}$. The magnitude of $B$ scales with the square root of the density, $B_x(z)=B_{x,0}\,[\rho(z)/\rho(z=0)]^{1/2}$. We do not introduce a random component of the magnetic field.

Feedback from stars is included as individual SNe exploding at a constant rate. We refrain from using sink particles and an accretion based star formation and SN rate \citep[e.g.][]{Schmidt1959} because of the uncertainties of how to treat the magnetic field at the interface between sink particles and the gas. By using the same SN rate and positioning, it is also easier to directly compare the different magnetisations. Based on the total gas surface density in the box we use the Kennicutt-Schmidt \citep{KennicuttSchmidt1998} relation to derive a star formation rate, which we convert into a SN rate using the stellar initial mass function of \citet{Chabrier2003}. For our surface density of $\Sigma=10\,\mathrm{M}_\odot\,\mathrm{pc}^{-2}$ this yields $15$ SNe per Myr. We apply the \emph{clustered} SN driving from \citet{SILCC1} and \citet{GirichidisSILCC2}, which we summarize briefly. The total SN rate is split into a fraction of 20\% type~Ia SNe, which explode with random $x$ and $y$ positions and a Gaussian distribution in $z$ with a scale height of $300\,\mathrm{pc}$. The remaining 80\% are SNe of type~II with a scale height of $50\,\mathrm{pc}$. We split the latter component into $2/5$ of individual SNe (representing run-away OB stars)  and $3/5$ of SNe, which are associated with clusters \citep{CowieSongailaYork1979, GiesBolton1986,Stone1991, HoogerwerfDeBuijneDeZeeuw2001, deWitEtAl2005, TetzlaffNeuhaeuserHohle2011} and are expected to be an important agent for driving a hot phase and superbubbles \citep[e.g.][]{MacLowMcCray1988,GattoEtAl2015,LiEtAl2015}. The clusters have an associated life time of $40\,\mathrm{Myr}$ and contain $7-20$ SNe, randomly drawn from a power law distribution $P\propto N^{-2}$, where $N$ is the number of SNe. All SNe of a cluster explode at the same position. We ensure a constant SN rate in the box, which requires the temporal spacing between SNe in individual clusters to vary. The random positions of \emph{all} SNe are precomputed beforehand and stored in a table, so that all simulations have exactly the same SN positions. If we resolve the Sedov-Taylor radius with at lest four cells at the highest refinement level ($r_\mathrm{SN, min}=4\Delta x_\mathrm{min}$), we inject $10^{51}\,\mathrm{erg}$ of thermal energy per SN into a dynamically computed volume with a total mass of $800\,\mathrm{M}_\odot$. If we do not meet this resolution criterion, we heat the gas in the minimum volume ($r_\mathrm{SN,min}$) to $10^4\,\mathrm{K}$ and inject momentum according to \citet{BlondinEtAl1998}, see also \citet{GattoEtAl2015}. The resolution in all simulations is high enough to resolve the Sedov-Taylor radius with at least 4 cells for more than $98\%$ of all SNe, which ensures that the efficiency of the explosions is not impaired by numerical over-cooling or errors in the momentum injection, e.g. for overlapping SN remnants with cancelling momenta \citep[see also the resolution requirements discussed by][]{KimOstriker2015}.

The base resolution is $128^3$ cells corresponding to a cell size of $4\,\mathrm{pc}$. We allow for one and two additional levels of refinement with cell sizes of $2\,\mathrm{pc}$ (name tag \texttt{2pc}) and $1\,\mathrm{pc}$ (\texttt{1pc}), respectively. The local adaptive refinement is based on two quantities: The density field following \citet{Loehner1987}, which is the standard in \textsc{Flash}, as well as on the fraction of molecular hydrogen. We refine a block if the molecular mass fraction of a cell exceeds $0.01$ and derefine if the values fall below $0.005$. For comparison we perform two additional runs with a ten times higher refinement/derefinement threshold, $0.1/0.05$. We perform eight different simulations listed in Table~\ref{tab:simulations}. The first column shows the name, which encodes the initial field strength (\texttt{B0}, \texttt{B3}, \texttt{B6}), the resolution (\texttt{2pc} and \texttt{1pc}) as well as the additional higher refinement threshold (\texttt{ht}), where needed. All simulations use the same SN rate as well as the same clustering and SN positioning. We run all simulations for $60\,\mathrm{Myr}$.

\section{Morphological evolution and global dynamics}
\label{sec:morphology-global-dynamics}

\begin{figure*}
  \includegraphics[width=0.85\textwidth]{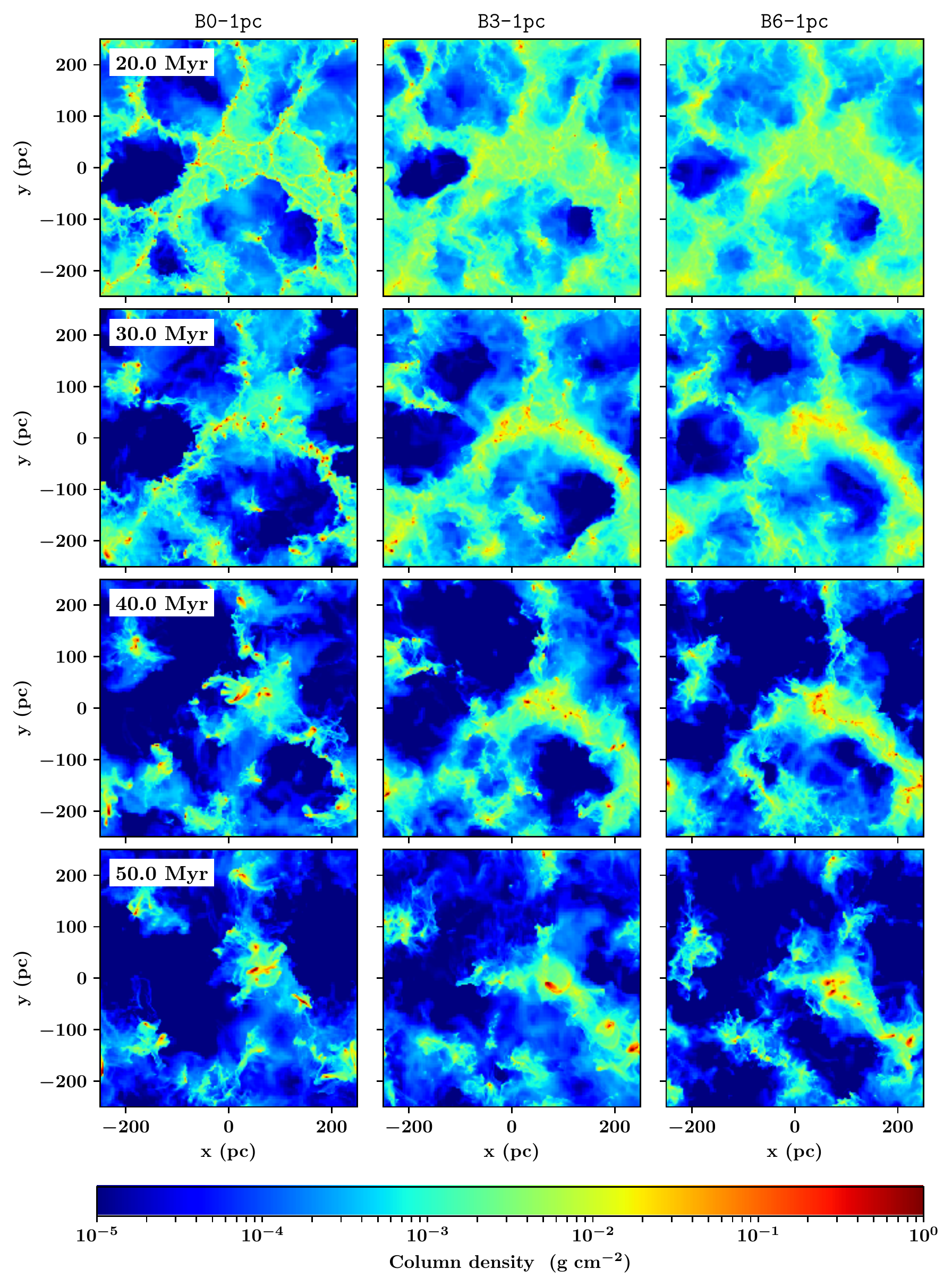}
  \caption{Column density projections face-on. We show simulations \texttt{B0-1pc}, \texttt{B3-1pc}, and \texttt{B6-1pc} (left to right) at $20$, $30$, $40$, and $50\,\mathrm{Myr}$ (top to bottom). Filaments and clumps start forming very early in the non-magnetic run. The stronger the magnetic field, the longer it takes to form dense structures. In addition, the magnetised clumps (centre and right-hand panels) are larger, less numerous and more strongly embedded in extended diffuse gas.}
  \label{fig:L7-coldens-face}
\end{figure*}

\begin{figure*}
  \includegraphics[width=0.85\textwidth]{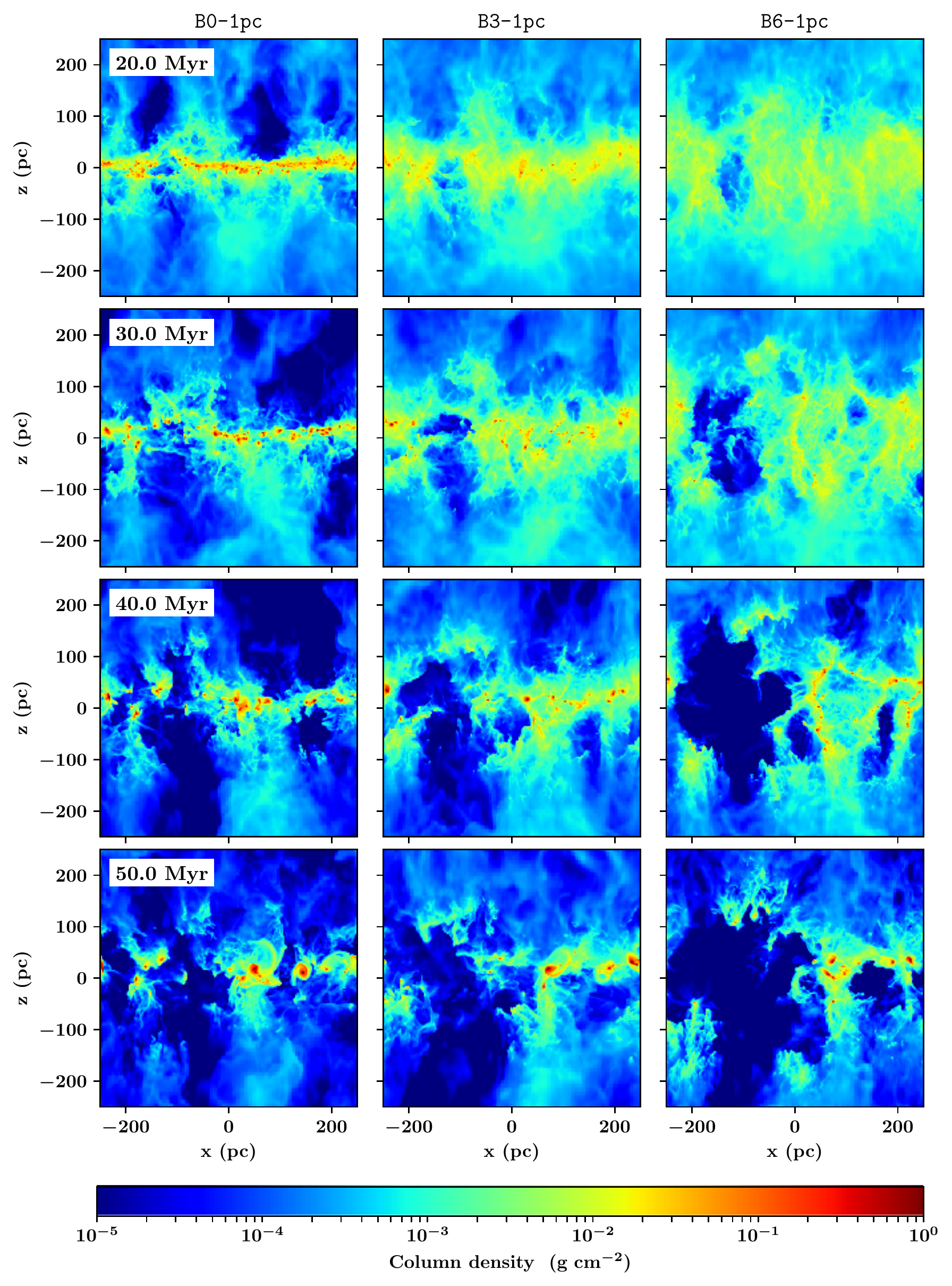}
  \caption{Column density projections edge-on. We show simulations \texttt{B0-1pc}, \texttt{B3-1pc}, and \texttt{B6-1pc} (left to right) at $20$, $30$, $40$, and $50\,\mathrm{Myr}$ (top to bottom). Magnetic fields delay the contraction of gas close to the midplane. As a result condensations of dense gas can form over a larger range in vertical height. At the end of the simulations the largest and most massive clouds are however always close to the midplane.}
  \label{fig:L7-coldens-edge}
\end{figure*}

\subsection{Fragmentation}

The morphological evolution differs early on when comparing the simulations of varying initial magnetic field strength. Changes of the maximum resolution and refinement threshold result in significantly smaller effects than the differences in magnetisation. We therefore concentrate here on the differences in $B$ and only discuss resolution and refinement details in the appendix. We show face-on column density plots in Fig.~\ref{fig:L7-coldens-face} for simulations \texttt{B0-1pc} (left), \texttt{B3-1pc} (centre) and \texttt{B6-1pc} (right) at simulation times ($20$, $30$, $40$, and $50\,\mathrm{Myr}$ from top to bottom) and the corresponding edge-on view in Fig.~\ref{fig:L7-coldens-edge}. In the non-magnetic runs, the SNe can quickly form over-dense regions, which cool and condense to filamentary structures and clouds near the midplane under the effect of the external potential and self-gravity. After $20\,\mathrm{Myr}$ the simulation reaches column density contrasts of more than six orders of magnitude ranging from $10^{-5}\,\mathrm{g\,cm}^{-2}$ in voids to $1\,\mathrm{g\,cm}^{-2}$ in numerous small dense clouds forming at the intersections of elongated filaments. After $30\,\mathrm{Myr}$ dense individual clouds have formed in the non-magnetic run with a sharp transition from the dense cloud to the low-density voids. Over time the small individual clouds in the non-magnetic run merge to eventually result in approximately 10 clouds after $50\,\mathrm{Myr}$ with masses of $\sim10^5\,\mathrm{M}_\odot$ each. In the first $20-30\,\mathrm{Myr}$ the two magnetised runs mostly form diffuse overdensities. The maximum column densities of $\sim1\,\mathrm{g\,cm}^{-2}$ are found in only a few clouds in \texttt{B3-1pc} and are barely reached in \texttt{B6-1pc}. In addition, the magnetised clouds are embedded in larger coherent structures with column densities of $10^{-3}-0.3\,\mathrm{g\,cm}^{-2}$. The field provides direct magnetic support in addition to thermal and turbulent pressure, which prevents or delays the condensation of gas into thin filaments and sheets. In addition, the magnetic pressure attenuates the impact of the SNe leading to lower maximum densities in the swept-up gas, which in turn results in less efficient cooling and more thermal pressure support. The delayed formation of dense gas has also been reported by \citet{HeitschMacLowKlessen2001} and more recently by \citet{HennebelleIffrig2014} and \citet{IffrigHennebelle2017} as a later accumulation of gas in sink particles and a lower star formation rate. The smaller number of clouds and the more diffuse envelope in the magnetic runs is consistent with similar simulations by \citet{PardiEtAl2017} in a periodic box. After $50\,\mathrm{Myr}$ the magnetised simulations end up with a similar number and mass distribution of clouds as the non-magnetic counterpart due to continuous merging of clumps. 

A common behaviour in all simulations is that early overdensities created by the first SNe continuously grow in mass and size. The SN feedback influences the formation process but does not provide an efficient destruction mechanism for the clouds. This means that we do not simulate the full cycle of a molecular cloud but focus on the formation process. To some extent the destruction can be enhanced by placing SNe always in the densest regions, which mimics the destruction of clouds from inside out. However, previous models like the \emph{peak} and \emph{mixed} driving runs in \citet{SILCC1} and \citet{GirichidisSILCC2} lead to an unrealistic composition of the ISM (\emph{peak driving}) and in general do not solve the problem of the indestructible clouds, see also Section~\ref{sec:discussion}. 

\subsection{Vertical structure}

\begin{figure}
\includegraphics[width=8cm]{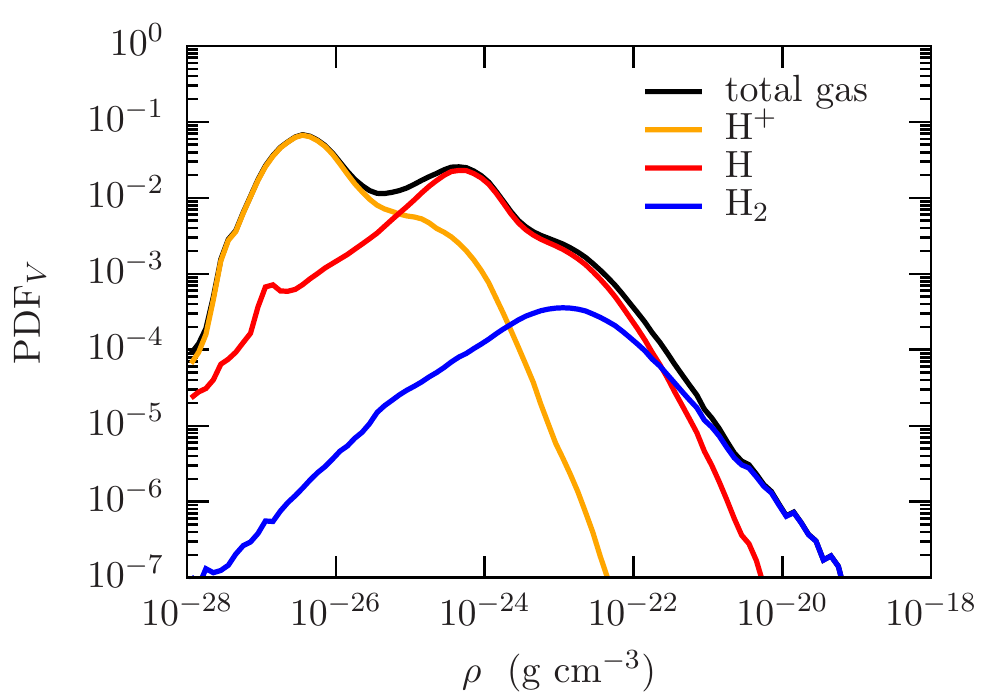}
\caption{Volume weighted density PDF for the total mass and the individual chemical species for simulation \texttt{B3-1pc} at $t=30\,\mathrm{Myr}$. The broad and widely overlapping distributions of H$^+$, H, and H$_2$ illustrate that it is problematic to define density thresholds for the individual species.}
\label{fig:pdf-dens-chem}
\end{figure}

\begin{figure}
\centering
\includegraphics[width=8cm]{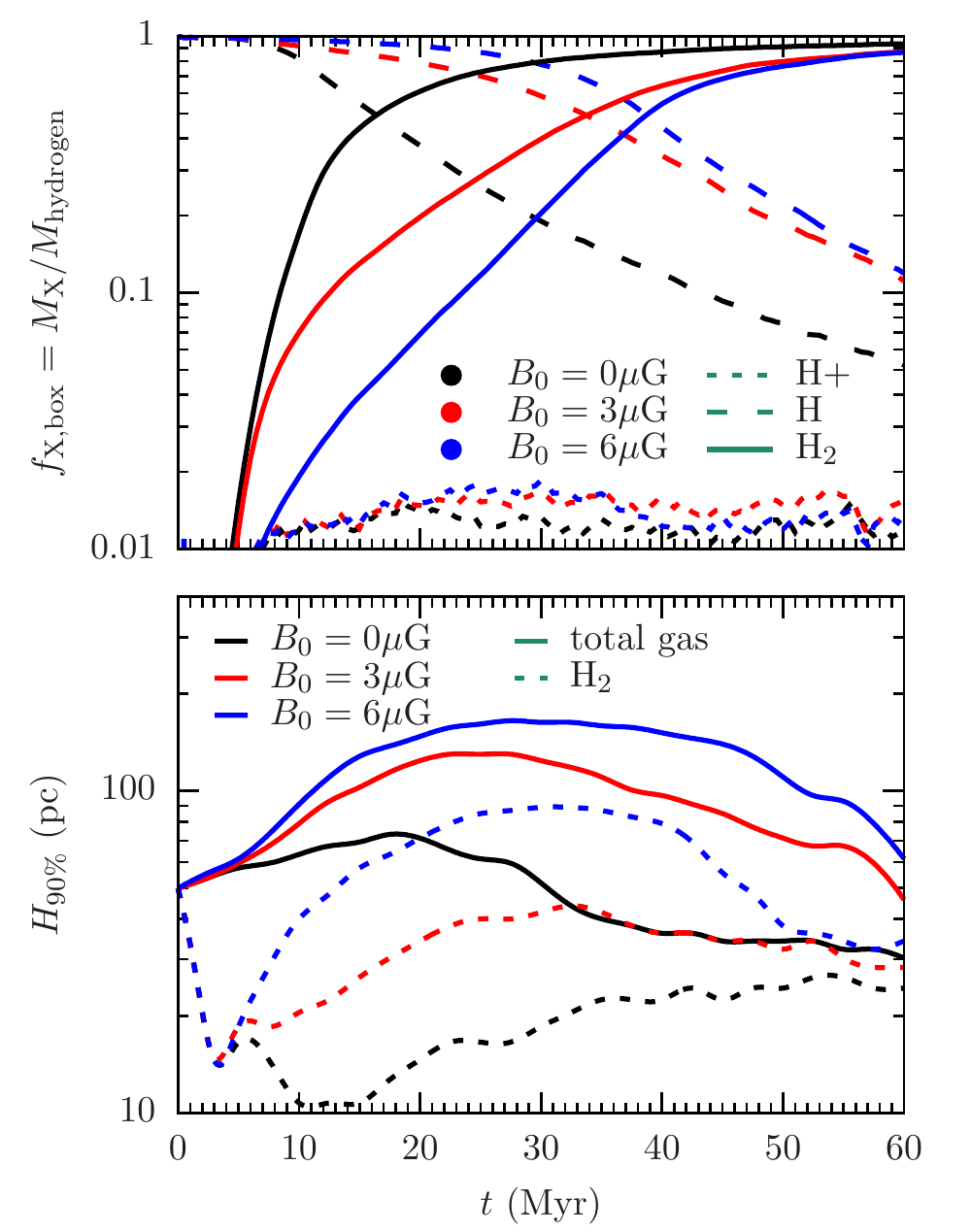}
\caption{Top: Time evolution of the mass fractions in all three hydrogen components. The initial atomic hydrogen condenses to for molecular gas after $\sim10-20\,\mathrm{Myr}$ depending on magnetisation. Ionized hydrogen only makes up for less than 2\% of the mass in all simulations. Bottom: Vertical heights of 90\% enclosed mass for all the gas (solid lines) and molecular hydrogen (dotted lines) over time. The stronger the magnetic support the thicker is the disc. For the total gas the height of 90\% enclosed mass reaches $\sim150\,\mathrm{pc}$. Molecular hydrogen can form at large heights (up to $80\,\mathrm{pc}$) in the case of strong fields ($6\mu\mathrm{G}$). For weak and no fields the dense clouds embedding H$_2$ form and reside close to the midplane ($\lesssim30\,\mathrm{pc}$).}
\label{fig:mf-scale-heights-time}
\end{figure}

\begin{figure}
\centering
\includegraphics[width=8cm]{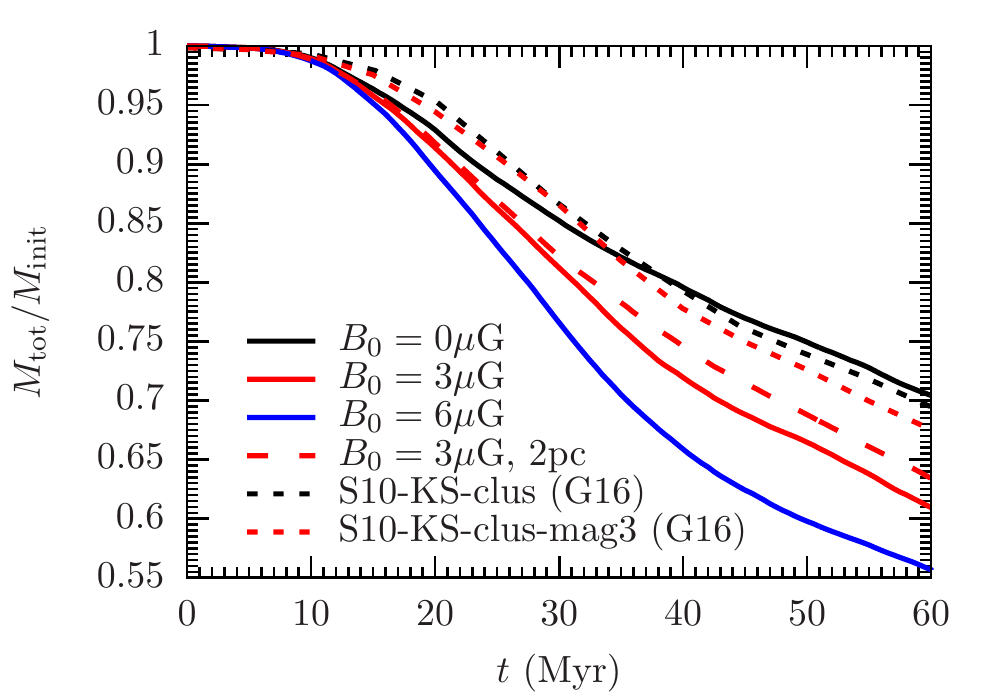}
\caption{Total gas mass over time for the main simulations of the current study (solid lines). For comparison we also include the gas mass in a box of $(500\,\mathrm{pc})^3$ of the corresponding simulations of \citet{GirichidisSILCC2} (dashed lines, G16). In all cases the simulations lose mass through the $z$ boundary at $\pm250\,\mathrm{pc}$. The stronger the magnetisation the more gas can reach the boundary. By the end of the simulation \texttt{B0-1pc} loses 30\% of the mass. For \texttt{B6-1pc} the total mass loss is $45\%$. The non-magnetic runs behave very similar; The magnetic run \texttt{B3-1pc} is more efficient in driving gas away from the midplane compared to S10-KS-clus-mag3.}
\label{fig:total-mass-time-evol}
\end{figure}

Fig.~\ref{fig:L7-coldens-edge} shows column densities edge-on and reveals the different evolution in the vertical structure. Without magnetic pressure support the overdensities in the intersections of swept-up SN shells can cool, lose their thermal pressure support and contract towards the midplane. After only $20\,\mathrm{Myr}$ the gas has formed a thin sheet with few vertical fluctuations. Including magnetic fields, the gas can delay the thermal compression by SNe as well as the gravitational attraction until the first dense clouds form at $\sim30\,\mathrm{Myr}$ for \texttt{B3-1pc} and at $\sim40\,\mathrm{Myr}$ for \texttt{B6-1pc}. The gas thus remains in a diffuse state for a longer time and provides a larger effective cross section for the SN blast waves, which results in a thicker disc with elongated structures of diffuse gas along the vertical direction. The dense clouds form later and partially in the extended elongated structures. As a consequence the dense gas has a larger spread in the vertical distribution. Once compact clouds have formed out of the diffuse gas and the effective cross section is reduced, the gravitational attraction dominates over the outward pointing thermal and magnetic support of the environment and the clouds move towards the midplane. This causes the vertical heights to decrease towards the end of the simulation.

Before quantifying the results of the simulations in terms of different density and chemical composition, we would like to illustrate the relation between them. Fig.~\ref{fig:pdf-dens-chem} presents the volume weighted density distribution for simulation \texttt{B3-1pc} at $t=30\,\mathrm{Myr}$. The lines show the total gas as well as the individual hydrogen species, H$^+$, H, and H$_2$. All chemical distributions widely overlap over several orders of magnitude \citep[see also discussion in][]{SILCC1}. In the following discussion we therefore investigate various quantities with weighting by mass, volume and chemical species separately.

Fig.~\ref{fig:mf-scale-heights-time} depicts the chemical mass fractions, $f_\mathrm{X,box}=M_\mathrm{X}/M_\mathrm{hydrogen}$ ($\mathrm{X}=\mathrm{H}^+,\,\mathrm{H},\,\mathrm{H}_2$) over time (top) as well as the vertical extent of the gas (bottom). Initially, all the gas is in atomic form, which starts condensing to form molecular hydrogen at different times depending on magnetisation, see Section~\ref{sec:optthick} for details. The sum of atomic and molecular gas makes up for more than 98 per cent of the total hydrogen mass. The lower panel shows the heights of 90\% enclosed mass for all the gas (solid lines) and 90\% of the molecular gas (dotted lines). The visual impression of the column density plots is well reflected in the time evolution plots. The vertical scales for the gas reach a few $10$s of parsecs in the non-magnetic runs. The magnetic field helps to lift the gas to heights of $\sim100\,\mathrm{pc}$ for the weak magnetic field and up to $\sim150\,\mathrm{pc}$ for the strong magnetic field. In the case of H$_2$ the vertical heights show the same trend but at a lower absolute height with $\sim20\,\mathrm{pc}$, $\sim30\,\mathrm{pc}$, and up to $\sim80\,\mathrm{pc}$ for \texttt{B0-1pc}, \texttt{B3-1pc}, and \texttt{B6-1pc}, respectively. We note that the scale height of the gas is always well below $250\,\mathrm{pc}$. Observations of the vertical mass distribution in the Milky Way based on CO (corresponding more to our molecular gas rather than the total gas) reveal an exponential scale height of $47\,\mathrm{pc}$ \citep{LangerPinedaVelusamy2014}, which would be more consistent with our magnetic runs. In other galaxies the scale height can range from $40-200\,\mathrm{pc}$ for the molecular gas \citep{CalduPrimoEtAl2013, YimEtAl2014}. This range is large and besides different physical effects from within the disc, global disc dynamics like warps or accretion onto the galaxy are likely to play a role for the vertical distribution of gas in individual systems. Ignoring the global effects, the observations seem to be more consistent with our magnetic runs. In the Milky Way giant molecular clouds (GMCs) tend to populate regions much closer to the midplane compared to low-mass clouds \citep{StarkLee2005}. This agrees well with our evolution of the cloud population in the magnetic runs. The smaller clouds form at larger altitudes and merge to GMCs close to the midplane, which causes the scale height to decrease along with the formation of massive clouds via accretion and merging.

The applied SN driving is expected to result in fountain flows that exceed the boundary at $|z|=250\,\mathrm{pc}$ \citep{HillEtAl2012, SILCC1, GirichidisSILCC2}. We notice outflowing gas in all simulations, which reduces the total amount of gas in the box. Fig.~\ref{fig:total-mass-time-evol} shows the total mass in the simulation box over time for the main simulations in this study (\texttt{B0-1pc}, \texttt{B3-1pc}, and \texttt{B6-1pc}; solid lines) as well as simulation \texttt{B3-2pc} (long-dashed line). In addition we plot the total mass of the same volume (up to the height of $250\,\mathrm{pc}$ above the midplane) in the clustered non-magnetic simulation \emph{S10-KS-clus} and the clustered magnetic run \emph{S10-KS-clus-mag3} of \citet[G16]{GirichidisSILCC2} with a resolution of $\approx4\,\mathrm{pc}$. In the non-magnetic run the mass loss is $30\%$ over the total simulation time in very good agreement with the previous simulation at a four times coarser resolution. The larger vertical extent of the gas in simulations \texttt{B3-1pc} and \texttt{B6-1pc} result in losses of $40$ and $45\%$, respectively. The comparison between \texttt{B3-1pc}, \texttt{B3-2pc} and \emph{S10-KS-clus-mag3} (all red lines) indicate that the mass loss through the boundary at $\pm250\,\mathrm{pc}$ is resolution dependent. With higher resolution we can resolve the Sedov-Taylor phase of more SNe, which allows for a more efficient stirring of the gas. In addition, higher resolution allows for a more efficient small-scale dynamo and a resulting stronger magnetic support, which helps lifting the gas above $250\,\mathrm{pc}$. We note that with the outflows from the box, we also lose magnetic flux through the $z$ boundary. However, we are primarily losing low-density gas through the boundary, which is weakly magnetised. As a result the accompanying loss of magnetic flux with the outflow is unlikely to affect the dynamics in the region around the midplane.

\subsection{Global dynamics}

\begin{figure}
\includegraphics[width=0.5\textwidth]{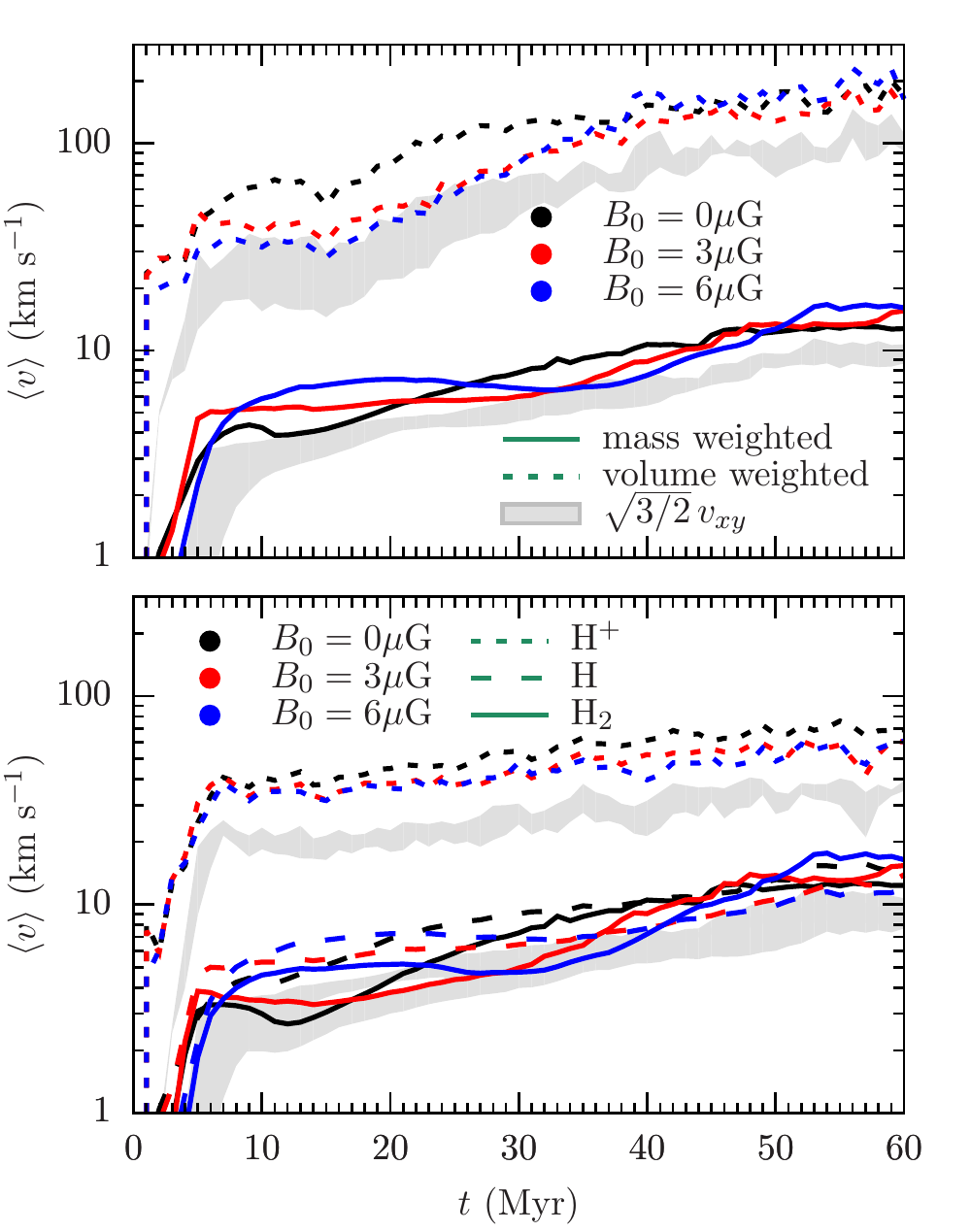}
\caption{Global weighted mean gas velocities over time. The top panel shows the weighting by mass and volume; the bottom one by chemical composition. The lines show the global velocities, the shaded area excludes outflow motions. In all cases the temporal variations are larger than the differences between the simulations. The weighting by mass, H$_2$ and H are basically indistinguishable with final values of around $10\,\mathrm{km\,s}^{-1}$. Weighting by H$^+$ results in velocities up to $\sim50\,\mathrm{km\,s}^{-1}$. The volume weighted numbers, which are dominated by the hot gas, yield velocities of $\gtrsim50\,\mathrm{km\,s}^{-1}$.}
\label{fig:gas-velocities-time}
\end{figure}

\begin{figure*}
\begin{minipage}{\textwidth}
\includegraphics[width=0.5\textwidth]{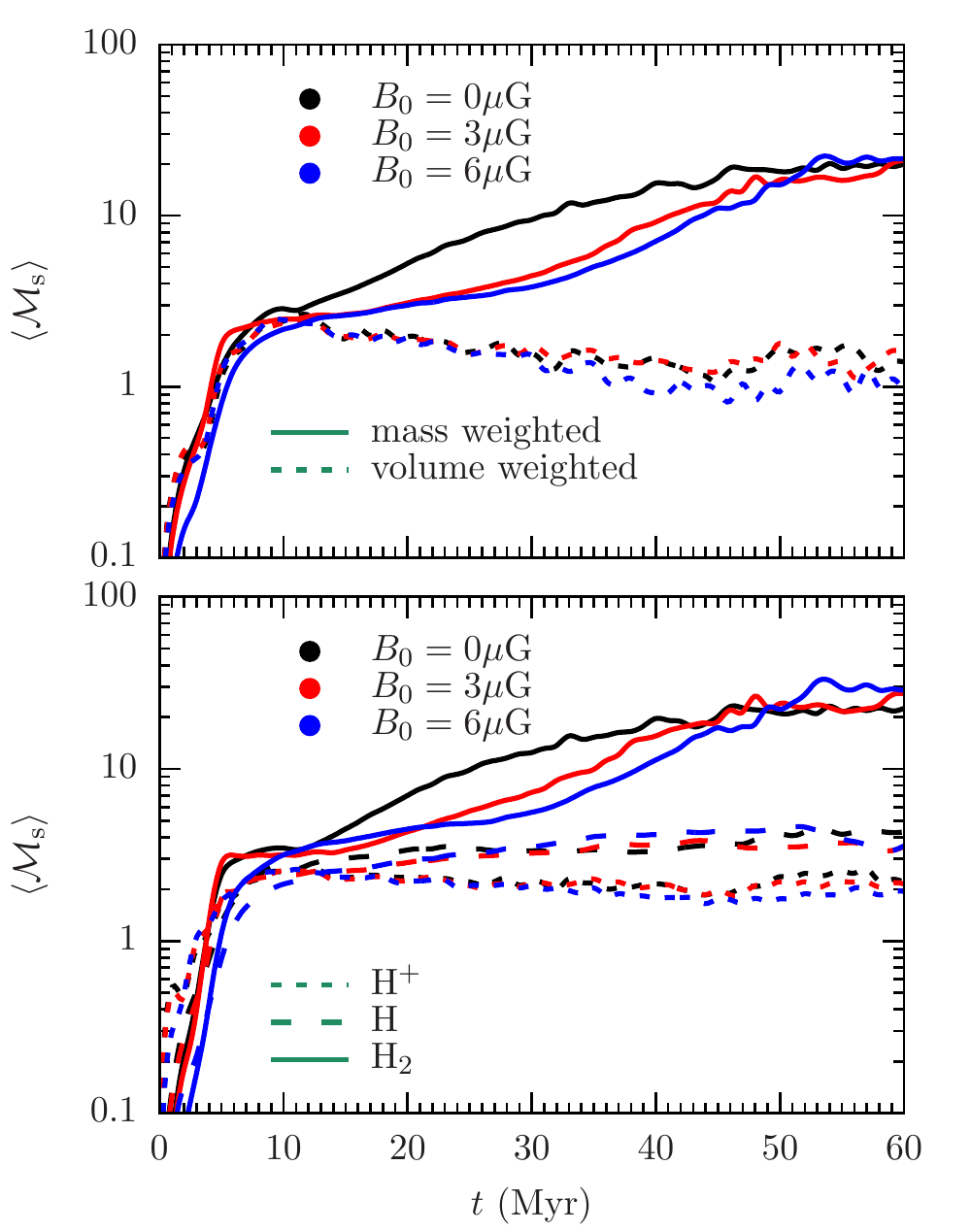}
\includegraphics[width=0.5\textwidth]{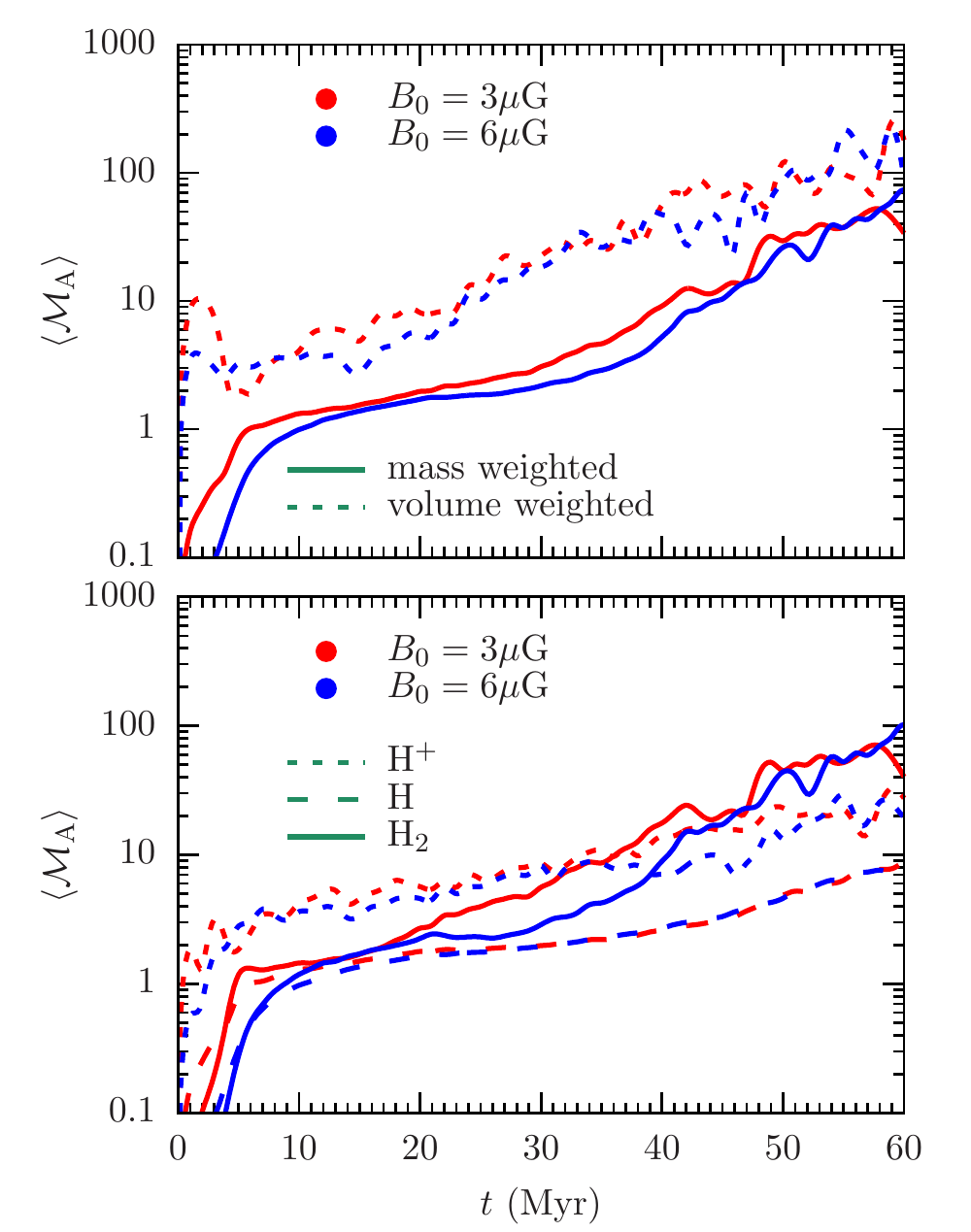}
\end{minipage}
\caption{Sonic (left) and Alfv\'{e}nic (right) Mach number over time. Basically all gas moves at supersonic speeds. There is a clear trend of low-density, ionised gas to have lower sonic Mach numbers, whereas warm atomic gas and molecular gas reach Mach numbers of order $5$ and $20$, respectively. The Alfv\'{e}nic Mach number increases over time, reflecting that the kinetic energy density increases, whereas the magnetic energy density does not increase significantly or even slightly decreases in the case of low-density gas (see Section~\ref{sec:magnetisation}).}
\label{fig:Mach-number-time}
\end{figure*}

\begin{figure*}
\begin{minipage}{0.8\textwidth}
\includegraphics[width=\textwidth]{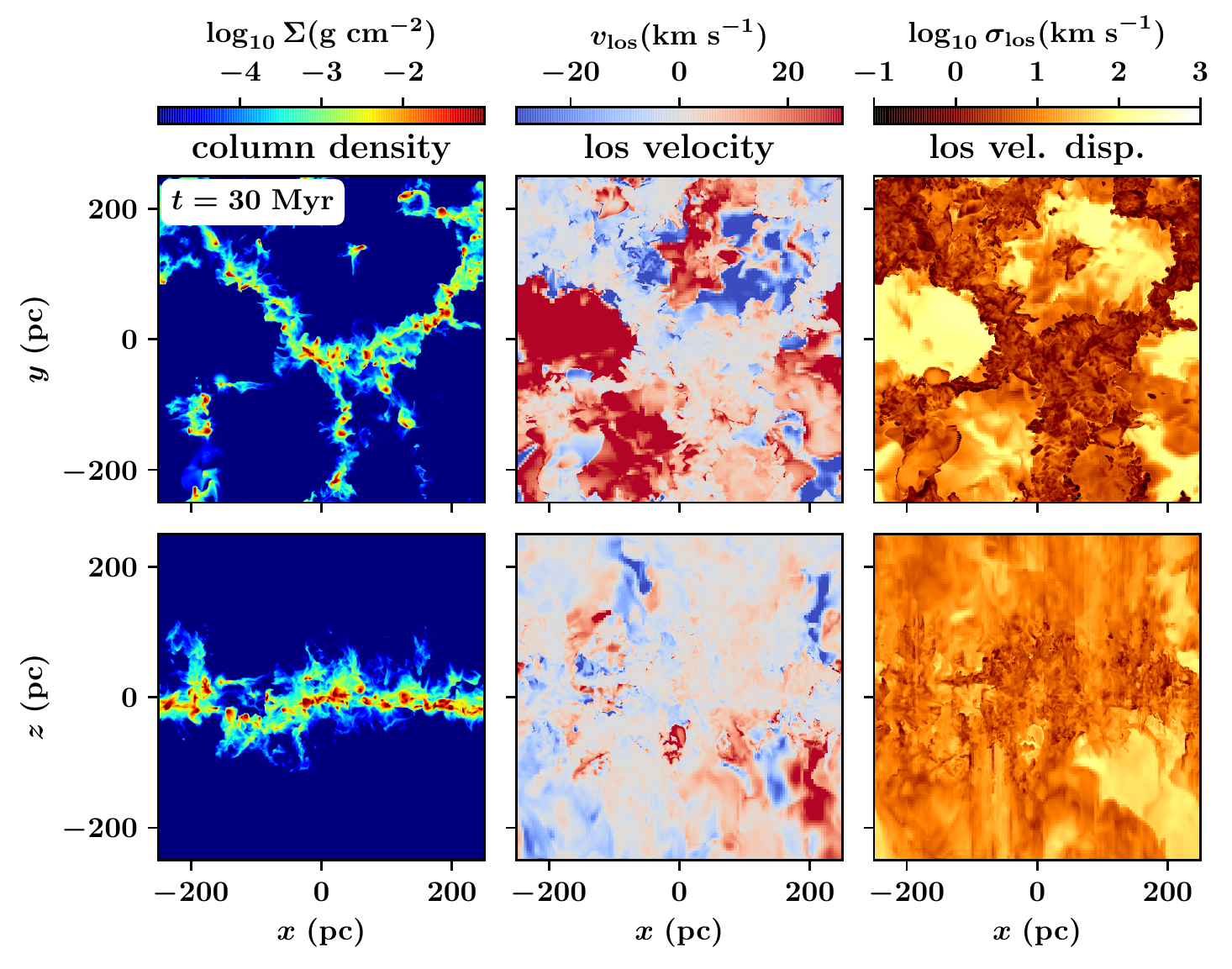}\\
\includegraphics[width=\textwidth]{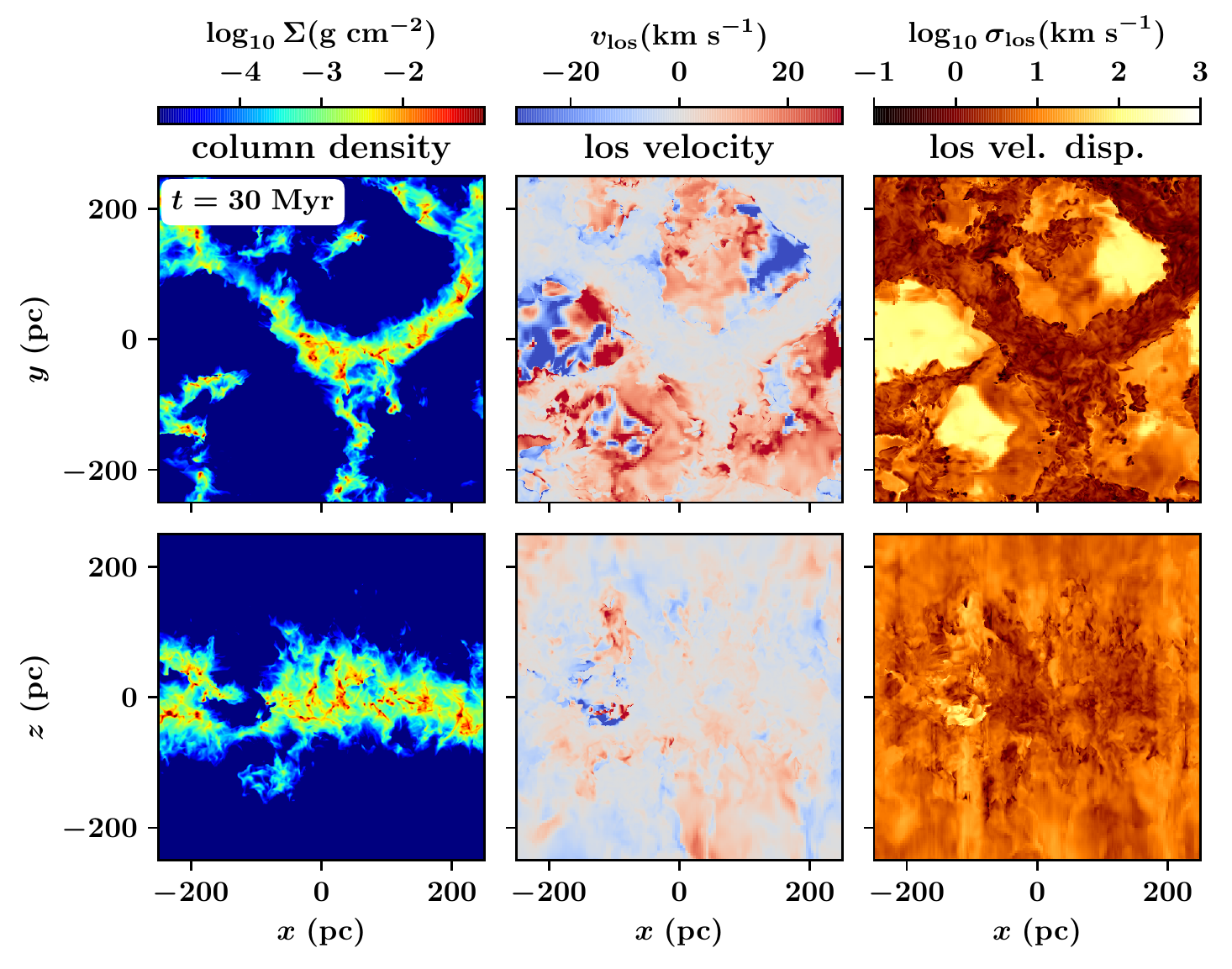}
\end{minipage}
\caption{From left to right we plot the column density, the line-of-sight velocity as well as the corresponding velocity dispersion viewed face-on (top row) and edge-on (bottom row) for \texttt{B0-1pc} (top plot) and \texttt{B3-1pc} bottom plot at $t=30\,\mathrm{Myr}$. For the projection along $z$ the velocity and the dispersion are smaller in dense regions, which is connected to the vertical acceleration. The edge-on projection mainly reflects the turbulent dynamics with weaker correlations between $v_{\mathrm{los},y}, \sigma_{\mathrm{los},y}$ and $\Sigma_y$.}
\label{fig:coldens-los-vel}
\end{figure*}

\begin{figure*}
\begin{minipage}{0.9\textwidth}
\includegraphics[width=\textwidth]{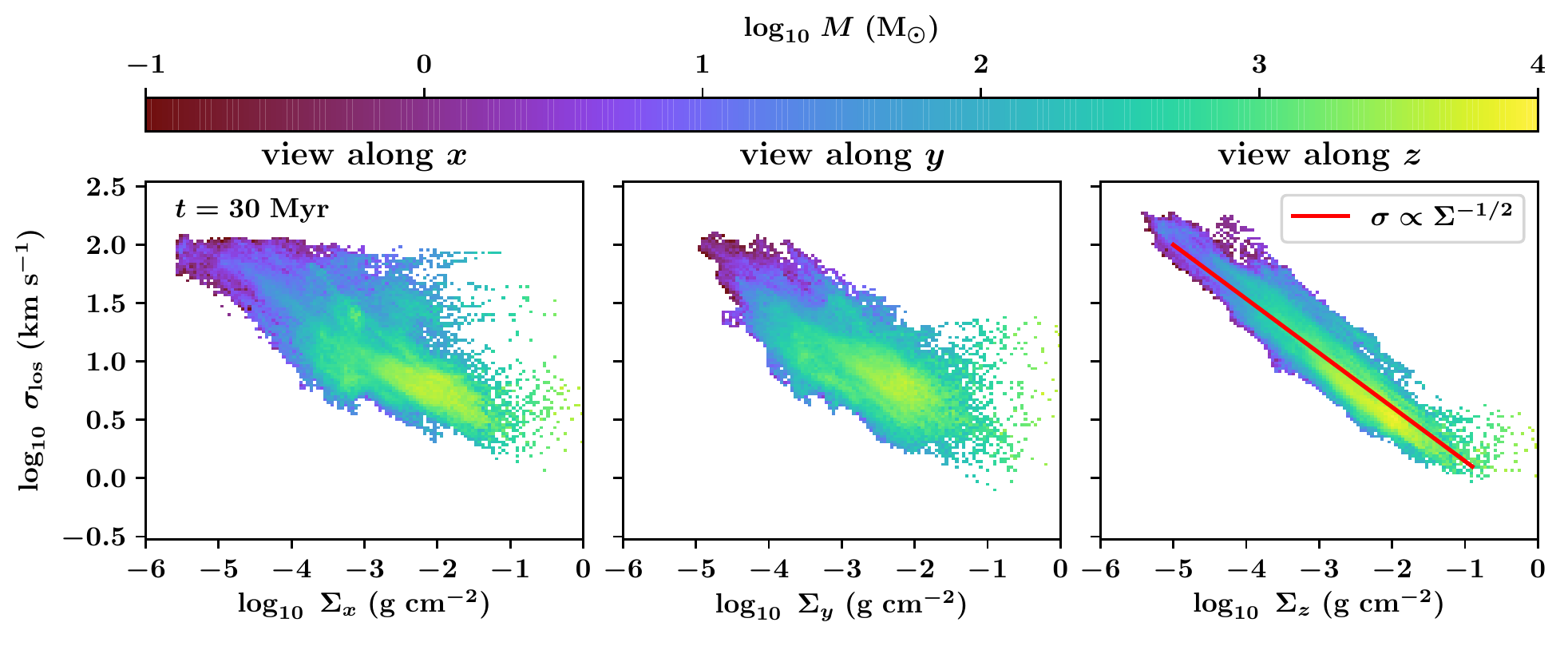}
\includegraphics[width=\textwidth]{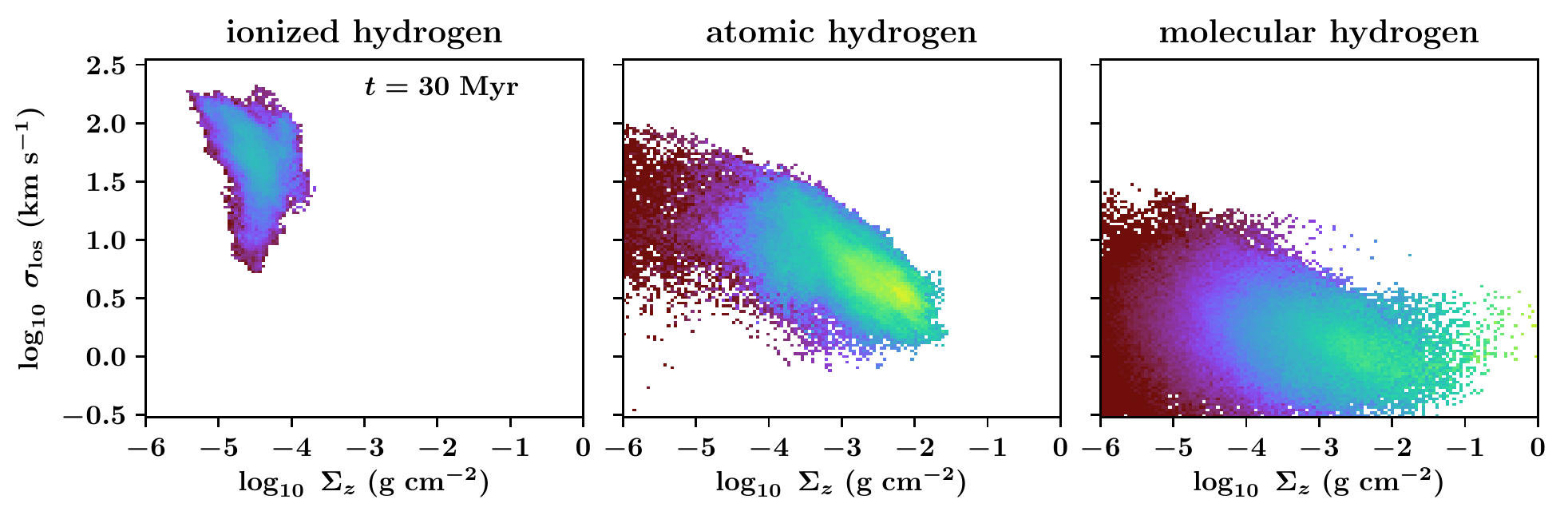}
\end{minipage}
\caption{Correlation of the line-of-sight velocity dispersion with the column density, projected along all principal orientations (top) and along $z$ split into chemical species (bottom) for simulation \texttt{B3-1pc} at $t=30\,\mathrm{Myr}$. For the total gas we note an anti-correlation of $\sigma$ with $\Sigma$, in particular along $z$. Split into chemical species, this correlation vanishes.}
\label{fig:coldens-los-correlation}
\end{figure*}

Before analysing the global dynamics in the box it is worth noting that the gas is initially at rest. The missing shear and the absent dynamical coupling to galactic dynamics in our setup results in little large scale motions besides the ones driven by the SNe and a vanishing centre-of-mass velocity.

We investigate the geometric weighted mean of the gas velocity
\begin{equation}
	\skl{v}_w = \exp\ekl{\rkl{\sum_iw_i}^{-1}\,\sum_i w_i\,\ln\rkl{|\mathbf{v}_i|}},
\end{equation}
where we use weights $w_i$ for the velocity of cell $i$, $v_i$. We distinguish between weighting by the cell mass ($w_i=m_i$), the cell volume ($w_i=V_i$), as well as by chemical mass fractions $X$, $w_i=m_i f_{\mathrm{X},i}$, for ionised, atomic and molecular hydrogen. We use the \emph{geometric} mean in order to avoid a disproportionately strong impact of a few cells with very large values. In addition we correct for outflow motions by investigating only the velocities in $x$ and $y$, $v_{xy} = (v_x^2+v_y^2)^{1/2}$, and computing an isotropic 3-dimensional velocity $v_\mathrm{3D,corr}=\sqrt{3/2}\,v_{xy}$. Fig~\ref{fig:gas-velocities-time} presents the time evolution of the gas velocities (lines) for mass and volume weighting in the top panel and chemical weights in the bottom. The corrected values excluding the outflow are shown in the grey area, which is bounded by the smallest and largest average of the three main simulations. The noticeable differences in the vertical structure and the amount of outflowing gas between the different magnetisations are not reflected in the globally averaged velocities. The temporal variations are larger than the differences between the different runs. The hot gas can increase the average speeds in the volume weighted case to values above $100\,\mathrm{km\,s}^{-1}$. Weighting by mass, atomic and molecular hydrogen shows a very similar evolution with a steady increase from a few to about $10\,\mathrm{km\,s}^{-1}$. The outflow motions have a noticeable effect on the global gas velocities. However, the denser gas (weighting by mass, H, and H$_2$) is not strongly affected, which causes the grey area to overlap with the lines. For the low-density gas (weighting by volume and H$^+$) the averages for the global and the corrected velocities vary by a factor of two.

In order to relate the velocities to the sonic and Alfv\'{e}nic speeds we also compute the Mach number,
\begin{equation}
	\skl{\mathcal{M}_\mathrm{s}}_w = \skl{\frac{|\mathbf{v}_i|}{c_{\mathrm{s},i}}}_w = \exp\ekl{\rkl{\sum_iw_i}^{-1}\,\sum_i w_i\,\ln\rkl{\frac{|\mathbf{v}_i|}{c_{\mathrm{s},i}}}},
\end{equation}
with the same weights $w_i$ as for the velocity, $\skl{v}_w$. Here, $c_{\mathrm{s},i}$ is the adiabatic sound speed of cell $i$, where we take the chemical composition into account and compute the cell-averaged thermal pressure according to the local abundances. The corresponding Alfv\'{e}nic Mach number reads
\begin{equation}
\label{eq:mach-alfven}
	\skl{\mathcal{M}_\mathrm{A}}_w = \skl{\frac{|\mathbf{v}_i|}{v_{\mathrm{A},i}}}_w = \exp\ekl{\rkl{\sum_iw_i}^{-1}\,\sum_i w_i\,\ln\rkl{\frac{|\mathbf{v}_i|}{v_{\mathrm{A},i}}}}
\end{equation}
with the Alfv\'{e}n speed $v_{\mathrm{A},i}=B_i/\sqrt{4\pi\rho_i}$ that depends on the magnetic field strength $B_i$ and the density $\rho_i$ in cell $i$. The weighting factors are again the same. Here, we directly use the velocities without subtracting a locally averaged mean velocity. The numbers thus include the effects of cloud motions and outflow velocities. The dynamics in the dense gas is investigated in Section~\ref{sec:clouds} in more detail. In Fig.~\ref{fig:Mach-number-time} we plot the geometric mean of the sonic Mach number (left) and the Alfv\'{e}nic Mach number (right) as a function of time. The upper panels show the mass and volume weighted quantities in the entire box. The lower panels depict the Mach numbers weighted by chemical mass fractions. The sonic Mach number indicates transsonic and supersonic motions in all cases. The volume weighted numbers indicate transsonic motions with small variations over time. The mass weighted sonic Mach number increases over time and flattens after $50\,\mathrm{Myr}$ at a value of $20$. Due to the time delay in the formation of clouds and dense gas the mass and H$_2$ weighted Mach numbers differ with magnetisation. The increase over time is relatively similar with a delay of the order of $10\,\mathrm{Myr}$ for the magnetic runs. After $50\,\mathrm{Myr}$ the curves for molecular gas flatten at a value of $20$. For atomic hydrogen the numbers slightly increase over time to reach $\mathcal{M}_\mathrm{s}\sim5$ with little variations. The ionised gas is slightly supersonic ($\mathcal{M}_\mathrm{s}\sim2-3$) with very little variations over time and between the non-magnetic and magnetic runs. The similarity between atomic and ionised hydrogen indicates that the faster velocity dispersion in H$^+$ compensates the hotter temperatures and the resulting sound speed. We note very similar numbers of $\mathcal{M}_\mathrm{s}$ for weighting by mass and molecular hydrogen as well as for the volume and the H$^+$ weighted mean.

We measure super-Alfv\'{e}nic motions with $\mathcal{M}_\mathrm{A}$ steadily increasing with time. For all weightings the differences between the initial magnetisation is not significant. The mass weighted values quickly reach unity and increase to $\sim50$ at the end of the simulation. The volume weighted numbers with overall lower magnetic fields and faster motions in the low-density cells start at values of order 5 and reach $\mathcal{M}_\mathrm{A}\sim100$ after $60\,\mathrm{Myr}$. Contrary to the sonic Mach number the relation between mass and volume weighted numbers and the chemistry weighted counterparts is more complicated. Atomic hydrogen has the lowest Mach number ranging from $1-8$. Molecular hydrogen corresponds to the densest gas with the strongest fields, which would lower $\mathcal{M}_\mathrm{A}$. However, the gravitational contraction and the corresponding acceleration as well as the inter-cloud motions outweigh the stronger field and the mass weighted Alfv\'{e}nic Mach number increases to $\sim50$. The non-negligible fraction of \emph{partially} ionised gas with stronger average fields compared to most of the volume results in Mach numbers for the ionised hydrogen of $\sim20$, between the values for molecular and atomic hydrogen. Concerning the volume and H$^+$ weighted Alfv\'{e}nic Mach numbers we may be limited by our small box compared to global disc models, which we discuss in Section~\ref{sec:discussion}.

As the simulations have multiple collapsing regions, we would like to separate motions related to the individual clouds and centres of collapse. In order to do so, we identify clumps and investigate their physical quantities in detail in Section~\ref{sec:clouds}. Here, we illustrate the local dynamics by means of the line-of-sight (los) velocity,
\begin{equation}
v_{\mathrm{los},k} = \rkl{\sum_i m_i}^{-1}\,\sum_i m_i v_{i,k},
\end{equation}
and the corresponding velocity dispersion
\begin{equation}
	\sigma_{\mathrm{los},k} = \ekl{\rkl{\sum_i m_i}^{-1}\,\sum_i m_i\rkl{v_{i,k} - v_{\mathrm{los},k}}^2}^{1/2}.
\end{equation}
We use a uniform grid with the maximum effective resolution of the simulation ($512^3$, i.e. a cell size of $0.98\,\mathrm{pc}$) and sum up the mass weighted ($m_i$) velocity ($v_{i,k}$) of cell $i$ along the $z$ and $y$ direction ($k=z, k=y$). The view along $x$ looks very similar despite the fact that the magnetic field is mainly oriented along the $x$ direction. Fig.~\ref{fig:coldens-los-vel} shows the los velocity, the dispersion and the correlation with the column density. The top plot shows simulation \texttt{B0-1pc}, the bottom one \texttt{B3-1pc} both at $t=30\,\mathrm{Myr}$. The left column depicts the column density projected along $z$ (top row) and $y$ (bottom row). The central panels show $v_\mathrm{los}$, the right ones $\sigma_\mathrm{los}$. The view along $z$ reveals that the largest line-of-sight velocities ($|v_\mathrm{los}|\gtrsim30\,\mathrm{km\,s}^{-1}$) are found in regions of low column density, which is consistent with the low-density gas being driven out of the midplane to form a fountain or outflow. Dense regions show smaller values of $|v_\mathrm{los}|<10\,\mathrm{km\,s}^{-1}$. For the view along $y$ we find overall lower numbers. This is not surprising as we average over multiple independent regions rather than along the stratification of the gas. The velocity dispersion shows a strong inverse correlation with the column density, in particular for the view along $z$. In low-density voids the values increase to $\sigma_\mathrm{los}\gtrsim100\,\mathrm{km\,s}^{-1}$. In the regions of molecular clouds $\sigma_\mathrm{los}$ can be as low as $\sim0.1\,\mathrm{km\,s}^{-1}$.

Fig.~\ref{fig:coldens-los-correlation} shows the correlation of the line-of-sight velocity dispersion with the column density of the gas for a projection along all three principal directions (top) as well as the projection along $z$ split into the three hydrogen components (bottom) for simulation \texttt{B3-1pc} at $t=30\,\mathrm{Myr}$. Over time more gas moves to higher column densities but the overall shape of the distribution does not change. The differences between the simulations are minor. The views along $x$ and $y$ look similar with a broad distribution and a weak anti-correlation of $\sigma_\mathrm{los}$ with $\Sigma_\mathrm{los}$. A wide spread is not surprising because we integrate over several individually contracting and independently moving regions. For the line-of-sight along $z$ we note a clear anti-correlation which we can approximate with a power-law,
\begin{equation}
\rkl{\frac{\sigma_\mathrm{los}}{\mathrm{km\,s}^{-1}}} \approx 0.3\,\rkl{\frac{\Sigma_\mathrm{los}}{\mathrm{g\,cm}^{-2}}}^{-1/2},
\end{equation}
independent of magnetisation and time, except for the highest column densities at late times, where gravitational collapse of the clouds keeps the velocity dispersion above $1\,\mathrm{km\,s}^{-1}$, see also Section~\ref{sec:clouds}. The scaling indicates a constant kinetic energy per unit area ($\Sigma\sigma^2=\mathrm{const.}$), which is likely to be connected to our SN driving scheme. We inject thermal energy at a constant rate and in an area-filling manner because the SNe have uniformly distributed random positions in $x$ and $y$. The fraction of clustered SNe is not dominating the positioning in the $xy$-direction. As a result, the energy injection rate is reflected in the vertical motions.

At the lowest column densities, which correspond to the hot low-density gas, we find $\sigma_\mathrm{los}\sim100\,\mathrm{km\,s}^{-1}$, which is in good agreement with the averaged volume-weighted box velocities (see Fig.~\ref{fig:gas-velocities-time}). At high column densities we find $\sigma_{\mathrm{los}}\sim1-10\,\mathrm{km\,s}^{-1}$ in good agreement with the velocity dispersion measured in individually identified clouds, see Section~\ref{sec:clouds}.

The strong correlation in $\Sigma_z-\sigma_z$ for the total gas is not observed, see for example \citet{HeyerEtAl2009}. However, there are several aspects that weaken this conflict. Firstly, our SN driving scheme is clearly idealised and simplified despite the fact that the spatial distribution and clustering properties are motivated by observations. Secondly, our box is not exposed to the larger galactic environment. Missing components like shear, galactic infall or warp of the disc reduce the overall dynamics within the box, which provides better conditions for a nicely correlated energy input by SNe. The projections along $x$ and $y$, which correspond to views much closer to the perspectives in our Galaxy and include integration effects through several independently moving regions, already indicate a weaker anti-correlation. Finally, observations rarely cover all the gas in a continuous fashion, but are rather biased towards a special component with possibly much weaker or vanishing correlation. In fact, split into individual chemical species the correlation is not visible any more. Ionised hydrogen occupies a relatively narrow low-$\Sigma$ regime with a high line-of-sight velocity dispersion. Atomic and molecular hydrogen cover a significantly broader range in $\Sigma$ with intermediate (H) and low (H$_2$) $\sigma_\mathrm{los}$, respectively.

\section{Magnetisation}
\label{sec:magnetisation}

\subsection{Global magnetic properties}

\begin{figure}
  \centering
  \includegraphics[width=8cm]{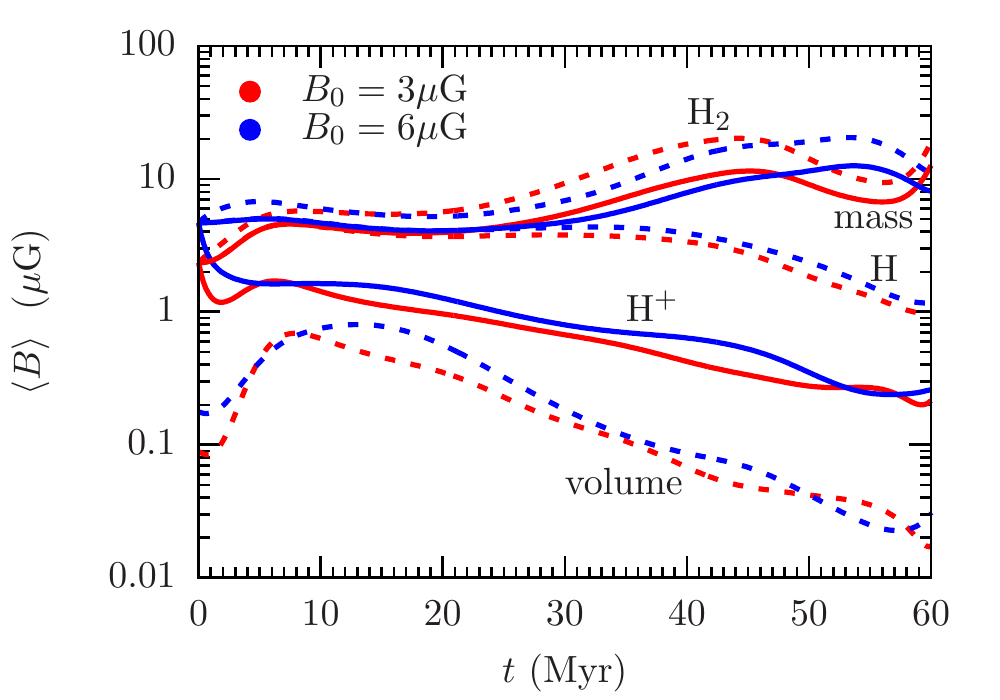}
  \caption{Time evolution of the magnetic field strength. The weighting by H$_2$ shows an increase of the field strength to $10-20\,\mu\mathrm{G}$. The H$^+$ weighted field decreases to $0.2\,\mu\mathrm{G}$ over time. The mean field in the atomic gas stays approximately constant.}
  \label{fig:L7-mag-field-strength}
\end{figure}

\begin{figure}
  \includegraphics[width=8cm]{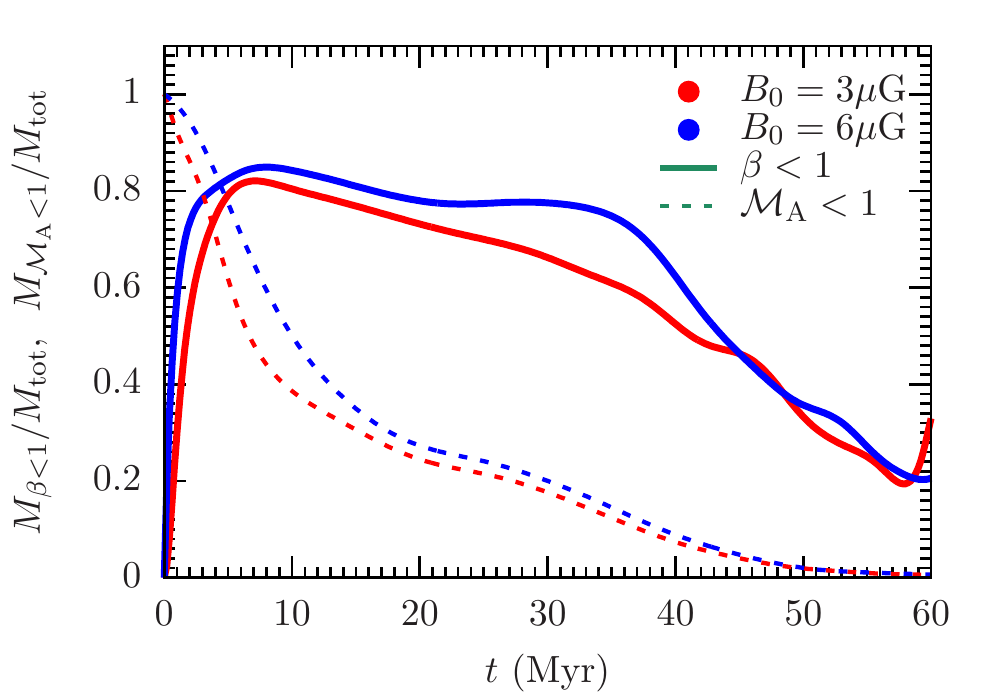}
  \caption{Time evolution of the the mass fraction of gas with $\beta<1$ and $\mathcal{M}_\mathrm{A}<1$. The fraction of low-$\beta$ gas slowly decreases from $80$ to $20\%$ over time. For the gas with low $\mathcal{M}_\mathrm{A}$ the curves decrease quickly during the first $10-20\,\mathrm{Myr}$ to $\sim30\%$ of the total mass. At the end of the simulation less than $5\%$ of the mass is at $\mathcal{M}_\mathrm{A}<1$.}
  \label{fig:L7-magdom-gas-comp}
\end{figure}

Fig.~\ref{fig:L7-mag-field-strength} depicts the time evolution of the globally averaged geometric mean of the magnetic field strength in the simulation box,
\begin{equation}
\skl{B}_w = \exp\ekl{\rkl{\sum_iw_i}^{-1}\,\sum_i w_i\,\ln |\mathbf{B}_i|}.
\end{equation}
The weights are again cell mass, cell volume and chemical mass fractions. The averaged mass weighted field starts at a few $\mu\mathrm{G}$ and slightly increases to values of $\sim10\,\mu\mathrm{G}$.  The differences in the initial field do not result in a systematic difference throughout the simulation. The H$_2$ weighted field behaves very similarly with slightly higher values. The field in the atomic hydrogen follows the curves of the mass weighted field and starts decreasing after $t\sim30\,\mathrm{Myr}$ to $1\,\mu\mathrm{G}$. H$^+$ and volume weighted fields are the weakest decreasing from $2$ to $0.2\,\mu\mathrm{G}$ and from $1$ to $\sim0.02\,\mu\mathrm{G}$ The combination of expanding volume filling SN remnants and the ideal MHD approximation are intuitive explanations for the decreasing values over time. However, as for the Alfv\'{e}nic Mach number, the small box and the missing large scale galactic dynamics might significantly alter the volume filling mean magnetic field, see Section~\ref{sec:discussion}. It is interesting to note that the morphological differences between the two magnetic runs are not reflected in the evolution of the magnetic field strength.


Initially, the gas in the midplane with a temperature of $5000\,\mathrm{K}$ is embedded in a magnetic field of $3$ and $6\,\mu\mathrm{G}$ for \texttt{B3-1pc} and \texttt{B6-1pc}, which corresponds to $\beta\sim8$ and $\beta\sim2$. Due to cooling most of the dense gas reaches $\beta<1$ after a few Myr, except for the hot SN remnants. The SNe create volume-filling high-$\beta$ regions in the box. Over time more volume is dominated by hot gas with $\beta\gg1$, only the dense regions remain at low $\beta$. The compression of the gas into dense regions and the accompanying amplification of the magnetic field results in a similar evolution for the gas with low $\mathcal{M}_\mathrm{A}$. However, the Alfv\'{e}n speed does not increase as much as the gas velocities, which results in very small regions of the simulation box with $\mathcal{M}_\mathrm{A}<1$. A more quantitative view is presented in Fig.~\ref{fig:L7-magdom-gas-comp} showing the total fraction of the gas with $\beta<1$ (solid lines) and $\mathcal{M}_\mathrm{A}<1$ (lashed lines). After $\sim10\,\mathrm{Myr}$ $80\%$ of the gas is cold enough to have $\beta<1$. This fraction slightly declines to $60-70\%$ at $30\,\mathrm{Myr}$ before it drops perceptibly to $20\%$ at the end of the simulation. The evolution is attributed to the effects of SNe, which do not just create low-density regions with little mass but also heat gas at intermediate densities which contains a noticeable fraction of the total mass and is dominated by thermal pressure. The regions with $\mathcal{M}_\mathrm{A}<1$ show a different evolution. The total fraction of gas with $\mathcal{M}_\mathrm{A}<1$ decreases quickly to 20 percent of the total mass after $30\,\mathrm{Myr}$ leaving a negligible fraction of the mass with sub-Alfv\'{e}nic motions ($<10^{-2}$) at the end of the simulation. However, we should not over-interpret this small fraction. As most of the gas is collected in a small region of the box with limited resolution, we are likely to underestimate the effects of a small-scale dynamo. In addition, the collapse motions around the molecular clouds are poorly resolved.

\subsection{Scaling properties with the density}

\begin{figure}
\includegraphics[width=8cm]{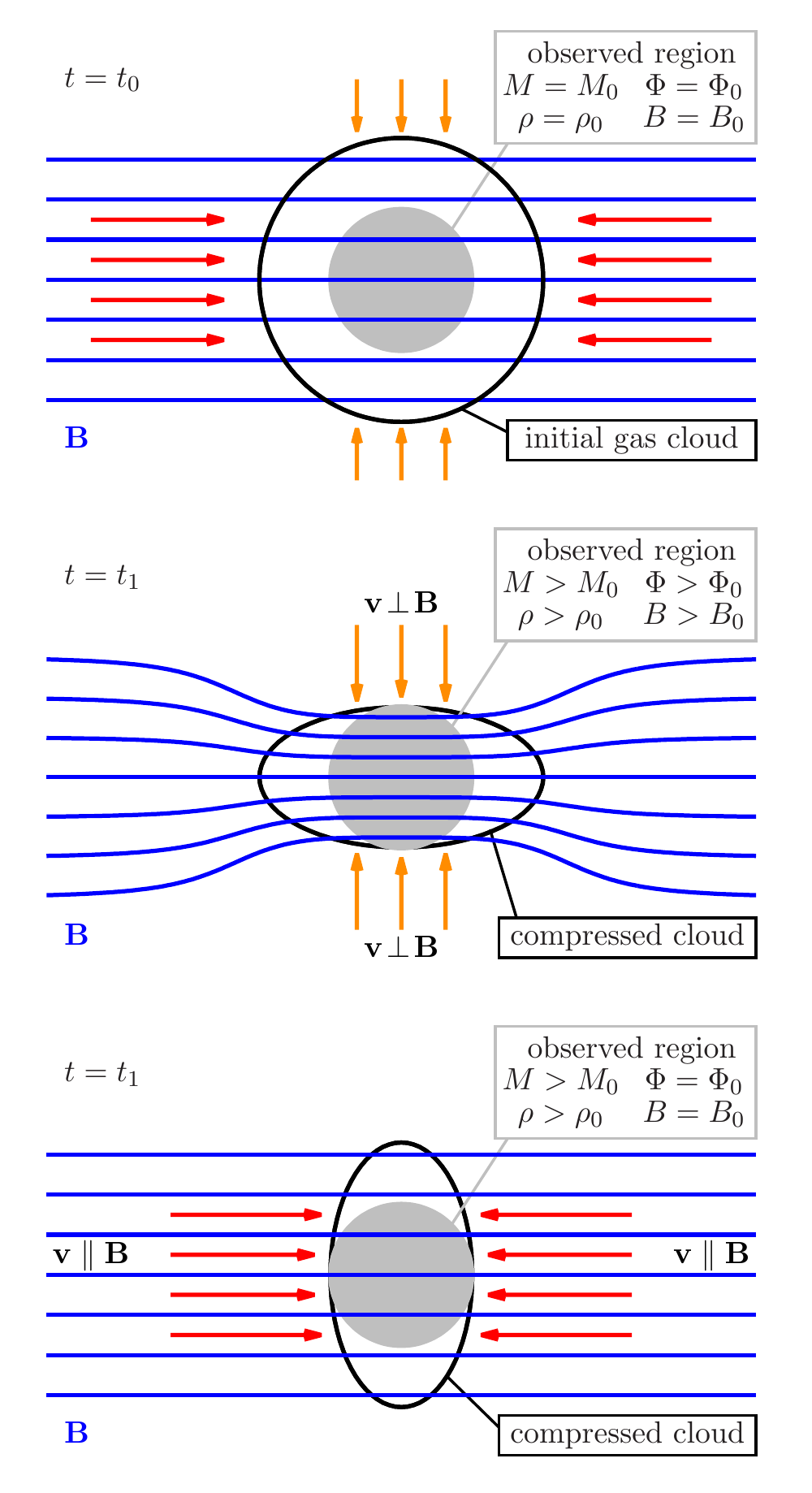}
\caption{Sketch of two modes of gas flow in the magnetised ISM. Top: Initial magnetic field configuration (blue lines), an initial gas cloud (black circle) and an investigated region (grey circle). Center: For contraction perpendicular to the magnetic field, the gas cloud is compressed together with the field lines. The observed region shows larger density and larger magnetic field strength. Bottom: Compression along the field lines increases the density in the observed region, but keeps the magnetic field strength constant.}
\label{fig:sketch-accretion-angle}
\end{figure}

\begin{figure}
  \includegraphics[width=8cm]{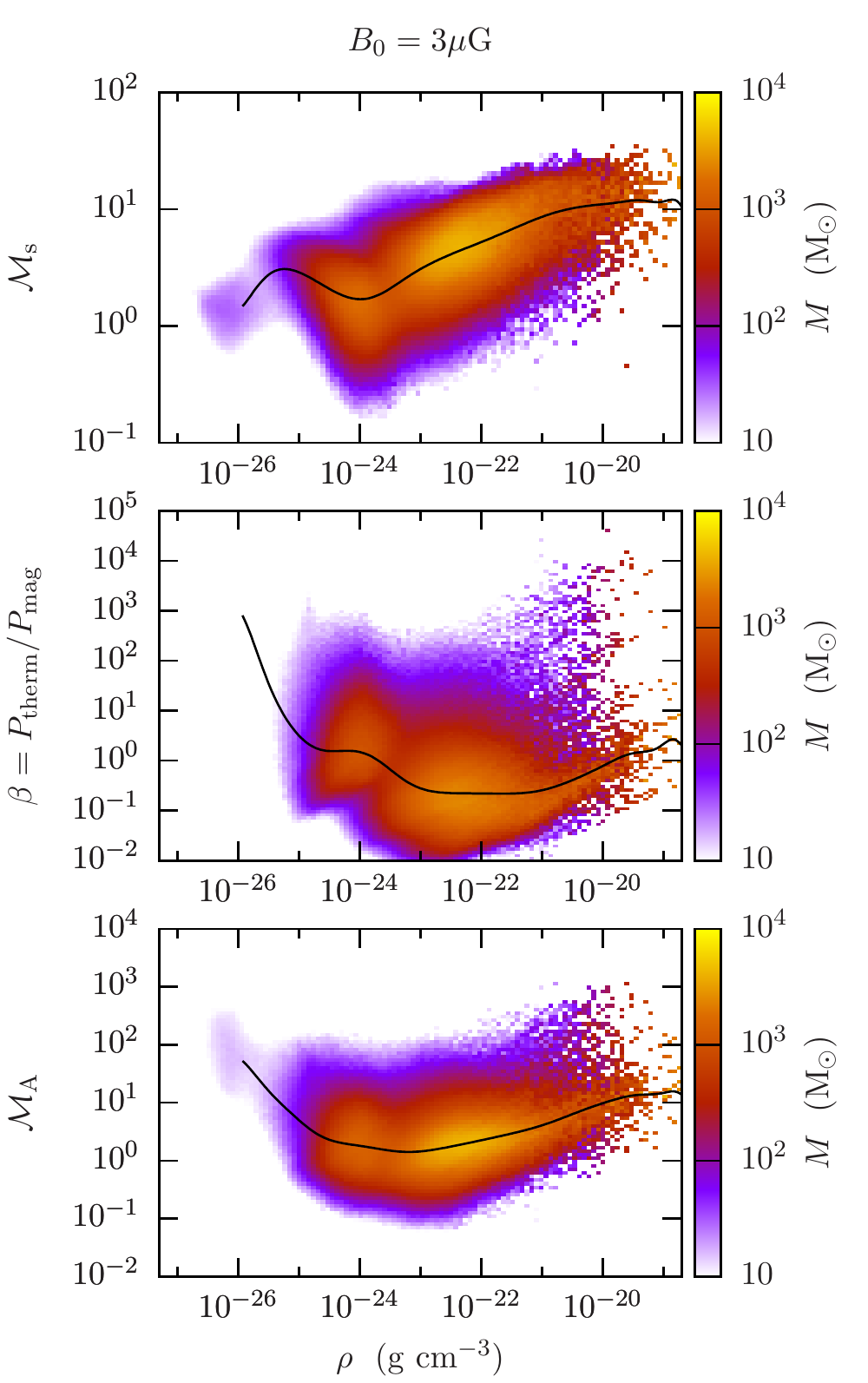}
  \caption{Scaling of different quantities with gas density. From top to bottom we show two-dimensional histograms of the sonic Mach number, the plasma $\beta$ parameter as well as the Alfv\'{e}nic Mach number for simulation \texttt{B3-1pc} at $t=30\,\mathrm{Myr}$. Colour-coded is the mass. The black lines show the median of each distribution. Most of the gas moves at supersonic and super-Alv\'{e}nic velocities with $\beta<1$.}
  \label{fig:L7-CM3-phase-plots}
\end{figure}

\begin{figure}
  \centering
  \includegraphics[width=8cm]{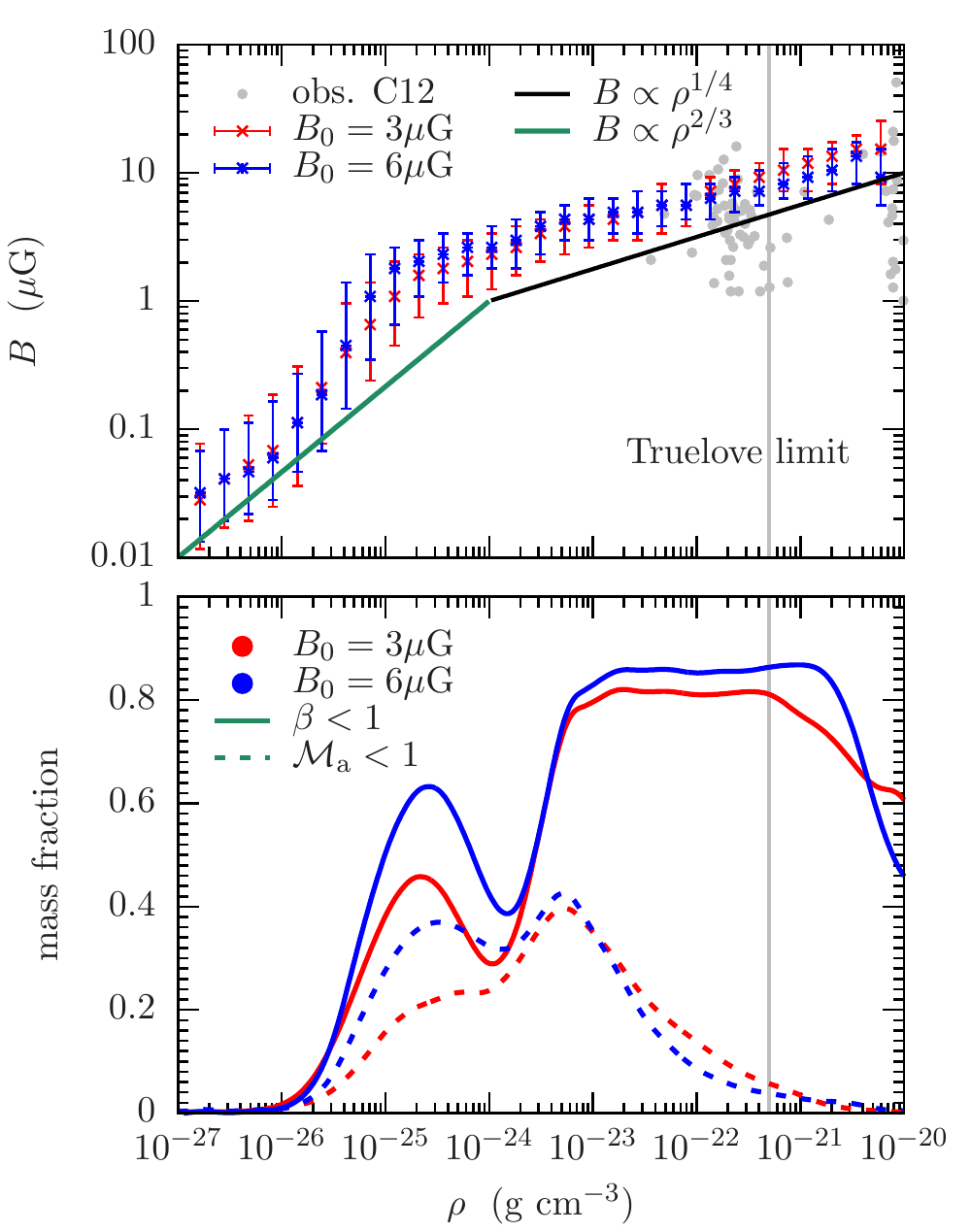}
  \caption{Top: Scaling of the magnetic field strength with the gas density at $t=30\,\mathrm{Myr}$. At low densities the scaling is steep, above $\rho\sim10^{-24}\,\mathrm{g\,cm}^{-3}$ a stable flat scaling of $B$ with $B\propto\rho^{1/4}$ develops. The grey points indicate the measurements by \citet{Crutcher2012}, which are in agreement with the simulations. Bottom: Fraction of the gas per density bin with $\beta<1$ (solid lines) and $\mathcal{M}_\mathrm{a}<1$ (dashed lines). The vertical lines marks the maximum density that we can resolve based on the Truelove criterion, see text.}
  \label{fig:L7-mag-field-scaling-density}
\end{figure}

Before discussing the scaling of the magnetic field with the density as well as the angle between the magnetic field vector and the velocity, we would like to stress an important property of the ideal MHD approximation. Fig.~\ref{fig:sketch-accretion-angle} illustrates two extreme cases of gas flow with respect to the magnetic field orientation. The top part of the figure presents the initial magnetic field configuration (blue lines), an initial gas cloud (black circle) and a region of fixed size that we investigate at different times (grey circle). Gas motions are indicated by the arrows, where we distinguish between flows along and perpendicular to the magnetic field lines. The centre part depicts the situation for a perpendicular flow. The field lines are compressed with the gas cloud. The observed region thus shows a larger density and magnetic field compared to the initial configuration ($\rho>\rho_0$, $B>B_0$) in that region. Compression along the field lines (bottom panel) results in a denser cloud with unaltered magnetic field configuration ($\rho>\rho_0$, $B=B_0$). For very strong magnetic fields the magnetic pressure will channel the flow and restrict it to the parallel configuration. Weak magnetic fields can easily be compressed by the gas flow. It is thus instructive to investigate the scaling of the field strength together with the energy ratios at different densities. Concerning the formation of dense clouds and cores, the distribution of the angle between gas velocity and magnetic fields is a quantitative measure of the accretion mode.

Two dimensional histograms of the gas distribution as a function of density are presented in Fig.~\ref{fig:L7-CM3-phase-plots} for \texttt{B3-1pc} at $t=30\,\mathrm{Myr}$. The distributions for \texttt{B6-1pc} are very similar. From top to bottom we plot $\mathcal{M}_\mathrm{s}$, plasma $\beta$ and $\mathcal{M}_\mathrm{A}$. Colour coded is the gas mass. The black lines are the median values along the ordinate. The sonic Mach number has a local minimum at $\rho\approx10^{-24}\,\mathrm{g\,cm}^{-3}$, which marks the low-density (high-temperature) end of the thermally unstable region. At lower densities the gas moves at larger velocities. For $\rho>10^{-24}\,\mathrm{g\,cm}^{-3}$ the median of the Mach number increases from transsonic motions to $\mathcal{M}_\mathrm{s}\sim10$, which is mainly due to cooling of the gas and the resulting decrease in the thermal sound speed. Plasma $\beta$ decreases from $\sim10^3$  at $\rho\sim10^{-26}\,\mathrm{g\,cm}^{-3}$ to unity at $\rho\sim10^{-24}\,\mathrm{g\,cm}^{-3}$ and below unity for densities up to $\rho\sim10^{-20}\,\mathrm{g\,cm}^{-3}$. At the highest densities the thermal pressure increases linearly with the density (isothermal behaviour of the gas). However, the magnetic pressure follows a flatter scaling with $\rho$ (see discussion below), which causes plasma $\beta$ to increase and exceed unity above $\rho\approx10^{-20}\,\mathrm{g\,cm}^{-3}$. The Alfv\'{e}nic Mach number (bottom panel) shows a similar overall behaviour. At low densities $\mathcal{M}_\mathrm{A}$ is of order 100. The region of the thermal instability marks the minimum with noticeable fractions of the gas located at sub-Alfv\'{e}nic speeds and a median of order unity. At high densities the strong increase in kinetic energy density outweighs the increase in magnetic pressure, which causes the ratio to increase again to reach median values of order 10.

We expect the scaling of the magnetic field strength to depend on the dynamical and thermal properties in different density regimes. In the case of very weak fields, turbulent motions and motions driven by thermal pressure (e.g. an expanding SN shell) dominate the evolution of the magnetic field. For compressive motions perpendicular to the field line the field strength in the MHD approximation scales directly with the density $B\propto\rho^\alpha$ with $\alpha=1$. For flows along the field line the strength is not altered, $\alpha=0$. Isotropic turbulence statistically results in a scaling of $B\propto\rho^{2/3}$ for super-Alfv\'{e}nic systems \citep{Mestel1966}. For sub-Alfv\'{e}nic motions the classical scaling law by \citet[][CF53]{ChandrasekharFermi1953} results in $\alpha=1/2$ or $\alpha\le1/2$ \citep[e.g.][]{MouschoviasCiolek1999}.

We analyse the strength of the magnetic field (Fig.~\ref{fig:L7-mag-field-scaling-density}, top) as well as the fraction of the gas that is magnetically dominated (bottom) as a function of density. The field strength is shown as the median values with errorbars indicating the 25 and 75 percentile of the distribution. The grey points show the low-density range of the observational estimates by \citet{Crutcher2012}, which are in good agreement with our simulated values. In the bottom panel we distinguish between gas with $\beta<1$ and gas with $\mathcal{M}_\mathrm{A}<1$. We also mark the maximum density we can resolve based on the Truelove criterion \citep{Truelove97}, which states that the Jeans length has to be resolved with at least 4 cells\footnote{We compute the maximum density using the isothermal sound speed, a mean molecular weight of 2.35 and a fiducial temperature of 50K. The cell size is taken at the highest level of refinement, i.e. $\Delta x=0.98\,\mathrm{pc}$.} in order to avoid artificial fragmentation. We compare simulations \texttt{B3-1pc} and \texttt{B6-1pc} at $t=30\,\mathrm{Myr}$. At low densities we note a steep scaling which can be considered approximately consistent with the statistical average for weak fields. In this density regime only a small fraction of the gas is magnetically dominated. At a density of $\rho\sim10^{-25}\,\mathrm{g\,cm}^{-3}$ the scaling becomes flatter with $\alpha\sim1/4$, which coincides with an increase in $B$-dominated gas. Above $\rho\sim10^{-23}\,\mathrm{g\,cm}^{-3}$ the relative fraction of gas with $\mathcal{M}_\mathrm{A}<1$ drops again, but with motions mainly parallel to the field line, so the field strength does not have to increase, see also next section. The simulations thus do not reproduce the classical scaling law of \citet{ChandrasekharFermi1953}, where $\alpha=1/2$. However, their model assumes sub-Alfv\'{e}nic turbulence, which is not the case in our simulations, where we have supersonic and super-Alfv\'{e}nic motions (see Fig.~\ref{fig:Mach-number-time}) and resulting MHD shocks, see also \citet{deAvillezBreitschwerdt2005}. Recent ISM simulations by \citet{PadoanEtAl2016} show a similar scaling behaviour compared to our runs. They find a very flat scaling exponent of $\alpha\approx0.13$ below $10^3\,\mathrm{cm}^{-3}$ and $\alpha\approx0.29$ above $10^3\,\mathrm{cm}^{-3}$, respectively. Fitting a power law to the data in the density range $10^{-24}<\rho/(\mathrm{g\,cm}^{-3})<10^{-20}$ yields scaling exponents between $\alpha\approx0.25-0.31$ for simulation \texttt{B3-1pc} and slightly smaller numbers of $\alpha\approx0.20-0.27$ for \texttt{B6-1pc}, all slopes increasing over time from $t=30-60\,\mathrm{Myr}$. A common feature in all of our simulations is a locally flatter scaling around the density of the thermal instability, $\rho\sim10^{-24}-10^{-23}\,\mathrm{g\,cm}^{-3}$. A fit over one order of magnitude around $\rho=10^{-23}\,\mathrm{g\,cm}^{-3}$ reveals scaling exponents that are $\approx0.06$ smaller with ranges of $\alpha=0.17-0.25$ for \texttt{B3-1pc} and $\alpha=0.13-0.21$ for \texttt{B6-1pc}, again measured from $t=30-60\,\mathrm{Myr}$. At ISM densities of around $10^{-22}\,\mathrm{g\,cm}^{-3}$ the median field strength is of order a few $\mu\mathrm{G}$ in our simulation, which is in good agreement with the observations by \citet{Crutcher2012}. Observationally, the field strength increases to values of up to a few hundred $\mu\mathrm{G}$ at molecular cloud densities of $10^{-18}\,\mathrm{g\,cm}^{-3}$, consistent with recent other numerical studies \citep{GentEtAl2013b, KimOstriker2015b, KoertgenBanerjee2015, PadoanEtAl2016, IffrigHennebelle2017, EvirgenEtAl2017} but not resolved in our setups.

We should add a word of caution with regard to the densest regions and the credibility of the strongest magnetic fields because of magnetic dissipation and numerical reconnection. Close to the resolution limit, i.e. in entities that are resolved with $\lesssim10$ cells \citep[see comments in][]{Hennebelle11}, the magnetic fields are prone to numerical diffusion and (turbulent) reconnection \citep{LazarianVishniac1999,KowalEtAl2009,KowalEtAl2012}, which causes flux to be lost. \citet{LazarianEsquivelCrutcher2012} conclude that the reconnection model is consistent with observations and that there is little evidence for subcritical self-gravitating clouds. Although there is gas with low Alfv\'{e}nic Mach numbers in our simulations, most of the gas is dominated by the kinetic energy density, both infall as well as turbulent motions. This means that the centres of the clouds at the resolution limit are likely to be affected. In which way is difficult to estimate as the loss of magnetic flux will be (partially) balanced by a magnetic dynamo, which is underresolved inside our clouds.

Our measured range in magnetic field strength is consistent with observations in the ISM. \citet{Crutcher2012} reports values ranging from $\sim1-100\,\mu\mathrm{G}$ for densities up to $\rho=10^{-20}\,\mathrm{g\,cm}^{-3}$. Stronger fields of up to a $\mathrm{mG}$ are only found in gas that is denser than what we simulate. At a density of $10^{-20}\,\mathrm{g\,cm}^{-3}$ the median field is $\sim20\,\mu\mathrm{G}$, which is consistent with recent observations of Taurus \citep{ChapmanEtAl2011}, Serpens South \citep{SugitaniEtAl2011}, clouds in the Goult Belt \citep{LiEtAl2013}, the Lupus~I  molecular cloud \citep{FrancoAlves2015} or the Polaris Flare \citep{PanopoulouPsaradakiTassis2016}.

\subsection{Alignment of the magnetic field with the velocity}
\label{sec:alignment-B-v}

\begin{figure}
\includegraphics[width=8cm]{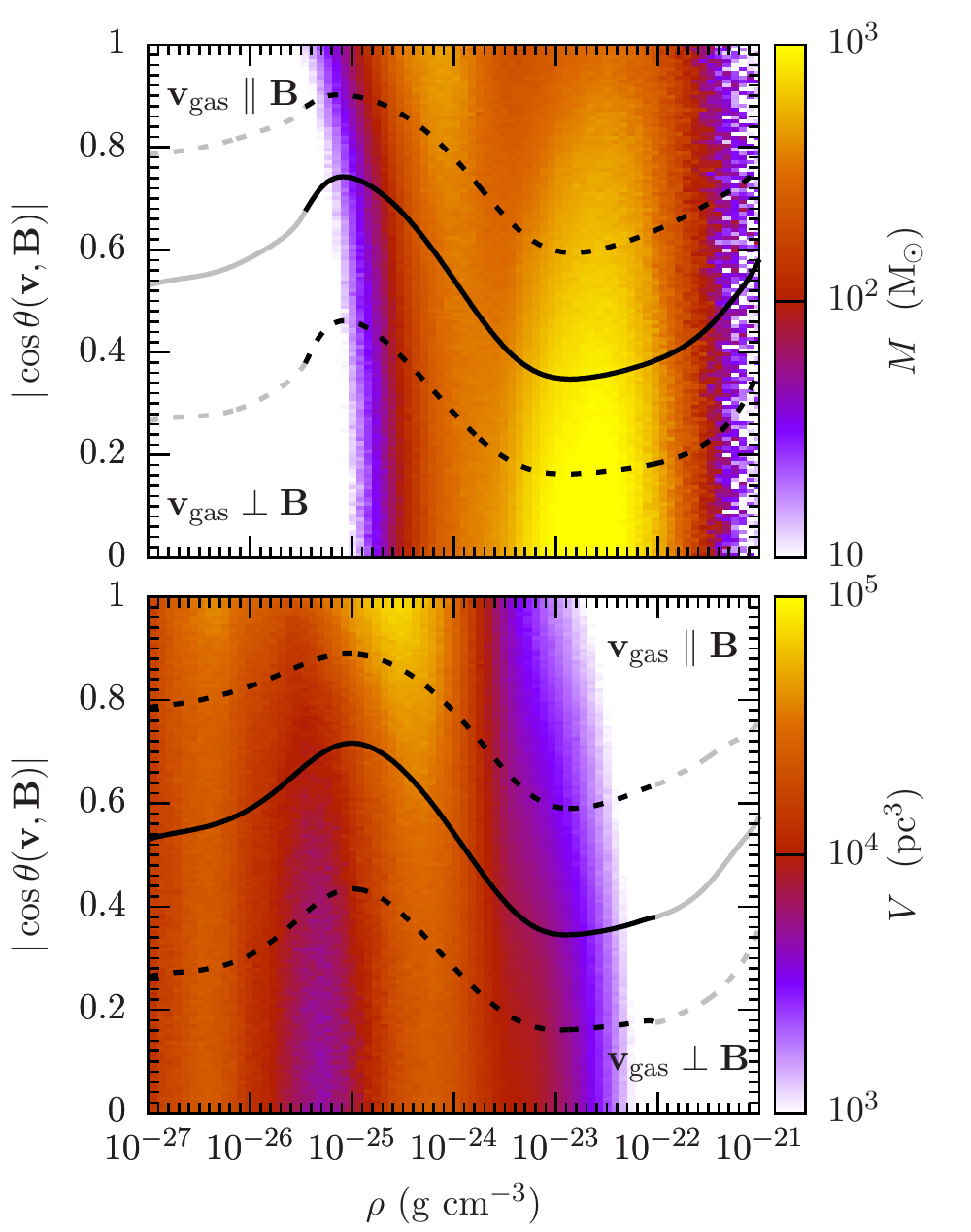}
\caption{Distribution of the angle between velocity and magnetic field vector at $t=20\,\mathrm{Myr}$ for simulation \texttt{B6-1pc} as a function of density. Colour-coded is the mass (top panel) and the volume (bottom panel). We also plot the median (solid lines) together with the 25 and 75 percentile (dashed lines) along the ordinate. The low-density gas at $10^{-26}<\rho/\mathrm{g\,cm}^{-3}<10^{-24}$ moves preferentially along the field lines, the high density gas ($10^{-24}<\rho/\mathrm{g\,cm}^{-3}<10^{-22}$) mainly perpendicular to the magnetic field.}
\label{fig:L7-angle-vB-global-density}
\end{figure}

In order to investigate the direction of the flow with respect to the magnetic field lines, we compute the normalised distribution function of the angle between velocity and magnetic field vector,
\begin{equation}
\cos\theta(\mathbf{v}, \mathbf{B}) = \frac{\mathbf{v}\cdot\mathbf{B}}{|\mathbf{v}|\,|\mathbf{B}|}.
\end{equation}
As in ideal MHD the sign of the orientation of the field lines does not matter -- a flow parallel to the field is numerically and physically the same as an anti-parallel flow -- we use this symmetry and plot $|\cos\theta|$. In Fig.~\ref{fig:L7-angle-vB-global-density} we show the distribution of angles for the entire simulation box for simulation \texttt{B6-1pc} at $t=20\,\mathrm{Myr}$. Colour coded are the mass (top panel) and the volume (bottom panel). The solid line marks the median of the distribution along the $|\cos\theta|$ distribution. The dashed lines are the 25 and 75 percentile. The median varies between 0.4 and 0.6 and becomes flatter over time (not shown). For simulation \texttt{B3-1pc} the variations are similar but less pronounced. Gas at the lowest and highest densities does not move with any preference with respect to the magnetic field orientation. In the range $10^{-26}<\rho/\mathrm{g\,cm}^{-3}<10^{-24}$ there is a tendency for a flow parallel to the field lines. The denser gas ($10^{-24}<\rho/\mathrm{g\,cm}^{-3}<10^{-22}$) moves with a slight preference perpendicular to the field lines. This is an effect that coincides mostly with the overall thickening of the disc, the launching of outflows at early times, and gravitational contraction of the disc at late times, where gas is moved along $z$ together with the field lines, which are mainly oriented along the $x$-direction. The evolution changes slightly over time with flatter distributions at later times, but the general trend remains. This behaviour is very similar to the results in \citet{IffrigHennebelle2017}, see their Fig.~11 and 12. We emphasise that there is a difference between the globally averaged distribution of angles and the distribution in dense clouds, where the overall effect of the thickening of the disc is not visible. There, most of the mass in the accretion flow is funneled along the field lines, which we investigate in more detail in Section~\ref{sec:clouds}. 

\section{Formation of shielded gas and molecular hydrogen}
\label{sec:optthick}

\begin{figure}
  \centering
  \includegraphics[width=8cm]{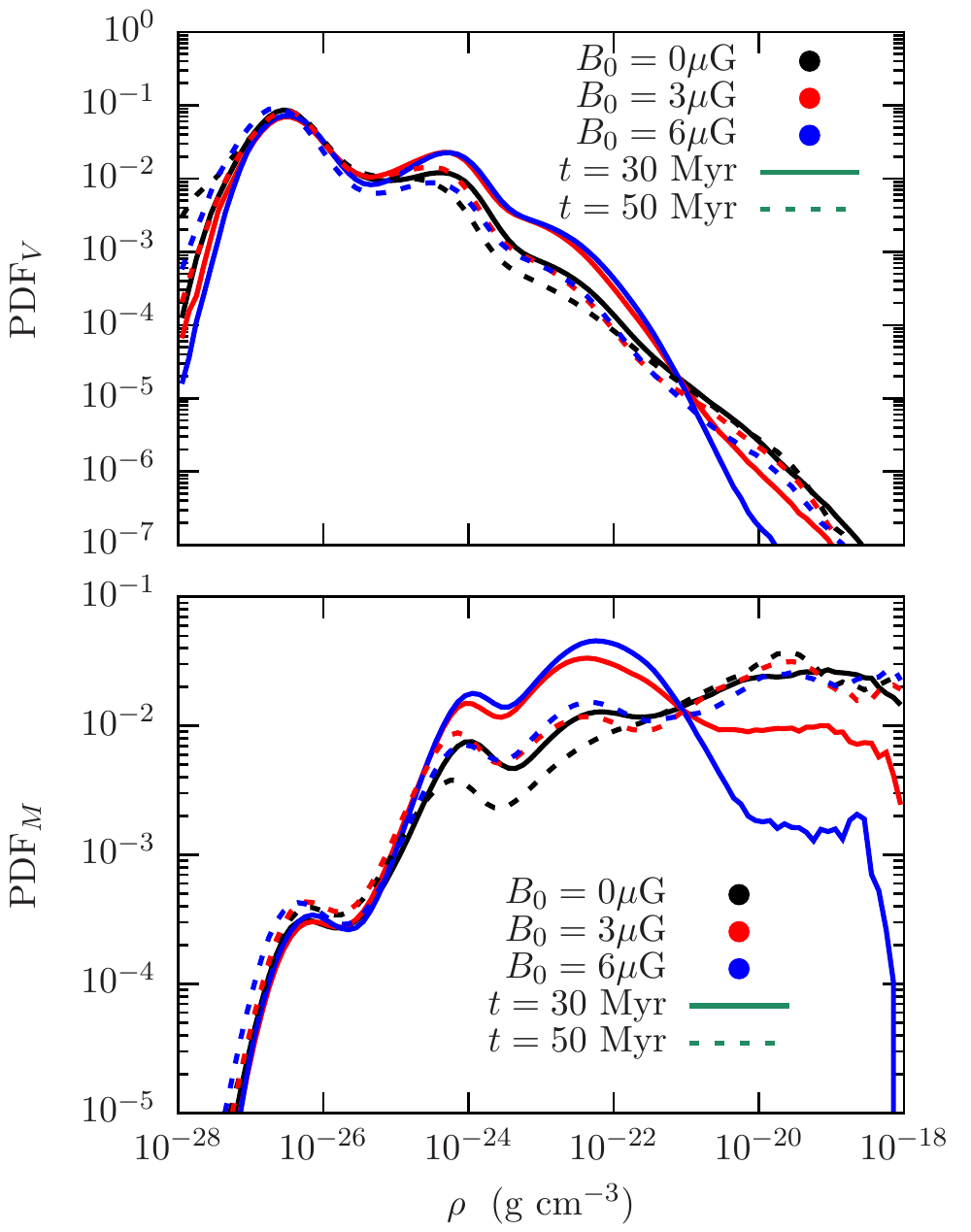}
  \caption{Volume weighted (top) and mass weighted density PDF (bottom) at $30$ (solid lines) and $50\,\mathrm{Myr}$ (dashed lines). The volume weighted PDF shows features of a lognormal distribution as well as a high-density tail. The mass weighted PDF emphasises the differences at high densities. The magnetic support leads to initial differences in the early evolutionary phase (solid lines). Once dense gas has formed the variations in the morphology and the vertical gas structure do not have a significant imprint on the density PDF (dashed lines).}
  \label{fig:L7-density-pdf}
\end{figure}

\begin{figure*}
  \includegraphics[width=0.85\textwidth]{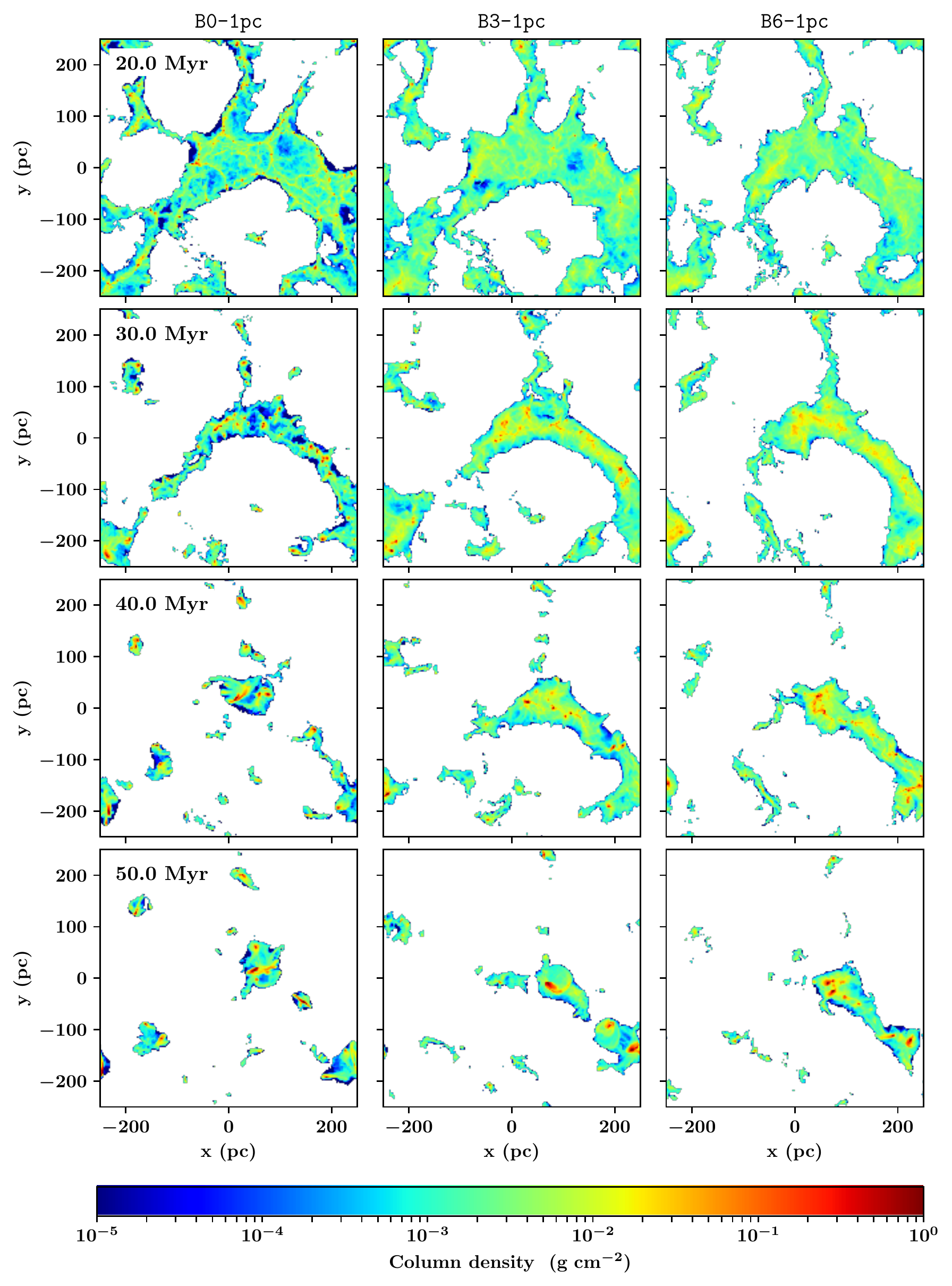}
  \caption{Column density of the shielded gas ($A_\mathrm{V}>0.3$) for simulations \texttt{B0-1pc} (left), \texttt{B0-1pc} (centre), \texttt{B0-1pc} (right) at $20$, $30$, $40$, and $50\,\mathrm{Myr}$ (top to bottom). The locations of $A_\mathrm{V}>0.3$ gas coincide with the locations of dense gas, see Fig.~\ref{fig:L7-coldens-face}.}
  \label{fig:L7-coldens-optically-thick}
\end{figure*}

\begin{figure}
\centering
\includegraphics[width=8cm]{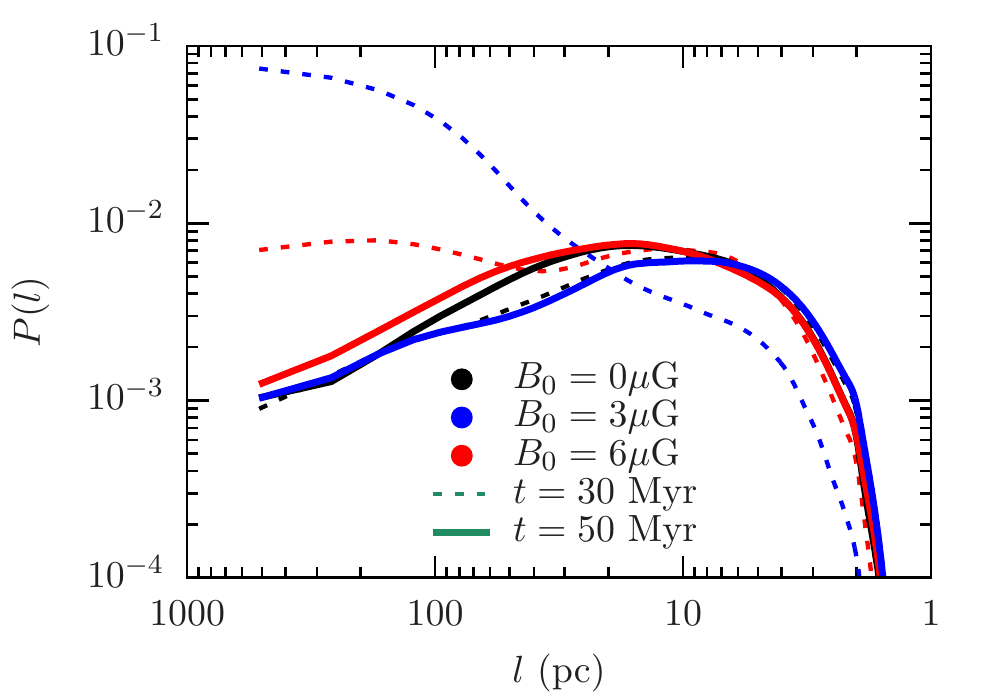}
\caption{Power spectra of the column density map in the $xy$-plane of gas with $A_\mathrm{V}>0.3$ at two different times. At $30\,\mathrm{Myr}$ (dashed lines) the magnetic field causes the regions to be more extended compared to the non-magnetic run. After $50\,\mathrm{Myr}$ (solid lines) all spectra are similar with a peak at around $20-30\,\mathrm{pc}$.}
\label{fig:L7-power-spectra-optically-thick}
\end{figure}

\subsection{Dense and shielded gas}

This section is dedicated to the details of the formation of dense gas, the correlation with shielding from the interstellar radiation field and the formation of molecular gas. The visual impression of the differences in the column density (cf. Fig.~\ref{fig:L7-coldens-face} and \ref{fig:L7-coldens-edge}) manifests in a significant delay in the formation of dense gas in the magnetised simulations. In Fig.~\ref{fig:L7-density-pdf} we present the volume weighted (top) and mass weighted (bottom) probability distribution function (PDF) of the total gas density at $30$ (solid lines) and $50\,\mathrm{Myr}$ (dashed lines). The volume weighted PDF shows typical features like the broad lognormal-like maxima at $\rho\sim10^{-27}$ and $\rho\sim10^{-25}\,\mathrm{g\,cm}^{-3}$, which is a natural result of turbulent gas dynamics \citep[e.g.][]{Vazquez94, PassotVazquezSemadeni1998}. In addition there is a high-density excess, which is attributed to the gravitational attraction and local collapse \citep[e.g.][]{Klessen2000, SlyzEtAl2005, KainulainenEtAl2009, SchneiderEtAl2012, GirichidisEtAl2014}. The earlier snapshot indicates that the distribution in the low-density gas shows similarities in all simulations. The non-magnetic run converts more gas into dense regions with $\rho>10^{-20}\,\mathrm{g\,cm}^{-3}$ indicated by the broad peak in the mass weighted PDF with an order of magnitude larger masses compared to \texttt{B6-1pc}. After $50\,\mathrm{Myr}$ of evolution the PDFs look very similar for all three simulations, so statistically the contraction to high densities seems to only experience a time delay because of magnetic fields, but no significant difference in the PDF after dense gas has started to form.

The formation of dense gas in clouds is accompanied by the formation of regions that are shielded from the interstellar radiation field. In the following we consider gas with an attenuation of $A_\mathrm{V}>0.3$ as \emph{shielded}, where $A_\mathrm{V}$ is computed in every grid cell by averaging over multiple lines of sight by means of the \textsc{TreeCol} algorithm. This is approximately the threshold for the formation of molecular hydrogen \citep{RoelligEtAl2007, GloverEtAl2010}. Fig.~\ref{fig:L7-coldens-optically-thick} shows the column density of the regions with $A_\mathrm{V}>0.3$ at different times for simulations \texttt{B0-1pc} (left), \texttt{B3-1pc} (centre), and \texttt{B6-1pc} (right). The maps reflect the impact of the magnetic field on the clumping properties. We can estimate the characteristic size of the regions by computing the two-dimensional power spectrum of the column density map in the $xy$-plane with $A_\mathrm{V}>0.3$, see Fig.~\ref{fig:L7-power-spectra-optically-thick} for the spectra at $30$ and $50\,\mathrm{Myr}$. The corresponding spectra for the $xz$-plane look very similar. In the magnetic runs the gas accumulates in larger coherent structures, which quantitatively becomes apparent in the power. At $30\,\mathrm{Myr}$ the strongly magnetised run is dominated by the power on the largest scales. For \texttt{B3-1pc} the spectrum is relatively flat. In the non-magnetic case the dense clumps show up as a broad peak at around $10\,\mathrm{pc}$. At $50\,\mathrm{Myr}$ the spectra look very similar with most of the power at spatial scales of $\sim20\,\mathrm{pc}$. We interpret this position as the characteristic size for molecular structures and as our fiducial \emph{cloud radius} for the subsequent analysis.

\subsection{Correlation between shielded and molecular gas}

\begin{figure}
  \centering
  \includegraphics[width=8cm]{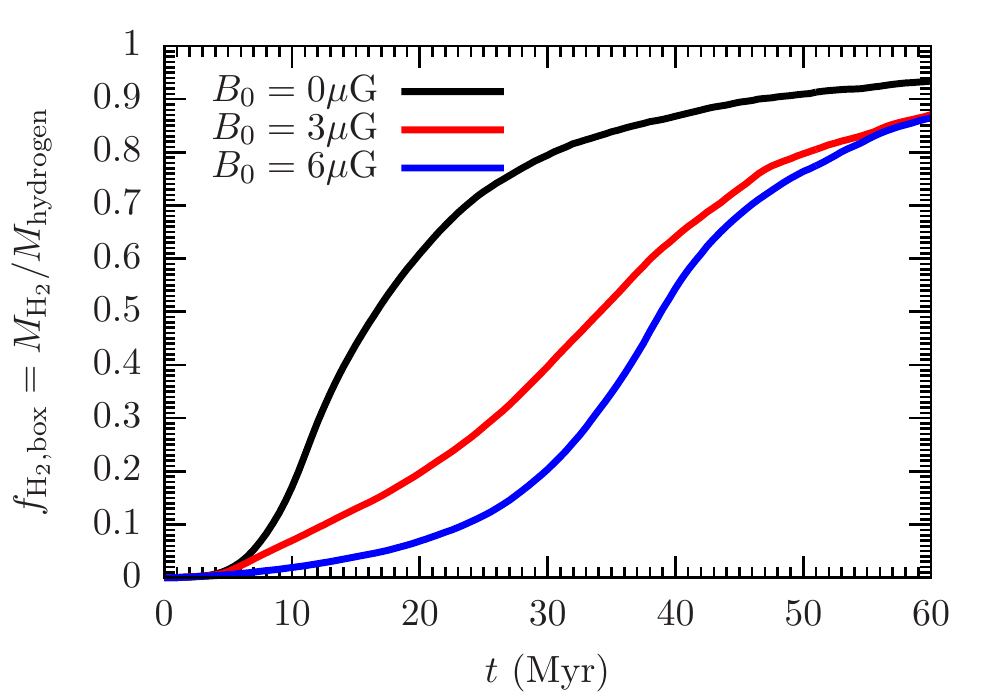}
  \caption{Time evolution of the fraction of molecular hydrogen. The magnetised runs show a delay in H$_2$ formation by approximately $25\,\mathrm{Myr}$. At the end of the simulation the majority of the gas is molecular with a slightly higher fraction for the non-magnetic run. The differences between the magnetic runs are minor and vanish at the end of the simulation.}
  \label{fig:L7-H2-time-evol}
\end{figure}

\begin{figure}
\centering
\includegraphics[width=8cm]{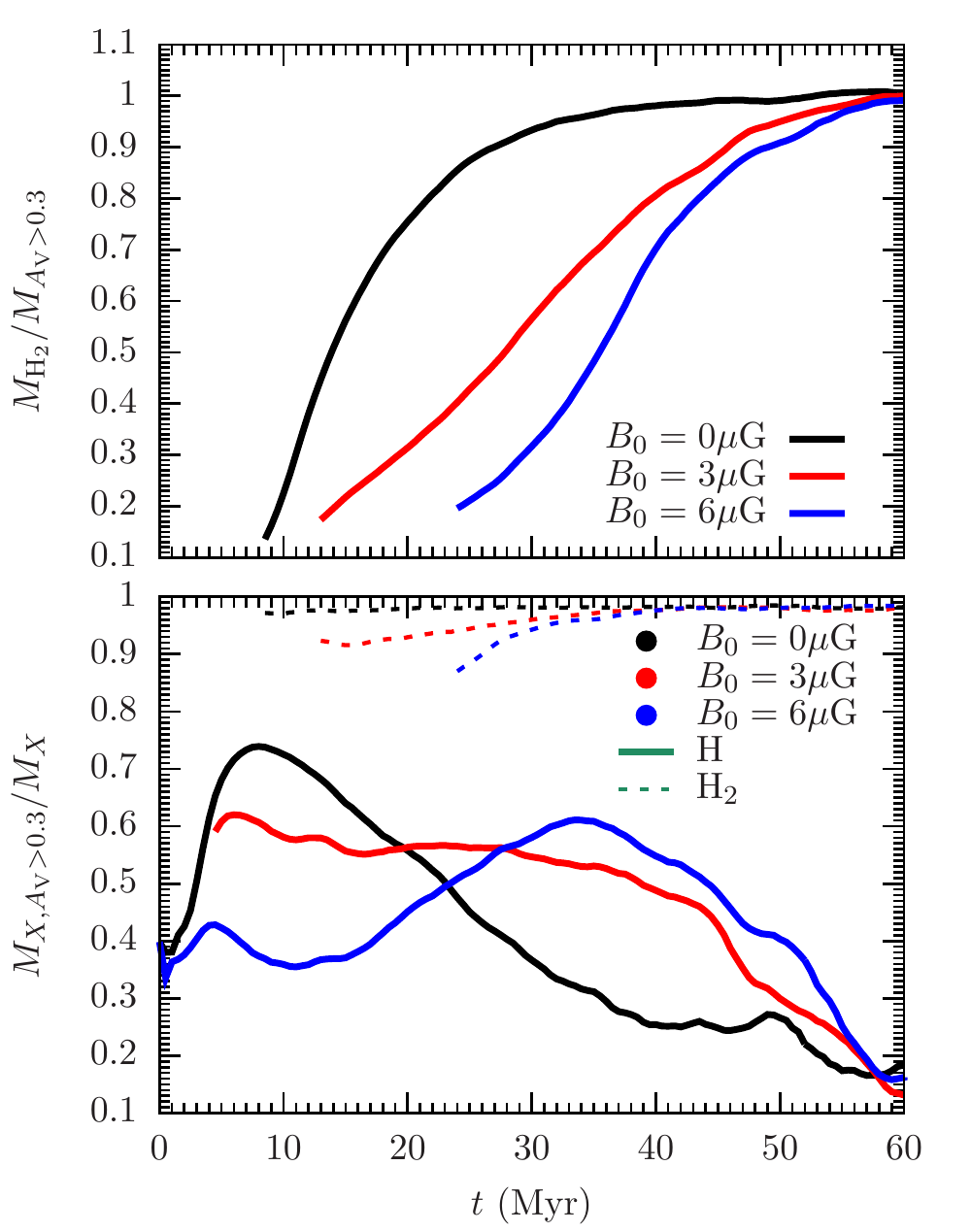}
\caption{Time evolution of the gas with $A_\mathrm{V}>0.3$ over time. We only plot the H$_2$ related data once their total mass fraction reached 10\% of the total mass in order to avoid numerical noise. Top: Time evolution of the ratio of molecular gas mass over total shielded gas mass. The total mass of gas at $A_\mathrm{V}>0.3$ increases faster then the total mass of molecular hydrogen. At late times the ratio approaches unity for all simulations. Bottom: Ratio of gas with $A_\mathrm{V}>0.3$ over total mass in each species (molecular and atomic gas). The fraction of shielded atomic gas varies between $20$ and $70\%$. In all simulations $\gtrsim90\%$ of the molecular gas is at positions with $A_\mathrm{V}>0.3$.}
\label{fig:L7-H2-mass-opt-mass-inv}
\end{figure}

The total fraction of molecular gas in the simulation box as a function of time is plotted in Fig.~\ref{fig:L7-H2-time-evol}. The stronger the magnetic field is the longer it takes to form molecular hydrogen. Towards the end of the simulation the majority of the gas in the box is in the form of H$_2$. The non-magnetic run reaches a fraction of $f_{\mathrm{H}_2,\mathrm{box}}\equiv M_{\mathrm{H}_2}/M_{\mathrm{hydrogen}}=0.95$. The magnetic runs show a delay in the formation of molecular gas, which vanishes after $\sim50\,\mathrm{Myr}$ to approach a total fraction of $f_{\mathrm{H}_2,\mathrm{box}}=0.85$.

Fig.~\ref{fig:L7-H2-mass-opt-mass-inv} (top) shows the ratio of molecular gas over total shielded gas. The bottom panel shows the fraction of shielded atomic ($M_\mathrm{H}(A_\mathrm{V}>0.3)/M_\mathrm{H}$) and molecular gas ($M_{\mathrm{H}_2}(A_\mathrm{V}>0.3)/M_{\mathrm{H}_2}$). We only plot the ratio if the mass fraction of molecular gas exceeds 10\% of the total mass in order to avoid numerical noise in the case of negligible H$_2$ gas. At early times the ratio (top) is small and increases to reach unity at the end of the simulations. This illustrates that first gas with $A_\mathrm{V}>0.3$ condenses out of the ISM before molecular gas forms within those regions. At late times the ratio approaches unity in all cases, which means that there is the same amount of molecular and gas with $A_\mathrm{V}>0.3$. The bottom panel shows the relative fractions of shielded gas in molecular and atomic gas. The lines for H$_2$ are close to unity, which emphasises that basically all molecular gas forms in shielded regions. The atomic gas has a more complicated evolution. The non-magnetic run forms $70\%$ of H with $A_\mathrm{V}>0.3$ at $t\approx10\,\mathrm{Myr}$. After that molecular gas forms out of the shielded gas and the fraction declines to give only $15\%$ at the end of the simulation. Weakly magnetised gas (\texttt{B3-1pc}) results in approximately half of the atomic hydrogen being shielded for about $40\,\mathrm{Myr}$ before dense condensations leave the majority of the atomic gas in a diffuse state that is not shielded. In the case of \texttt{B6-1pc} the fraction of atomic gas with $A_\mathrm{V}>0.3$ is initially lower ($40\%$) and increases to a peak fraction of $60\%$ at $35\,\mathrm{Myr}$, which coincides with the onset of H$_2$ formation. At the end of the simulations the differences are small for all runs.

\section{Fragmentation and Cloud Statistics}
\label{sec:clouds}

\subsection{Number of fragments}

\begin{figure}
  \includegraphics[width=8cm]{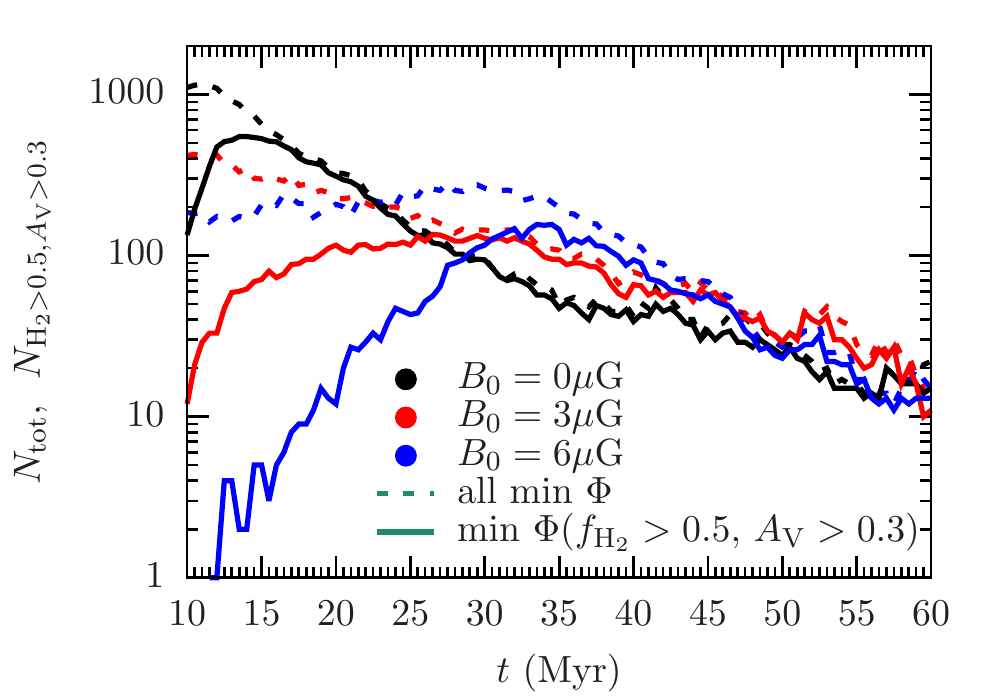}\\
  \includegraphics[width=8cm]{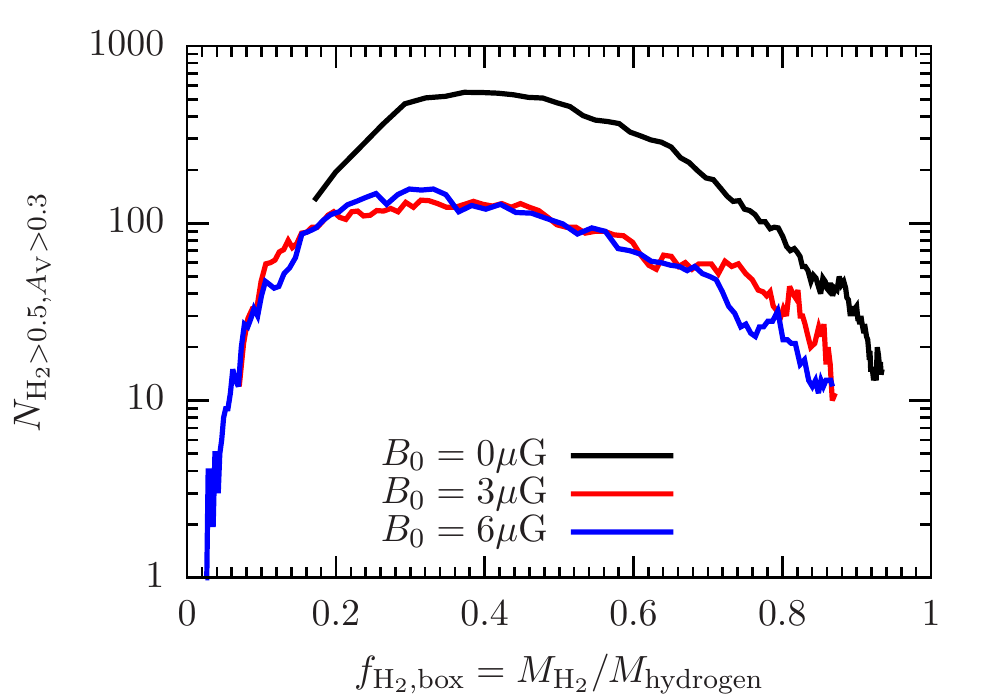}
  \caption{Time evolution of the total number of local minima of the gravitational potential (top, dashed lines) and the number with the additional constraints of $A_\mathrm{V}>0.3$ and $f_{\mathrm{H}_2}>0.5$ (top, solid lines). After $20$ (\texttt{B0-1pc}), $30$ (\texttt{B3-1pc}), and $45\,\mathrm{Myr}$ (\texttt{B6-1pc}) basically all clouds are molecular and shielded from the interstellar radiation field and both lines converge. Bottom: Number of molecular clouds as a function of the global H$_2$ mass fraction. The non-magnetic run forms up to 500 molecular clouds. The numbers in both magnetic runs are almost indistinguishable with a maximum of 100 at $f_{\mathrm{H}_2,\mathrm{box}}=0.3$.}
  \label{fig:L7-clump-number}
\end{figure}

We identify clouds by finding the local minima in the gravitational potential. In order to avoid effects due to initial transient fluctuations, we exclude the first $10\,\mathrm{Myr}$ from this analysis. In addition to this necessary criterion, we also require that at the position of the local minima the gas is shielded ($A_\mathrm{V}>0.3$) and molecular ($f_{\mathrm{H}_2}>0.5$) in order to be identified as a \emph{molecular} cloud. We have chosen the gravitational potential to identify clouds because it is less prone to temporal fluctuations compared to the density field \citep{SmithClarkBonnell2009}. There is a variety of other algorithms to identify clouds and connected structures, among which \textsc{Clumpfind} \citep{WilliamsDeGeusBlitz1994} is probably the most popular one, however, with weaknesses \citep{PinedaRosolowskyGoodman2009, Berry2015}. We apply \textsc{Clumpfind} to our data sets and compare the differences in Section~\ref{sec:defining-clouds}. However, as it has more free parameters and shows strong dependencies of the results on them, we refrain from using \textsc{Clumpfind} for the main analysis.

Fig.~\ref{fig:L7-clump-number} shows the total number of minima in the potential (top, dashed lines) as well as the number with the additional restrictions (top, solid lines). The non-magnetic run quickly develops numerous ($\sim1000$) local minima in the potential, which merge during the evolution. The magnetised runs show fewer clouds with only $\sim400$ and $\sim200$ for \texttt{B3-1pc} and \texttt{B6-1pc}. In simulation \texttt{B6-1pc} the number of minima remains constant for about $25\,\mathrm{Myr}$. Due to dissolution and merging of the individual condensations the number decreases during the simulation and drops to about $10-20$ with very little difference between the runs. The number of \emph{molecular} gravitational minima strongly varies between the runs in the first half of the simulated time. This is expected because the magnetic runs need a longer time before H$_2$ forms, so the secondary selection criteria are not fulfilled. After $t\sim35\,\mathrm{Myr}$ the difference between the runs vanishes, see also Fig.~\ref{fig:L7-coldens-edge}. We also note that after the formation of even a small fraction of molecular gas the additional constraints are not very restrictive, i.e. the majority of all gravitational minima are molecular and have $A_\mathrm{V}>0.3$ after $t\sim35\,\mathrm{Myr}$.

The differences in the dynamical evolution become apparent if we show the number of molecular gravitational minima as a function global H$_2$ fraction in the box (see Fig.~\ref{fig:L7-clump-number}, bottom). The two magnetic runs show a very similar behaviour concerning the total number as well as the evolution with increasing fraction of molecular gas. We note a peak in the distribution at around $f_{\mathrm{H}_2,\mathrm{box}}=0.4$ with $\sim100$ clouds. In the non-magnetic run, in which the gas is less filamentary and forms more spherical clouds, the peak is at almost 500 molecular clouds.   

\subsection{Fiducial cloud radius}

In order to analyse the \emph{formation} of molecular clouds from the very beginning we only use the gravitational minima without the other restrictions of optical thickness and molecular fractions unless explicitly indicated. We investigate a spherical volume of $20\,\mathrm{pc}$ radius around the minima of the gravitational potential as the fiducial cloud size. The motivation for the size is the peak in the spectra of the column density (cf. Fig.~\ref{fig:L7-power-spectra-optically-thick}). However, as the peaks are very broad and we would like to include not only the densest parts of the clouds, we increase the investigated volume and use the spectral peak at $20\,\mathrm{pc}$ as the \emph{radius} instead of the \emph{diameter} of the analysis volume. In addition we also probe radii of $10$ and $30\,\mathrm{pc}$, which are presented and discussed in the appendix (Section~\ref{sec:clump-radius-variation}). We would like to stress that clouds can encompass other local minima of the gravitational potential besides the one in their centre. Despite variations in cloud sizes, probing radii from $10-30\,\mathrm{pc}$ seems to be a reasonable choice consistent with observations \citep{SolomonEtAl1987, BolattoEtAl2008, MivilleDeschenesMurrayLee2017}. Simulations using a very similar setup to ours \citep{IbanezMejiaEtAl2016, IbanezMejiaEtAl2017} identify cloud radii with more sophisticated methods and find similar sizes ($\sim5-30\,\mathrm{pc}$). In addition, the temporal changes of the cloud radii in \citet{IbanezMejiaEtAl2016} are less than a factor 2 for an evolution of more than 2 free-fall times after they switched on self-gravity, see dotted lines in their Fig.~7. Also simulations of a global disc \citep[e.g.][]{TaskerWadsleyPudritz2015} find a distribution of cloud radii with a prominent peak at $20\,\mathrm{pc}$.

\subsection{Cloud masses and dynamics}

\begin{figure}
\centering
\includegraphics[width=8cm]{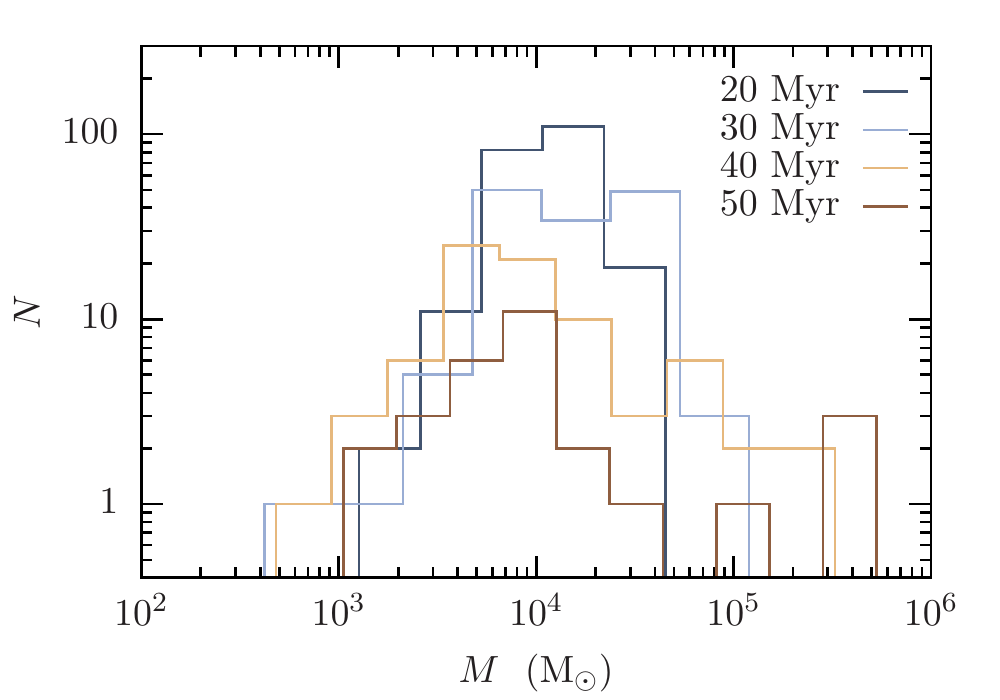}
\caption{Mass distribution of clouds for simulation \texttt{B3-1pc} at different times. The peak of the distribution is at around $10^4\,\mathrm{M}_\odot$. Over time the mass function shifts to more massive clouds. The mass of the most massive MC only increases.}
\label{fig:L7-clump-mass-function}
\end{figure}

\begin{figure}
\centering
\includegraphics[width=8cm]{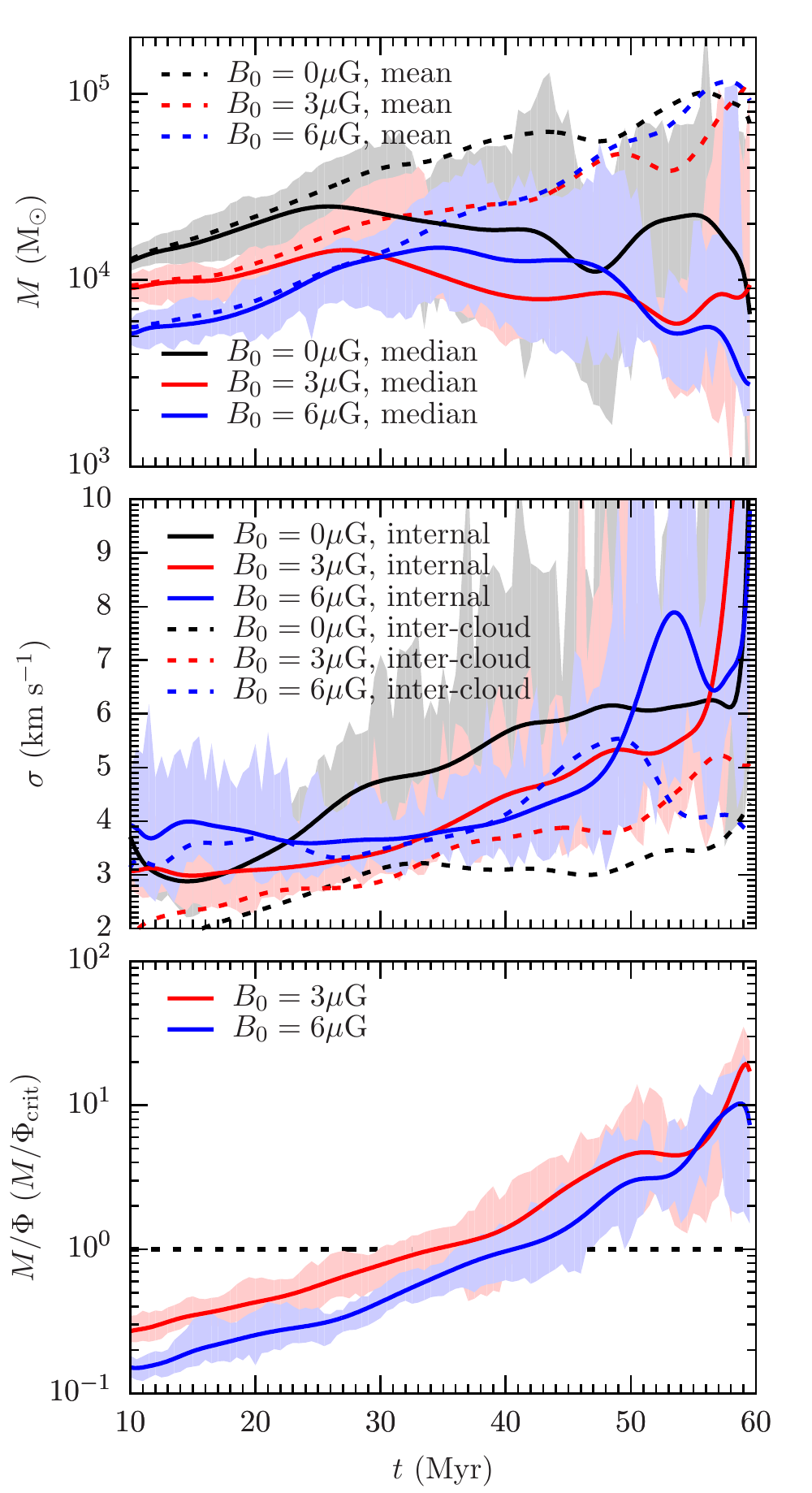}
\caption{Median and mean cloud mass (top), internal velocity dispersion (centre), and mass-to-flux ratio (bottom) over time for simulations \texttt{B0-1pc}, \texttt{B3-1pc}, and \texttt{B6-1pc}. The mean cloud mass increases due to the formation of a few very massive clouds, cf. Fig.~\ref{fig:L7-clump-mass-function}. The internal velocity dispersion increases due to accretion and merging of clouds. The mass-to-flux ratio in units of the critical value indicates that the clouds start as subcritical objects and turn into supercritical clouds.}
\label{fig:L7-clump-time-evol}
\end{figure}

Fig.~\ref{fig:L7-clump-mass-function} shows the mass function of MCs for simulation \texttt{B3-1pc} at different times. The evolution for the other simulations are similar. At $t=20\,\mathrm{Myr}$ the distribution is narrow ranging from $10^3-4\times10^4\,\mathrm{M}_\odot$ and widens to $\sim10^3-5\times10^5\,\mathrm{M}_\odot$ after $50\,\mathrm{Myr}$. Clouds merge and massive clouds accrete gas yielding final masses of up to $5\times10^5\,\mathrm{M}_\odot$. Overall the number of clouds is relatively small in particular at the end of the simulation, so a comparison of features of the distribution is inappropriate. Also, the entire gas mass in our simulations is $2.5\times10^6\,\mathrm{M}_\odot$, which is lower than the mass of observed massive GMCs. However, a few global numbers seem to agree with observational data. \citet{MivilleDeschenesMurrayLee2017} find median cloud masses of $\sim4\times10^{4}\,\mathrm{M}_\odot$ for all clouds with more massive clouds being located closer to the Galactic centre. Observations of clouds in the Goult Belt show as well median masses of the order of a few $10^4\,\mathrm{M}_\odot$ \citep{Loren1989, Cambresy1999, PinedaEtAl2010, LeeEtAl2015}.

Fig.~\ref{fig:L7-clump-time-evol} shows the arithmetic mean and median mass of all clouds (top), the mass weighted internal and inter-cloud velocity dispersion (centre) as well as the mass-to-flux ratio (bottom) as a function of time. In all three panels the solid lines are the median of the distribution. The shaded area is bounded by the 25 and 75 percentile. The mean mass in the top panel (dashed lines) is dominated by the most massive clouds and increases from $10^4\,\mathrm{M}_\odot$ to $\sim10^5\,\mathrm{M}_\odot$. The formation of low-mass clouds influences the median mass, which increases by a factor of $2-3$ in the first half of the simulation time. The merging -- inparticular of massive clouds in the second half of the simulation time -- causes the media to decrease approximately to the initial values at the end. For the velocity dispersion we distinguish between the mass weighted internal dispersion and the inter-cloud velocity dispersion. The former one is computed as
\begin{equation}
\sigma_\mathrm{internal} = \sqrt{\sum_k\sigma_{\mathrm{int},k}^2}.
\end{equation}
Here, $\sigma_{\mathrm{int},k}$, $k\in x, y, z$ are the mass weighted velocity dispersions in each direction,
\begin{equation}
\sigma_{\mathrm{int},k} = \frac{1}{M_\mathrm{cloud}}\sum_i m_i(v_{i,k}-v_{\mathrm{com},k})^2,
\end{equation}
where we sum over all the cells $i$ within a radius of $20\,\mathrm{pc}$ around the local gravitational minimum with $m_i$ being the cell mass, $M_\mathrm{cloud}$ the cloud mass, and $\mathbf{v}_\mathrm{com}$ is the centre-of-mass velocity of the cloud. The inter-cloud dispersion is computed analogously,
\begin{equation}
\sigma_\mathrm{inter-cloud} = \sqrt{\sum_k\sigma_{\mathrm{ic},k}^2}
\end{equation}
and
\begin{equation}
\sigma_{\mathrm{ic},k} = \rkl{\sum_j M_{\mathrm{cloud},j}}^{-1}\sum_j M_{\mathrm{cloud},j}(v_{\mathrm{com},j,k}-v_{\mathrm{com-all},k})^2.
\end{equation}
Here we weigh by the individual cloud masses $M_{\mathrm{cloud},j}$ and measure the dispersion with respect to the centre-of-mass velocity of all clouds, $v_{\mathrm{com-all},k}$. The internal velocity dispersions are similar for the magnetic and non-magnetic runs with values of the order of $3-4\,\mathrm{km\,s}^{-1}$ at $t=10\,\mathrm{Myr}$ and $5-6\,\mathrm{km\,s}^{-1}$ at $t\sim50\,\mathrm{Myr}$. These averages over the identified clumps with a radius of $20\,\mathrm{pc}$ are in good agreement with the simple line-of-sight estimates (cf. Fig~\ref{fig:coldens-los-vel}). At $t=60\,\mathrm{Myr}$ the velocity dispersion increases dramatically indicating the dominating impact of gravitational collapse, see also the results of high-resolution cloud MC1 in \citet{SeifriedEtAl2017}. The inter-cloud velocity dispersions are overall smaller than the internal values. We find the biggest difference between $\sigma_\mathrm{internal}$ and $\sigma_\mathrm{inter-cloud}$ in the non-magnetic run, which has the largest internal and the lowest intra-cloud velocity dispersion over the majority of the simulated time. This is in agreement with the morphological evolution in the simulation. The fast formation of dense cold clouds allows for locally more efficient gravitational contraction and stronger accretion flows. However, the clouds are all close to the midplane and the external potential does not accelerate them towards the midplane.

The fact that the inter-cloud velocity dispersion is smaller than the internal one is not surprising considering that we start with the gas being at rest and that the simulations do not include the effects of differential rotation of the galactic disc. However, including the galactic environment still seems to be consistent with our results. Simulations of an entire disc by \citet{DobbsPringleDuarteCabral2015} find cloud merging velocities of $\sim2-10\,\mathrm{km\,s}^{-1}$, which in agreement of what we find without galactic shear. In addition, those values are computed for directly merging clouds and are likely to be higher than the average inter-cloud velocities. Observations reveal slightly higher velocities ($7-20\,\mathrm{km\,s}^{-1}$) for interacting clouds \citep{FurukawaEtAl2009, OhamaEtAl2010, ToriiEtAl2011, ToriiEtAl2015, FukuiEtAl2014, FukuiEtAl2015, FukuiEtAl2016, SaigoEtAl2017}. However, in those observations there is also a focus on cloud-cloud collisions, which possibly marks the higher end of inter-cloud velocities. In fact, computing the maximum of all pairwise relative velocities reveals that the clouds are approaching each other with velocities in the range of $10-30\,\mathrm{km\,s}^{-1}$, in good agreement with the observational values.

The bottom panel shows the ratio of mass to magnetic flux in units of the critical ratio \citep{MouschoviasSpitzer1976,NakanoNakamura1978},
\begin{equation}
\rkl{\frac{M}{\Phi}}_\mathrm{crit} = \frac{1}{2\pi G^{1/2}},
\end{equation}
where $G$ is the gravitational constant. Objects with a ratio below the critical value are supported by magnetic pressure and are expected not to collapse. Above the critical number gravitational forces dominate. For the two magnetic runs the clouds start as subcritical objects with lower values for stronger magnetic fields as expected. After $t\sim35-40\,\mathrm{Myr}$ the median mass-to-flux ratio of the clouds exceeds the critical value and grows to values of order $10$ at $t=60\,\mathrm{Myr}$. From first-order theoretical considerations of the ideal MHD approximation, the mass-to-flux ratio can only change by mass flowing along the magnetic field lines. Mass flow perpendicular to the field lines compresses the field and increases the magnetic flux. However, magnetic dissipation and turbulent reconnection can provide an efficient loss of magnetic flux in dense turbulent clouds, which can increase the mass-to-flux ratio without significant accretion. Observationally, there seem to be more supercritial than subcritial clouds \citep{Crutcher2012}. This is in agreement with the cloud properties in the second half of the simulation. In the first $35\,\mathrm{Myr}$ the majority of our clouds is subcritical. This apparent disagreement might stem from the different analysis focus in our simulations compared to the observations. Diffuse structures that clearly appear as gravitational minima in the simulation might still be too diffuse to be classified as proper clouds and thus be excluded as observational targets.

\subsection{Mass accretion}

\begin{figure}
\centering
\includegraphics[width=8cm]{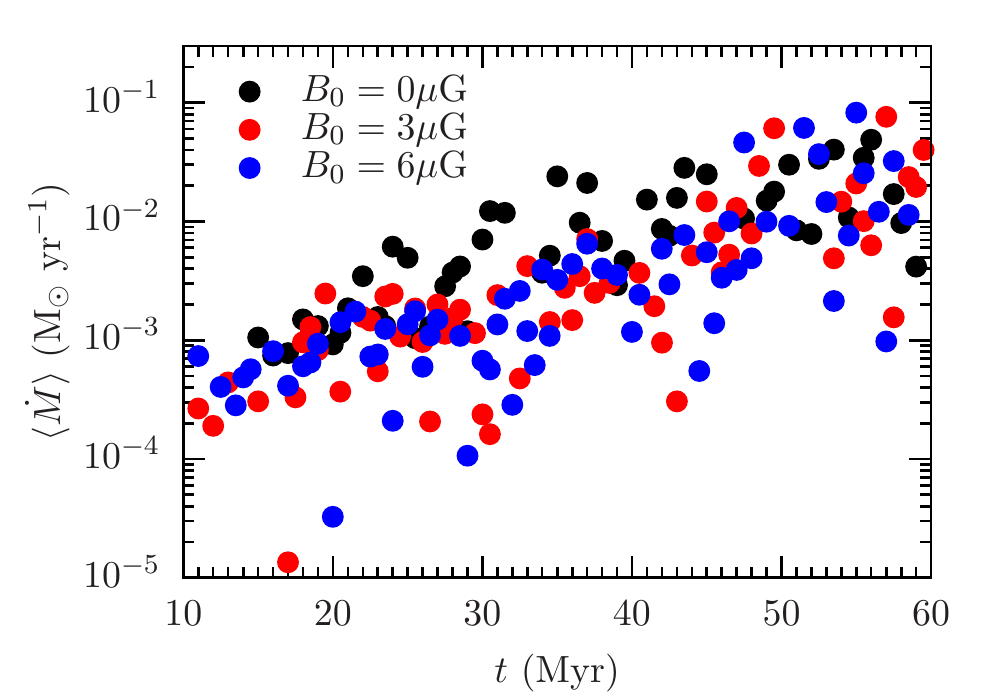}
\includegraphics[width=8cm]{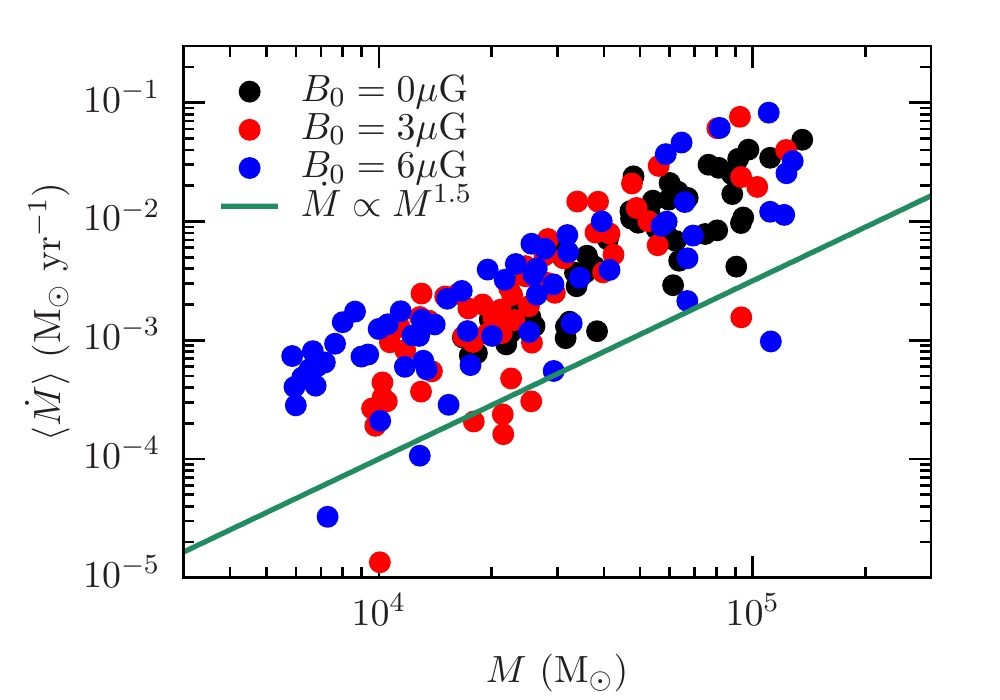}
\caption{Mean accretion rate as a function of time (top) as well as a function of cloud mass (bottom) for simulations \texttt{B0-1pc}, \texttt{B3-1pc}, and \texttt{B6-1pc}. The accretion rate increases from $10^{-4}$ to $10^{-1}\,\mathrm{M}_\odot\,\mathrm{yr}^{-1}$ over time. The accretion rate approximately scales with $\dot{M}\propto M^{1.5}$.}
\label{fig:L7-clump-mass-accretion}
\end{figure}

\begin{figure}
\centering
\includegraphics[width=8cm]{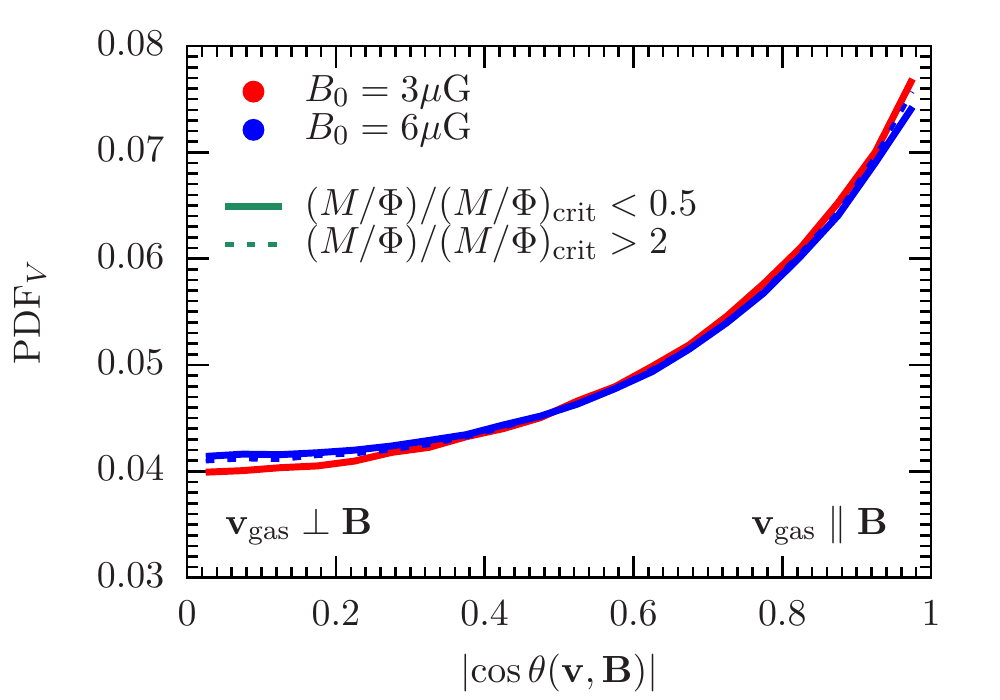}
\caption{Normalised distribution function of the angle between the velocity and the magnetic field inside the cloud radius ($r_\mathrm{cloud}=20\,\mathrm{pc}$) at $30\,\mathrm{Myr}$ for both magnetic runs. The distributions reveal a stronger flow along the field lines than perpendicular to $B$. This means that the accretion onto the cloud is predominantly along the field lines and is thus not strongly correlated with an increase in field strength. The variations over time are very small.}
\label{fig:L7-clump-angle-vB}
\end{figure}

\begin{figure}
\includegraphics[width=8cm]{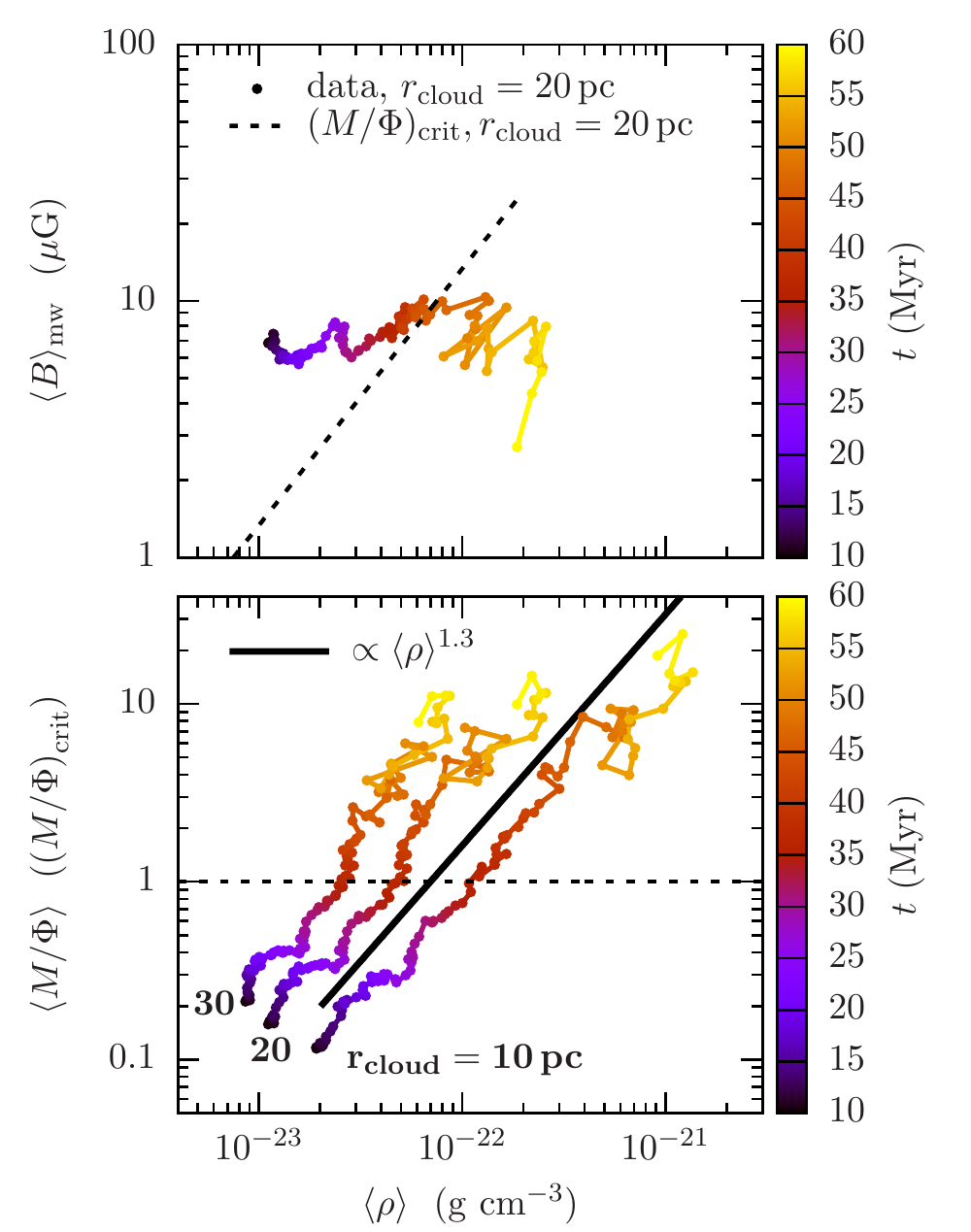}
\caption{Evolution of the mass weighted mean magnetic field (top) and the mass-to-flux ratio (bottom) as a function of mean cloud density for simulation \texttt{B6-1pc}. Colour coded is the simulation time. The top panel only includes the clouds with an analysis radius of $20\,\mathrm{pc}$. The bottom panel shows all different radii. The mass weighted mean magnetic field does not change perceptibly during the evolution. The transition from magnetically dominated to gravitationally dominated clouds occurs at the same time ($\approx35\,\mathrm{Myr}$) for all analysis radii. The mass-to-flux ratio scales super-linearly with the density.}
\label{fig:L7-MPhi-time-evol}
\end{figure}

Fig.~\ref{fig:L7-clump-mass-accretion} presents the mean value for the accretion rate as a function of time (top) and as a function of cloud mass (bottom) starting from $t=10\,\mathrm{Myr}$ to avoid spurious fluctuations at the beginning of the simulation. The accretion rates of individual clouds show strong fluctuations over time with positive and negative values. Averaged over $\Delta t=5\,\mathrm{Myr}$ the mean accretion rate of the cloud ensemble,
\begin{equation}
\skl{\dot{M}} = \frac{1}{N}\sum_{j=1}^N\rkl{\frac{M_{\mathrm{cloud},j}(t)-M_{\mathrm{cloud},j}(t-\Delta t)}{\Delta t}},
\end{equation}
slowly increases over time for all simulations with values from $10^{-4}-10^{-1}\,\mathrm{M}_\odot\,\mathrm{yr}^{-1}$, which again agrees well with the high-resolution simulations by \citet{SeifriedEtAl2017} as well with observations of star-forming regions \citep{FukuiEtAl2009, KawamuraEtAl2009, PerettoEtAl2013}. The non-magnetic clouds show noticeably higher accretion rates, which is consistent with the larger mean cloud mass as well as the higher internal velocity dispersion. Plotted as a function of cloud mass (bottom) reveals a scaling of the accretion rate with the cloud mass of approximately $\dot{M}\propto M^{1.5}$, approximately consistent in all simulations. At this point we need to emphasise that the effective accretion rate computed as stated above includes the gain of gas by merging of the clouds and thus cannot be understood as a smooth gas accretion model like, e.g. Bondi accretion \citep{Bondi44, Bondi52}, where $\dot{M}\propto M^2$. Consequently, we cannot apply this scaling to the time evolution of the mass of an \emph{individual} cloud but have to understand it as the scaling of the ensemble, i.e. statistically a more massive clouds accretes more gas with a relation following $\dot{M}\propto M^{1.5}$, but not smoothly over time.

We use again the angle between the velocity and the magnetic field vector to investigate the accretion flow onto the clouds. A uniform distribution of angles would result in a uniform distribution of $|\cos\theta|$. Fig.~\ref{fig:L7-clump-angle-vB} shows the volume weighted distribution of $|\cos\theta|$ at $30\,\mathrm{Myr}$ for the two magnetic runs, divided into clouds that are subcritical with $(M/\Phi)/(M/\Phi)_\mathrm{crit}<0.5$ as well as the supercritical clouds with $(M/\Phi)/(M/\Phi)_\mathrm{crit}>2.0$. Temporal variations as well as variations for different cloud radii are not significant. The distributions show a very similar structure with a peak at $\cos\theta=1$ indicating that the gas flow is stronger along the field lines than perpendicular to them. The variations across the distribution in $|\cos\theta|$ are not very large, but nonetheless reveal that the effective integrated accretion from $0.5<|\cos\theta|\le1$ is $40\%$ higher than the one in the range of $0<|\cos\theta|\le0.5$. Combined with the fact that the accretion rates strongly vary with also temporally negative rates, this could support the scenario, in which the clouds need some time to effectively accrete gas along this \emph{parallel channel} before becoming supercritical. However, accretion perpendicular to the field lines in combination with magnetic flux loss can as well increase the mass-to-flux ratio. Fig.~\ref{fig:L7-MPhi-time-evol} shows the mass weighted magnetic field strength (top) and the mean mass-to-flux ratio (bottom) of the clouds as a function of their mean density for simulation \texttt{B6-1pc}. Colour coded is the simulation time. The upper panel only includes the data points for a cloud analysis radius of $20\,\mathrm{pc}$ and includes the theoretical threshold line for the critical mass-to-flux ratio. We note that the clouds evolve towards higher masses without significantly increasing the mean magnetic field strength. This supports the accretion scenario, in which the gas flow mainly follows the magnetic field lines. The bottom panel shows the evolutionary tracks for all cloud radii. The critical ratio for $M/\Phi$ is indicated by a dashed line. The differences in the cloud analysis radii mainly manifest in different mean densities. The transition from subcritical to supercritical clouds occurs at approximately the same time ($\sim35\,\mathrm{Myr}$).

\subsection{Magnetic flux}

An important hint on the dominant accretion scenario is given by the scaling of $M/\Phi$ as a function of $\langle\rho\rangle$, which is equivalent to the same scaling with mass $M$ for our fixed cloud radii, $M/\Phi\propto\langle\rho\rangle^q$. Accretion perfectly along the field lines results in a linear scaling, $q=1$, because the flux does not change. A flatter scaling with $q<1$ requires a stronger increase in the flux than in the mass. Significant  flux losses would dramatically increase $q$. For $r_\mathrm{cloud}=10\,\mathrm{pc}$ the scaling is approximately $\propto\langle\rho\rangle^{1.3}$, which is significantly stepper than in previous simulations results, $(M/\Phi)/(M/\Phi)_\mathrm{crit}\propto M^{0.4}$, \citep{BanerjeeEtAl2009, InoueInutsuka2012, IffrigHennebelle2017} and still steeper than recent numerical simulations by \citet{Hennebelle2017}, who report a slightly sub-linear scaling. We can relate the effective flux to the mass as $\Phi\propto M^{1-q}$ and the time derivative
\begin{equation}
\dot{\Phi}\propto (1-q)M^{-q}\dot{M}.
\end{equation}
We approximate the time evolution of the cloud mass and the accretion rate as
\begin{align}
M(t) &= 10^4\,\exp\rkl{\frac{t}{20\,\mathrm{Myr}}}\,\mathrm{M}_\odot,\\
\dot{M}(t) &= 5\times10^{-4}\,\exp\rkl{\frac{t}{20\,\mathrm{Myr}}}\,\mathrm{M}_\odot\,\mathrm{yr}^{-1}.
\end{align}
Assuming a constant $q$ we find for the temporal change in the effective magnetic flux
\begin{equation}
\dot{\Phi}\propto\exp\rkl{\frac{t(1-q)}{20\,\mathrm{Myr}}},
\end{equation}
which gives an e-folding time of $67\,\mathrm{Myr}$ for $q=1.3$. This means that the effective flux in our clouds drops by approximately to $1/e$ over the simulated time.

\section{Discussion and Caveats}
\label{sec:discussion}

\subsection{Outflows and mass loss}

Due to the diode boundary conditions in $z$ (gas can leave the box but not enter) we include the effects of outflows but fountain flow activity is thus not captured in this setup. However, our previous studies \citep{GirichidisSILCC2} as well as other studies \citep[e.g.][]{HillEtAl2012, KimOstriker2018} indicate that the fountain flow cycle occurs on time scales larger than $100\,\mathrm{Myr}$. With our comparably short evolution time of $60\,\mathrm{Myr}$ we are not able to capture this effect. In our previous studies the formation of a steady outflow takes $\sim40\,\mathrm{Myr}$ for clustered SN driving, which is the most realistic of our employed SN driving modes. \citet{GattoEtAl2017} also report outflows after approximately $30-40\,\mathrm{Myr}$. This is consistent with the onset of mass loss that we measure in the setups discussed here. Due to the short simulation time we do not simulate the larger box as in previous studies and refrain from investigating outflow properties like temperature, outflow velocities and chemical composition.

\subsection{Continuous increase of cloud masses}

Clouds in our simulations mainly grow, but are not destroyed by the SN feedback. This is partially due to the positioning of the SNe (60\% clustered SNe and 40\% individual SNe) at random positions. Thus there is no correlation between the local gas density and the SNe. Therefore, there are no (or very few) explosions that occur directly in the centre of molecular clouds. However, previous studies with a very similar setup \citep{SILCC1, GirichidisSILCC2} show that the clustered SN driving results in the most realistic ISM conditions. Studies by \citet{LiEtAl2015} elucidate that the fraction and distribution of runaway stars justifies a SN positioning that is not directly related to the parental density peaks of the corresponding molecular cloud core.

Apart from the positioning problem of SNe our study is missing other physical effects like direct radiation from massive stars and stellar winds. Both effects have been included into similar setups using sink particles \citep{HennebelleIffrig2014, GattoEtAl2017, PetersEtAl2017a}. All simulations including sink particles are able to (partially) disperse molecular clouds. However, the interaction between the gas and sink particles should cover at least a few cells in radius on the highest refinement level, which corresponds to a region of $\sim10\,\mathrm{pc}$ with our current resolution. In particular the lack of a reliable prescription for the magnetic interaction between the Eulerian grid and the Lagrangian sink particles dissuades us from including sink particles in this study.

Numerical experiments on scales of a full galaxy also find similar effects concerning the evolution of GMCs. \citet{TaskerTan2009}, \citet{Tasker2011}, \citet{RenaudEtAl2013}, \citet{BournaudEtAl2014} and \citet{TaskerWadsleyPudritz2015} simulate full galactic discs with varying ISM and feedback models. They all conclude that feedback is not able to disperse massive condensations.

Detailed simulations of isolated clouds including internal feedback processes show that the destruction or dispersal of molecular clouds strongly depends on the cloud and feedback details. \citet{BanerjeeKlessenFendt2007} show that jets and outflows from individual sources are not able to disrupt clouds because the energy injection is primarily in compressive modes that are damped too efficiently. Outflows from multiple sources stir the gas more efficiently but still do not provide enough energy transfer to disrupt the clouds \citep{WangEtAl2010, NakamuraLi2014, PetersEtAl2014}. Stellar winds and ionising radiation are more efficient in removing gas from molecular clouds. However, ionising radiation and the resulting thermal pressure are able to change the internal structure but are unlikely to be able to disrupt clouds more massive than $\sim10^4\,\mathrm{M}_\odot$ \citep{KrumholzMatzner2009, MurrayQuataertThompson2010}. Numerical simulations by \citet{WalchEtAl2012} and \citet{DaleErcolanoBonnell2012} confirm this limited dynamical effect. Simulations including stellar winds allow to unbind $10-30\%$ of the mass \citep{DaleEtAl2013, DaleEtAl2014} but without completely disrupting them on time scales of $\sim10-20\,\mathrm{Myr}$. Both radiation and stellar winds are able to perforate the cloud's envelope, see e.g. the one-dimensional study by \citet{RahnerEtAl2017}. As a result, SNe placed in the centre of molecular clouds have a limited effect in disrupting them because of the high porosity and efficient radiative cooling \citep{RogersPittard2013, WalchNaab2015}. The missing central stellar feedback in our simulations might thus change the structures in and around the clouds \citep[see also][]{GattoEtAl2017, PetersEtAl2017a, ButlerEtAl2017}, but would likely allow the massive clouds to survive.

\subsection{Missing galactic shear}

We do not use shearing boundary conditions in our setup. For the evolution over long time scales shear might be an important effect for both GMC evolution as well as the global magnetic field properties. However, estimates based the Rossby number for this setup in \citet{GirichidisSILCC2} suggests that the simulated setup is not dominated by shear effects. Concerning the magnetic dynamo properties, simulations by \citet{PiontekOstriker2005},  \citet{GentEtAl2013b}, and \citet{KimOstriker2015b} in a similar simulation box including shear indicate that the time scales for an effective dynamo is much longer than the simulated $60\,\mathrm{Myr}$.

\subsection{Influence of the galactic environment}

More important than the dynamo effect due to large scale galactic dynamics might be the missing galactic large-scale structure that anchors the field at the boundaries of our box. The efficient SN driving can easily drive magnetic flux out of the box, which would be more difficult in a full galactic context. Although this effect is unlikely to change our results for the dense regions and the cloud analysis, it might help the magnetic field to persist in the low-density environment. As a result the field strengths and the Alfv\'{e}nic Mach numbers in the low-density gas (volume weighted curves of $B$ in Fig.~\ref{fig:L7-mag-field-strength} and $\mathcal{M}_\mathrm{A}$ in Fig.~\ref{fig:Mach-number-time}) might be lower and upper limits, see also observations by \citet{HeilesTroland2005} who find magnetism and turbulence to be in approximate equipartition. Observations summarized by \citet{Haverkorn2015} also report strong fields above the disc in the Milky Way, which would correspond to significantly lower Alfv\'{e}nic Mach numbers assuming that the overall dynamics in our simulations is comparable to the observations.

\subsection{Long-term effect of magnetic fields}

We primarily note differences between the magnetic and non-magnetic runs in the first half of the simulation time. Towards the end the differences significantly reduce. That suggests that the magnetic fields might not have a long-term effect on the evolution of the ISM. However, the initial differences are significant. In particular the formation of dense gas and the onset of the first gravitationally collapsing clouds are significantly delayed if magnetic fields are included. As we are not following the star formation based on dense or collapsing gas but apply a constant SN rate, we are not including possibly reduced star formation and SN rates in our model. Whether (temporally) reduced feedback will enhance or further reduce the impact of magnetic fields will depend on the details of the feedback like the positioning of the SN and cannot be answered with the current model.

\section{Conclusions}
\label{sec:conclusions}

We perform three-dimensional simulations of the SN-driven magnetised interstellar medium with focus of the formation of molecular clouds in the presence of magnetic fields. The simulated volume covers a stratified gas distribution along $z$ with a magnetic field strength of $0$, $3$, and $6\,\mu\mathrm{G}$ in the midplane ($z=0$). The total volume covers $(500\,\mathrm{pc})^3$ with a gas surface density of $10\,\mathrm{M}_\odot\mathrm{pc}^{-2}$. We employ a chemical network including ionised, atomic and molecular hydrogen as well as CO, C$^+$ and free electrons. The effects of (self-)shielding and the attenuation of the interstellar radiation field are taken into account via the \textsc{TreeCol} algorithm. We evolve the system for a total time of $60\,\mathrm{Myr}$ with two different effective resolutions ($256^3$ and $512^3$). The results can be summarised as follows. 

\begin{itemize}
\item In the presence of magnetic fields the morphological evolution of the ISM differs from non-magnetic environments. The magnetic pressure thickens the galactic disc leading to much larger scale heights ($\sim100-150\,\mathrm{pc}$ for the total gas, $\sim80-100\,\mathrm{pc}$ for molecular gas) compared to the non-magnetic environments, where 90\% of the total gas is confined to $30-40\,\mathrm{pc}$ and 90\% of the molecular gas is located at a height of only $20\,\mathrm{pc}$ from the disc midplane. As a result the non-magnetic simulation forms small clouds with a spherical shape close to the midplane. The thick magnetised disc allows the formation of structures at larger heights above the midplane. In addition the magnetised structures form as more massive and elongated entities.

\item Both the direct support by magnetic pressure as well as the resulting lower central densities due to the extended disc structure retard the formation of dense structures and molecular clouds. In the non-magnetic simulation molecular gas starts forming after only $10\,\mathrm{Myr}$ and reaches a total mass fraction of 0.5 after $\sim20\,\mathrm{Myr}$. In contrast the simulation with an initial magnetic field strength of $6\,\mu\mathrm{G}$ needs $\sim35\,\mathrm{Myr}$ before significant amounts of H$_2$ form. Once the formation of molecular hydrogen sets in the formation rates are comparable. At the end of the simulation we do not notice a difference in total H$_2$ fraction between the non-magnetic and magnetic runs.

\item Most of the gas is moving at supersonic and super-Alfv\'{e}nic velocities. The average Mach numbers in the ionised hydrogen are around $2$ with very little temporal variation and negligible difference between the magnetic and non-magnetic simulations. The Mach numbers in atomic hydrogen increase over time from $\mathcal{M}_\mathrm{s}\sim2$ to values of $4$, again with little dependence on the magnetisation. The highest Mach numbers ($\sim20-30$) are reached in the molecular gas. The final values are very similar for all simulations but due to the different formation times of molecular gas, the intermediate evolution differs with temporally lower values for the magnetic runs. The globally averaged Alfv\'{e}nic Mach numbers indicate super-Alfv\'{e}nic motions for both magnetic simulations. Over time $\mathcal{M}_\mathrm{A}$ increases to values of $\sim50-80$ for the molecular gas and $\sim20-30$ for H$^+$. The lowest Alfv\'{e}nic Mach numbers develop in atomic hydrogen with $\mathcal{M}_\mathrm{A}\sim3-8$.

\item The mean magnetic field increases from the initial values of $3$ and $6\,\mu\mathrm{G}$ to $\sim10-20\,\mu\mathrm{G}$ in the dense gas. The low-density environment loses magnetic intensity to minimum values of $\sim0.1\mu\mathrm{G}$. The low-density gas with weak magnetisation shows a scaling of the field strength with density of order $B\propto\rho^\alpha$ with $\alpha\approx2/3$ as expected from dimensional arguments for a negligible magnetic field impact. At densities of order the mean ISM density ($\rho\sim10^{-24}\,\mathrm{g\,cm}^{-3}$) the field becomes stronger such that $\sim30\%$ of the mass is in regions with $\mathcal{M}_\mathrm{A}<1$ and a fraction of $\sim80\%$ with $\beta<1$. Above this density the scaling of the field becomes flatter with $\alpha\approx1/4$, which indicates that the magnetic field is able to channel the flow along the field lines suppressing the compressional effect.

\item The non-magnetic simulations form significantly more molecular clouds (peak of 500 at a global molecular gas fraction of $0.4$) compared to the magnetic simulations with a maximum of $100$ clouds at approximately the same total molecular gas fraction. Over time the clouds merge and accrete material, which allows them to grow from an average mass of $10^4\,\mathrm{M}_\odot$ to $10^5\,\mathrm{M}_\odot$. The mean accretion rate scales approximately with the mass of the cloud as $\dot{M}\propto M^{1.5}$, independent of the magnetisation of the gas. The magnetised initial clouds form with a subcritical mass-to-flux ratio ($\sim0.2-0.3$ in units of the critical value), i.e. are supported by $B$ against gravitational collapse. Accretion flows that are stronger along the field lines than perpendicular to them allows the clouds to transition from sub to supercritical clouds at a time of $\sim30-40\,\mathrm{Myr}$. At the end of the simulation the clouds reach median mass-to-flux ratios of more than $10$ times the critical value. This transition is a potentially important process to delay the collapse of clouds and the formation of stars and might alter the star formation efficiency.

\item The internal velocity dispersion of the clouds ranges from $\sim3-7\,\mathrm{km\,s}^{-1}$ and constantly increases over time. The inter-cloud velocity velocity dispersion is slightly lower with values from $\sim2-5\,\mathrm{km\,s}^{-1}$ and also increases. The non-magnetic run shows the largest internal and smallest inter-cloud values. For the simulation with strong magnetic fields both quantities are the same within the temporal scatter.
\end{itemize}

\section*{Acknowledgements}
We thank the referee for helpful comments and suggestions that improved the paper.
PG, DS, SW, TN, SCOG, and RSK acknowledge support from the DFG Priority Program 1573 {\em Physics of the Interstellar Medium}. PG acknowledges funding from the European Research Council under ERC-CoG grant CRAGSMAN-646955.
DS and SW acknowledges the support of the Bonn-Cologne Graduate School, which is funded through the Excellence Initiative and the German Science Foundation (DFG) for funding through the Collaborative Research Center 956 \emph{The conditions and impact of star formation}, project C5. SW further acknowledges the support from the European Research Council through the ERC Starting Grant RADFEEDBACK (project number 679852) under FP8.
TN acknowledges support from the DFG cluster of excellence \emph{Origin and Structure of the Universe}.
RW acknowledges support by the Albert Einstein Centre for Gravitation and Astrophysics via the Czech Science Foundation grant 14-37086G and by the institutional project RVO:67985815 of the Academy of Sciences of the Czech Republic.
RSK and SCOG thank the DFG for funding via the SFB 881 The Milky Way System (subprojects B1, B2, and B8). RSK furthermore acknowledges support from the European Research Council under the European Community's Seventh Framework Programme (FP7/2007-2013) via the ERC Advanced Grant STARLIGHT (project number 339177).
The authors thank the Max Planck Computing and Data Facility (MPCDF) for computing time and data storage.
The software used in this work was developed in part by the DOE NNSA ASC- and DOE Office of Science ASCR-supported Flash Center for Computational Science at the University of Chicago. Parts of the analysis are carried out using the \textsc{YT} analysis package (\citealt{TurkEtAl2011}, \url{yt-project.org}).





\appendix
\section{Numerical Resolution \& Refinement Threshold}

\begin{figure}
\centering
\includegraphics[width=8cm]{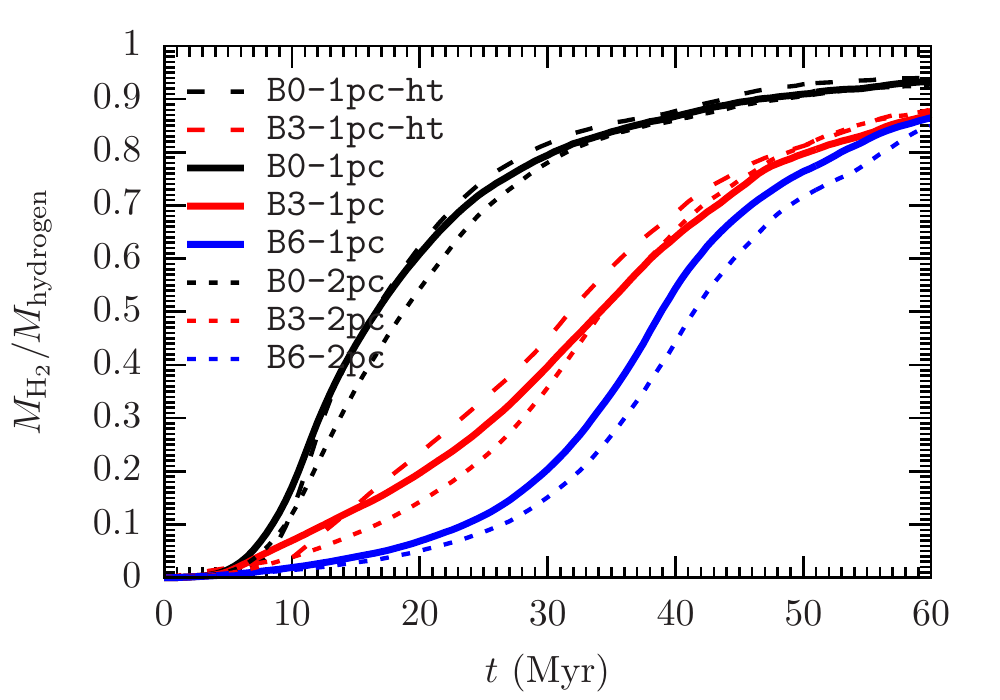}
\caption{Time evolution of the molecular hydrogen for different maximum refinement levels and different threshold H$_2$ fractions for refinement. The differences in global H$_2$ content are small compared to the effect of the magnetic field.}
\label{fig:H2-comparison-level-thresh}
\end{figure}

\begin{figure}
\centering
\includegraphics[width=8cm]{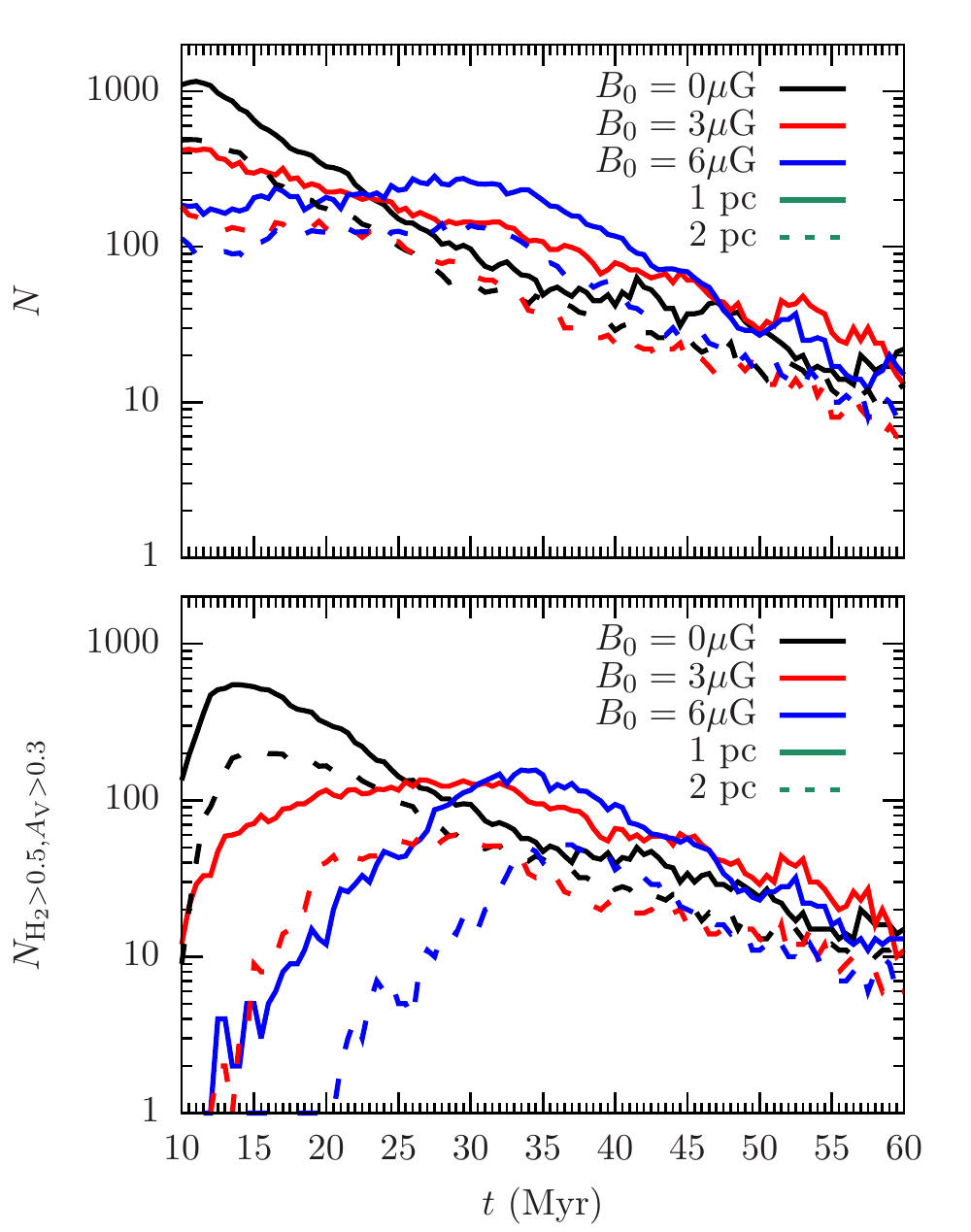}
\caption{Fragmentation for different spatial refinements. A higher resolution by a factor of two results in more fragments. Considering all minima of the gravitational potential (top) the number of fragments is $2-3$ times as high. If restricted to molecular clouds, the numbers vary by up to an order of magnitude in the fist half of the simulations and reduce to negligible differences at the end.}
\label{fig:resolution-fragmentation-comparison}
\end{figure}

We compare two different maximum refinement levels, corresponding to a minimum cell size of $2\,\mathrm{pc}$ and $1\,\mathrm{pc}$. In all simulations we refine based on the density as well as on the mass fraction of H$_2$. For the simulations \texttt{B0-1pc}, \texttt{B3-1pc} and \texttt{B6-1pc} we refine the grid at a threshold H$_2$ fraction of $0.01$ and derefine at a value of $0.005$. We perform two control runs (\texttt{B0-1pc-ht}, \texttt{B3-1pc-ht}), where we use the values of $0.1$ for refinement and $0.05$ for derefinement. Fig.~\ref{fig:H2-comparison-level-thresh} shows the total H$_2$ content in the box over time for all simulations. In all cases the differences due to the magnetic field are significantly larger compared to the refinement criterion or the maximum refinement level. However, we note that the interior structure of the molecular clouds is not resolved. The weak differences for the two spatial resolutions is thus not comparable to the much larger resolutions in \citet{SeifriedEtAl2017} in a very similar setup because their study resolves structures down to $0.06\,\mathrm{pc}$.

Fig~\ref{fig:resolution-fragmentation-comparison} shows the number of fragments (number of local gravitational minima) as in Fig.~\ref{fig:L7-clump-number} for the $1\,\mathrm{pc}$ (solid lines) and $2\,\mathrm{pc}$ (dashed lines) resolutions. In all cases the higher resolution results in approximately twice as many local gravitational minima (upper panel). Including the restrictions of optical thickness ($A_\mathrm{V}>0.3$) and molecular composition ($f_{\mathrm{H}_2}>0.5$) does not fundamentally change the behaviour (bottom panel). At early times the magnetic high resolution runs form up to an order of magnitude more molecular clouds but this difference reduces at later stages of the simulations.

\section{Defining Clouds}
\label{sec:defining-clouds}

\begin{figure*}
\includegraphics[width=13.5cm]{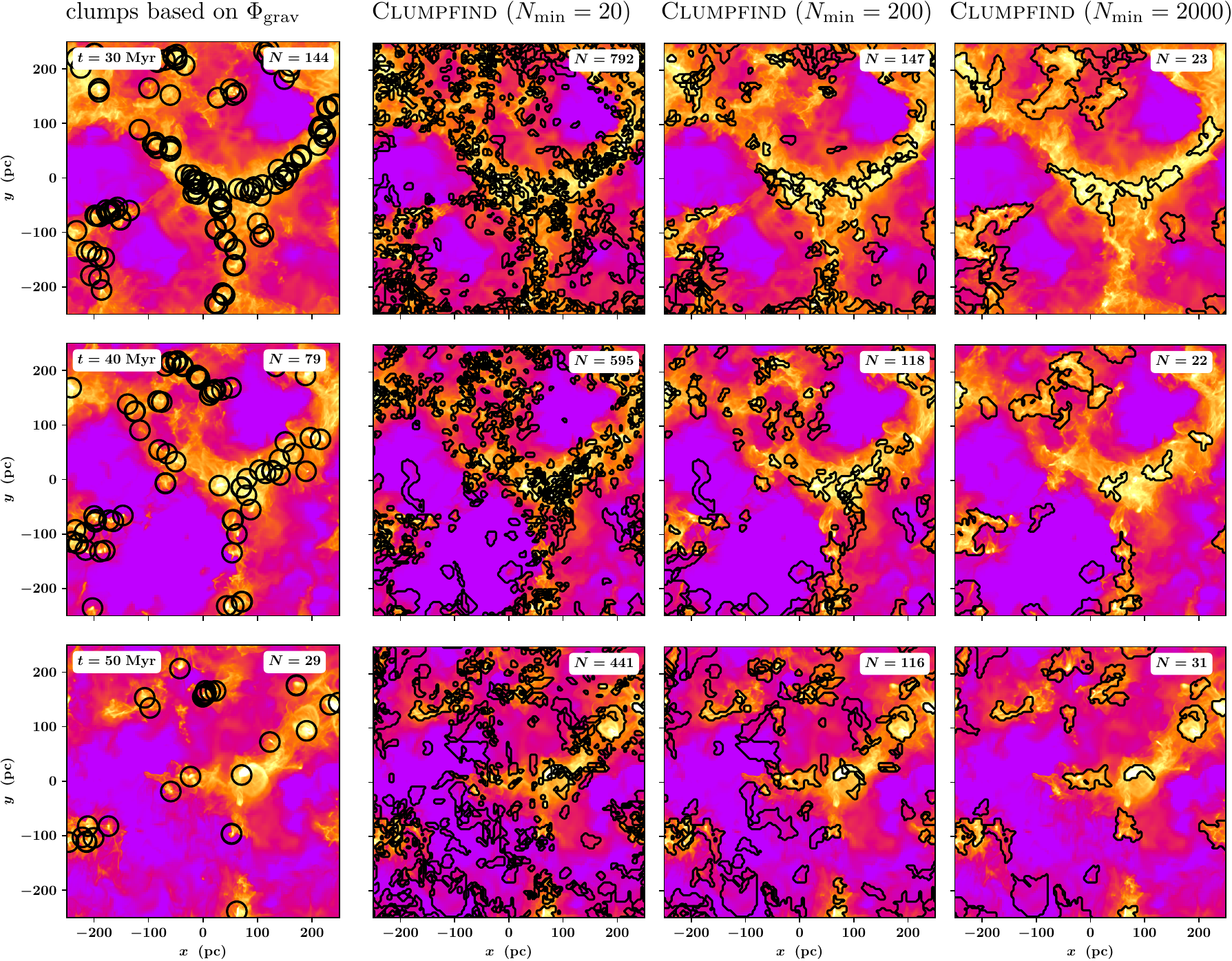}
\includegraphics[height=10.5cm]{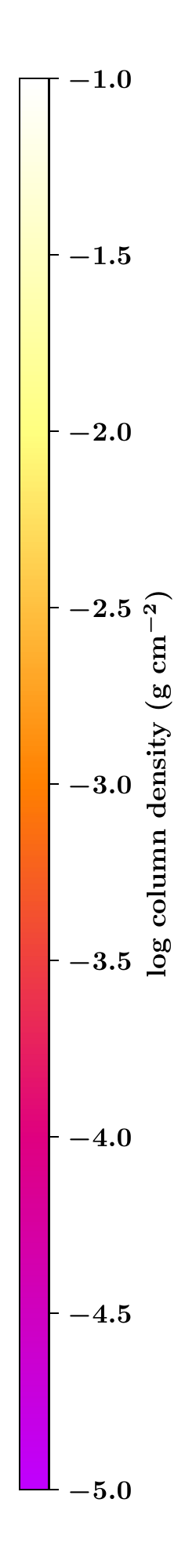}\\[0.5cm]
\includegraphics[width=13.5cm]{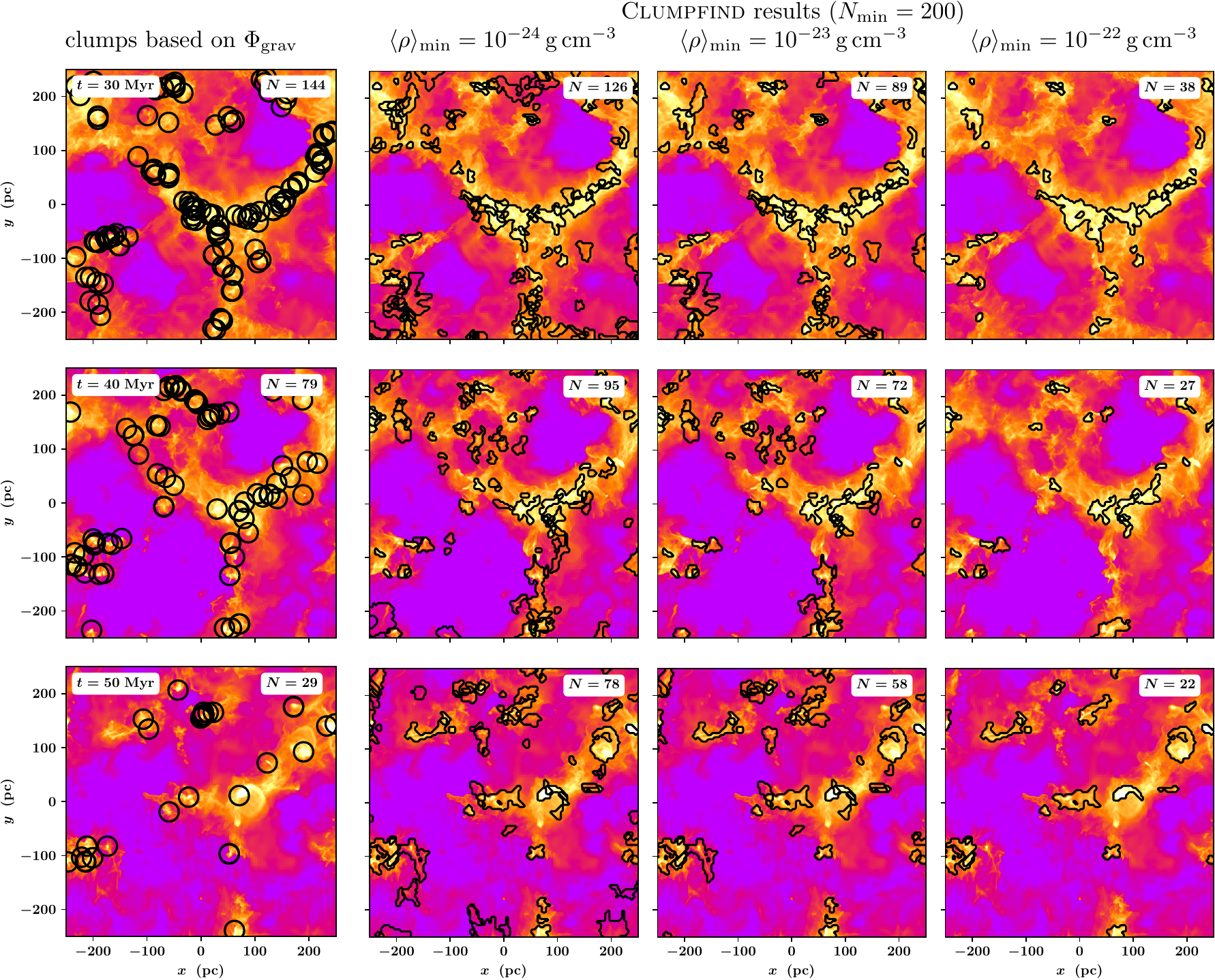}
\includegraphics[height=10.5cm]{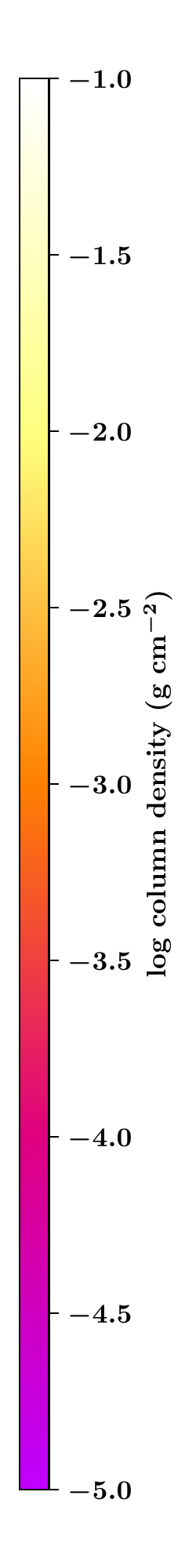}
\caption{Comparison of identified clumps. Colour-coded is the column density. From top to bottom we show the time evolution. The left-hand column shows the clouds based on the local minima of the gravitational potential with a fixed radius of $r_\mathrm{cl}=20\,\mathrm{pc}$. The three right-hand columns highlight the \textsc{Clumpfind} results for a different minimum number of cells (top plot, $N_\mathrm{min}=20, 200, 2000$ from left to right) as well as the results for different minimum average clump densities (bottom plot, $\langle\rho\rangle_\mathrm{min}=10^{-24}, 10^{-23}, 10^{-22}\,\mathrm{g\,cm}^{-3}$) using a minimum number of cells of $N_\mathrm{min}=200$.}
\label{fig:clump-finder-params-coldens}
\end{figure*}

\begin{figure*}
\includegraphics[width=0.48\textwidth]{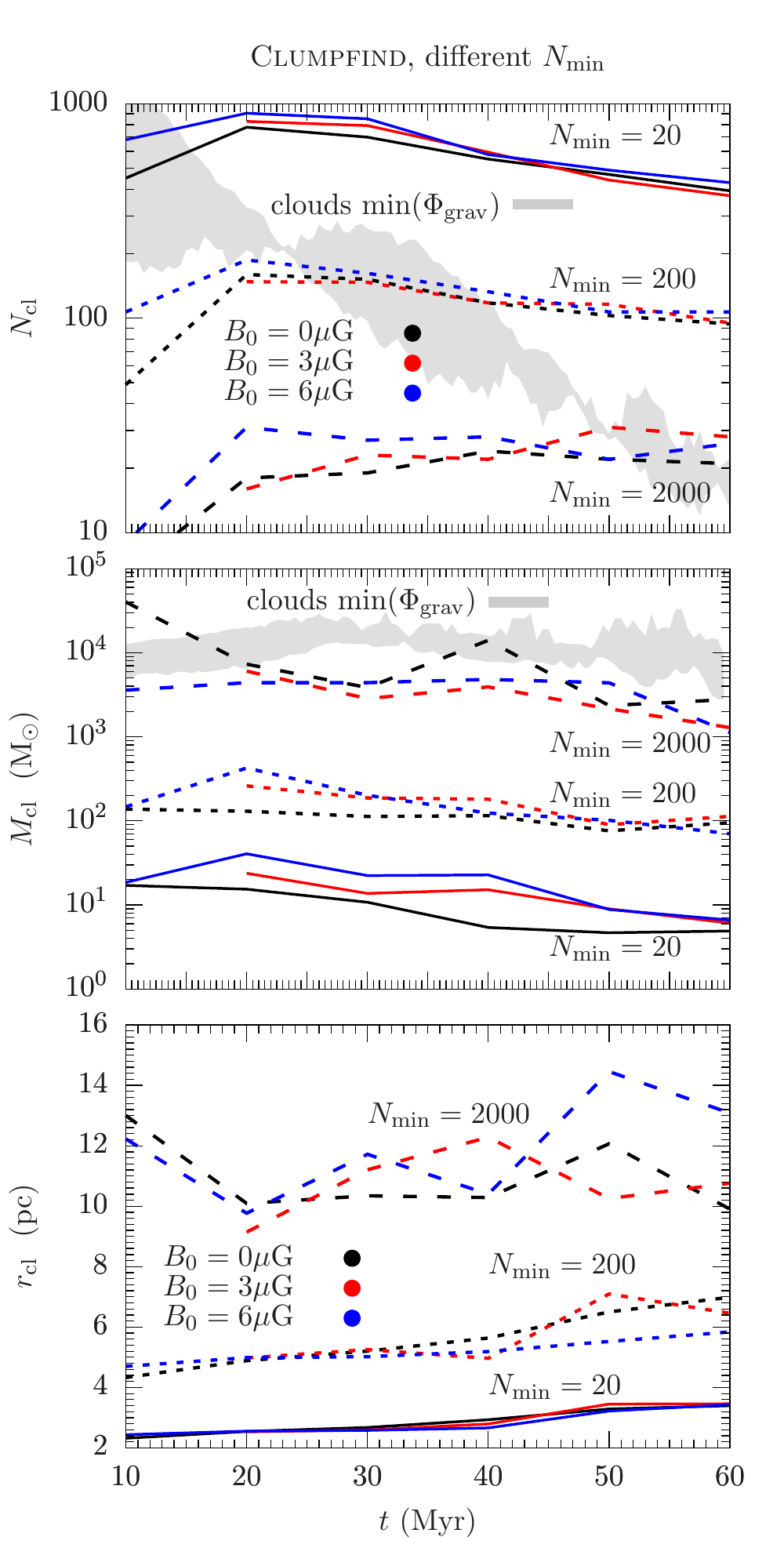}
\includegraphics[width=0.48\textwidth]{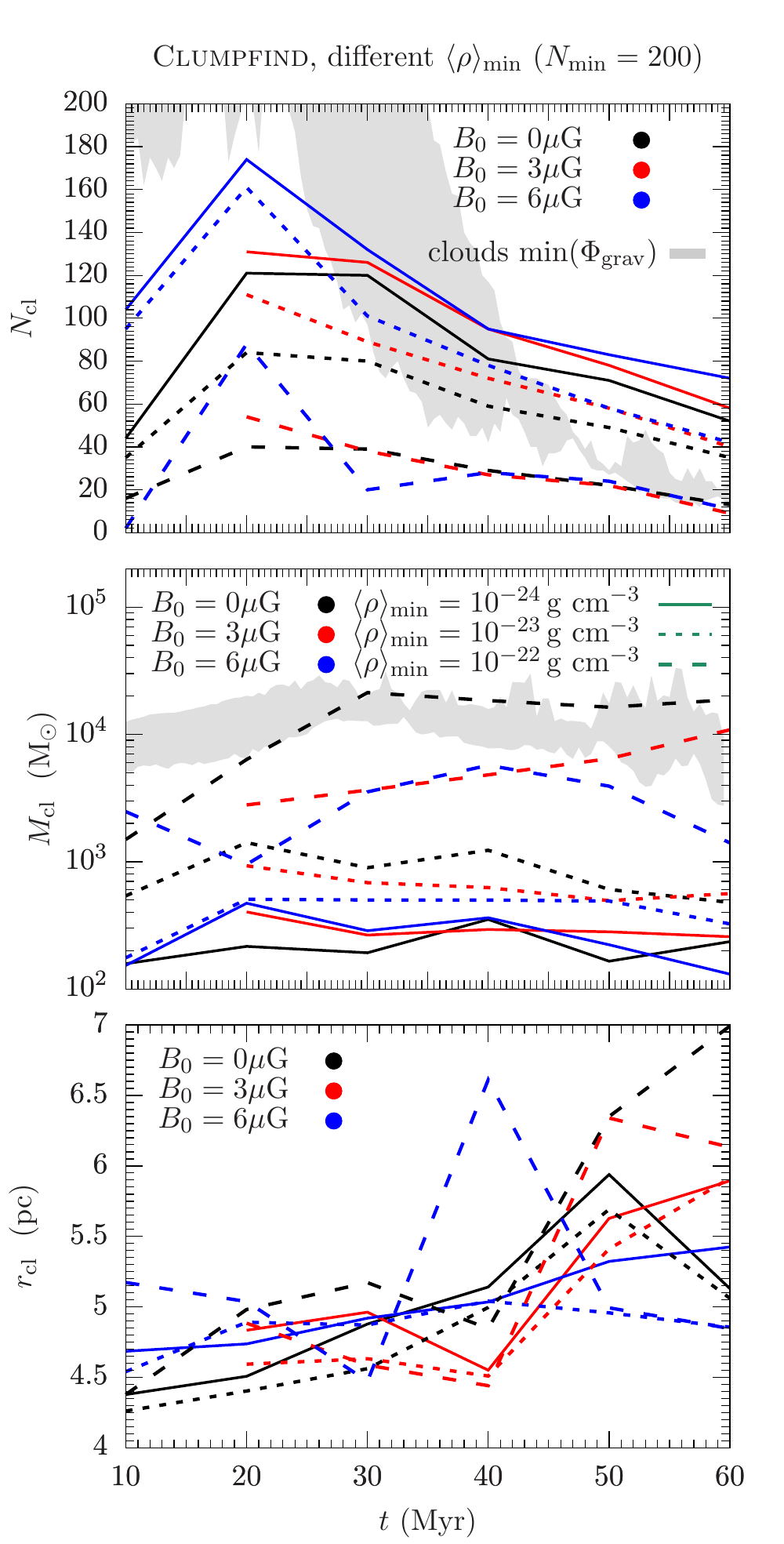}
\caption{Comparison of the clump properties over time. From top to bottom we show the total number of identified clumps, the median mass and median effective radius based on the volume assuming spherical symmetry ($r_\mathrm{cl}=\rkl{3V/(4\pi)}^{1/3}$). The lines are the \textsc{Clumpfind} results, the shaded area indicates the numbers for the gravitationally selected clouds. The left-hand panels show the properties for different $N_\mathrm{min}$; the right-hand ones are for different $\langle\rho\rangle_\mathrm{min}$ and a minimum number of cells of $N_\mathrm{min}=200$. The parameter $N_\mathrm{min}$ has a large impact on all quantities. Excluding low-density clumps has a much smaller effect.}
\label{fig:clump-finder-params-numbers}
\end{figure*}

\begin{figure*}
\includegraphics[width=0.48\textwidth]{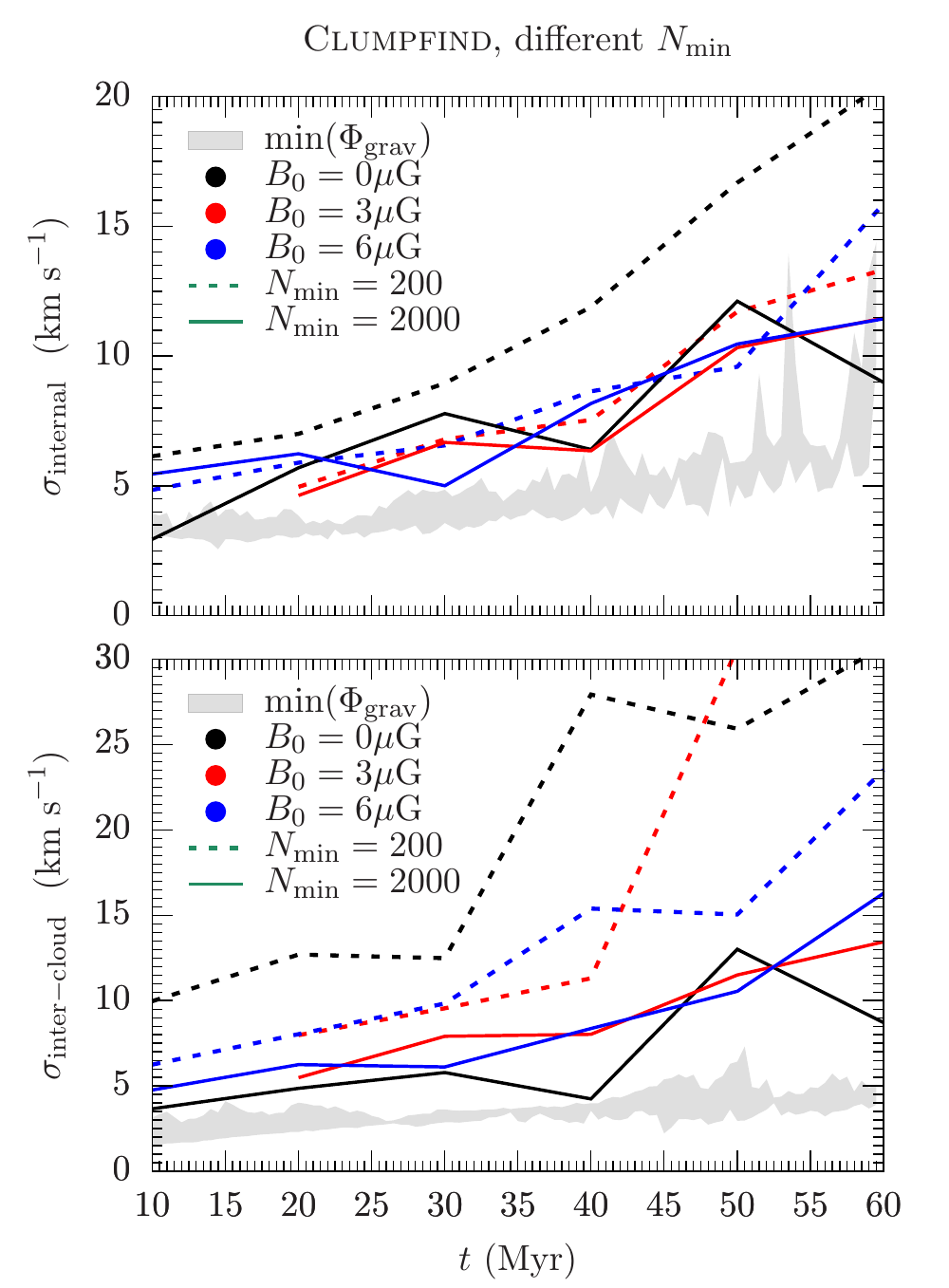}
\includegraphics[width=0.48\textwidth]{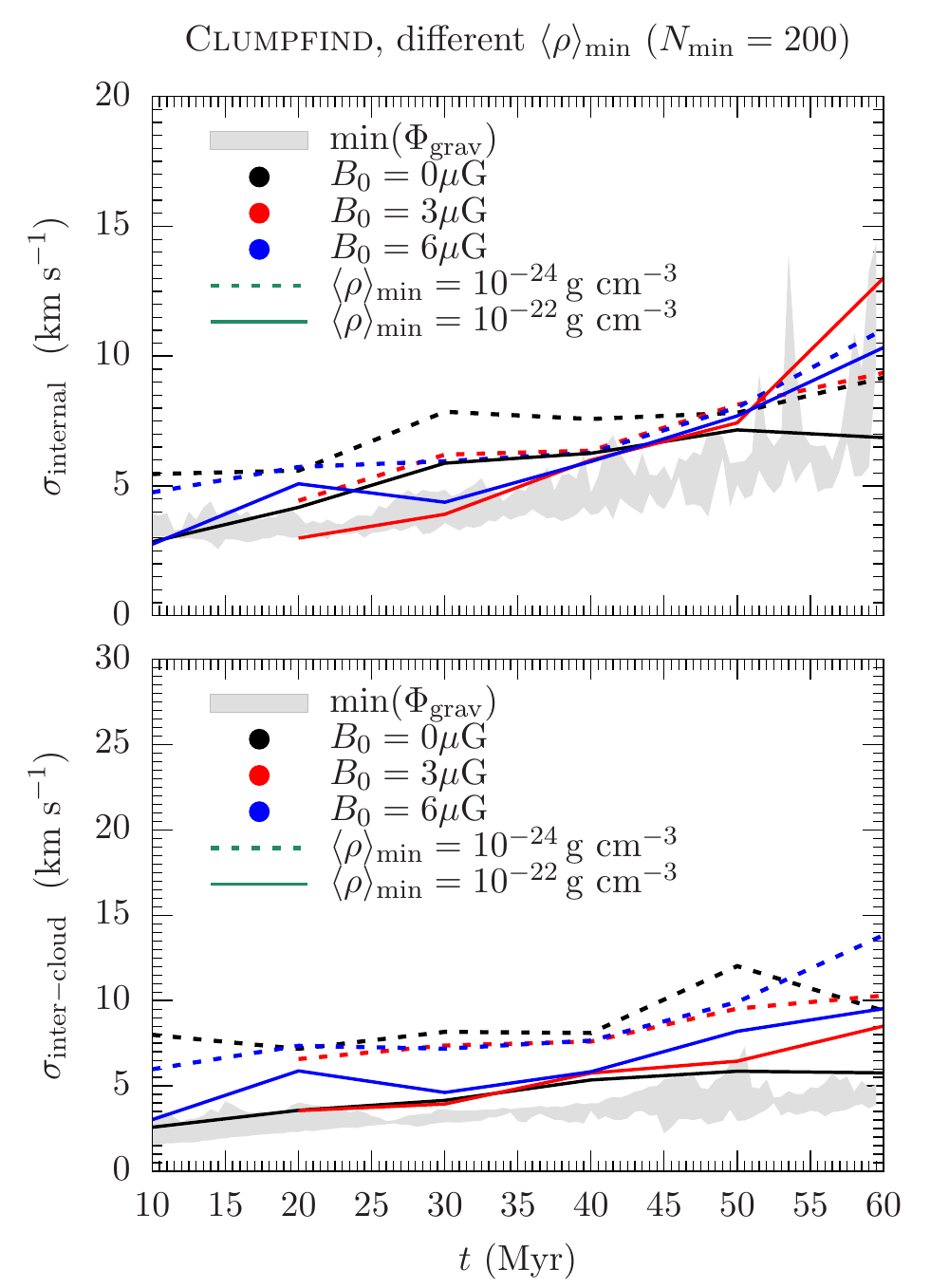}
\caption{Comparison of the internal (top) and inter-cloud (bottom) velocity dispersion for the identified clumps over time (lines for \textsc{Clumpfind}, shaded area for the $\Phi_\mathrm{grav}$ clouds). The left-hand panels show the properties for different $N_\mathrm{min}$; the right-hand ones are for different minimum average densities using $N_\mathrm{min}=200$. Overall, there are smaller differences between the \textsc{Clumpfind} clouds and our preferred method compared to the variations in the number of clouds, the mass and the cloud radius.}
\label{fig:clump-finder-sigma-sigma}
\end{figure*}

We have used local minima of the gravitational potential to identify clouds. However, there are also other methods for defining clouds. One popular method is \textsc{Clumpfind} \citep{WilliamsDeGeusBlitz1994}, which investigates contours of the density to define connected clouds. Alternatively, friends-of-friends methods or dendrograms have been used to find groups of connected regions. One difficulty with other methods are the additional free parameters. In this section we discuss the differences between our method and the classical \textsc{Clumpfind} algorithm.

For \textsc{Clumpfind}, we vary the minimum number of cells that are needed to identify clumps ($N_\mathrm{min}$) as well as the minimum average density of the clump ($\langle\rho\rangle_\mathrm{min}$). In all cases we set the density ratio between the contours to 10. Fig.~\ref{fig:clump-finder-params-coldens} shows the column density of the gas together with the identified clumps. The top plots compare different $N_\mathrm{min}$ required to be identified as clump. The bottom one depicts the differences based on minimum average densities, $\langle\rho\rangle_\mathrm{min}$, with a fixed $N_\mathrm{min}=200$. The left-hand column in both parts shows the clumps based on the local minima of the gravitational potential and is identical in both parts. The three right-hand columns in the top plot show the clumps for $N_\mathrm{min}=20$, $200$, and $2000$. The number of identified clouds has a large range from $\sim800-20$. The bottom part shows the number of clouds with $N_\mathrm{min}=200$ and the additional constraint of $\langle\rho\rangle_\mathrm{min}=10^{-24}$, $10^{-23}$, and $10^{-22}\,\mathrm{g\,cm}^{-3}$ from left to right. The number of clumps is much less dependent on the minimum average density. For $N_\mathrm{min}=200$ and $\langle\rho\rangle_\mathrm{min}=10^{-23}\,\mathrm{g\,cm}^{-3}$ there is reasonable agreement between the clumps identified by the gravitational potential and \textsc{Clumpfind} for the most-massive clumps.

Fig.~\ref{fig:clump-finder-params-numbers} shows the time evolution of the total number of clouds (top), the median mass (centre) and the derived size (bottom) based on spherical symmetry, $\rkl{r_\mathrm{cl}=\rkl{\frac{3V}{4\pi}}^{1/3}}$. The lines indicate the \textsc{Clumpfind} results; the shaded area shows the numbers for the gravitationally found clumps and is bound by the smallest and largest median of the three main simulations. The left-hand panels are for different $N_\mathrm{min}$, the right ones for different $\langle\rho\rangle_\mathrm{min}$ with a fixed $N_\mathrm{min}=200$. The number of clumps is a strong function of $N_\mathrm{min}$, which indicates that most of the identified structures are small density fluctuations. None of the parameters yields a comparable evolution to $N_\mathrm{cl}$ found for the gravitationally detected clumps. The median masses even vary by three orders of magnitude when changing $N_\mathrm{min}$ from $20$ to $2000$. Here, $N_\mathrm{min}=2000$ is comparable to our preferred method within a factor of a few. The radius also reflects the strong dependence on $N_\mathrm{min}$, which is smaller than our fiducial radius of $20\,\mathrm{pc}$. To investigate the effects of $\langle\rho\rangle_\mathrm{min}$ we set $N_\mathrm{min}=200$, which corresponds to a radius of $r_\mathrm{cl}\approx3.5\,\mathrm{pc}$ at the highest level of refinement. The variations for different minimum densities are overall smaller. The total number of clouds shows a similar qualitative time evolution as for our main method. The masses are smaller except for the highest minimum density, which only extracts the densest gas as clouds. On one hand this suggests that the properties the found clumps are not a strong function of $\langle\rho\rangle_\mathrm{min}$. On the other hand this reflects that all numbers are again set by $N_\mathrm{min}$.

Finally, in Fig.~\ref{fig:clump-finder-sigma-sigma}, we compare the internal (top) and inter-cloud (bottom) velocity dispersions for the clumps. Again, the lines are the \textsc{Clumpfind} results; the shaded regions correspond to the range of the median for the $\Phi_\mathrm{grav}$ clouds. In the left-hand panel we set $N_\mathrm{min}=200$ and $2000$, in the right-hand one we set $N_\mathrm{min}=200$ and select clouds with a minimum density of $\langle\rho\rangle_\mathrm{min}=10^{-24}$ and $10^{-22}\,\mathrm{g\,cm}^{-3}$. We note that the numbers show much less variations for different \textsc{Clumpfind} parameters compared to the numbers in Fig.~\ref{fig:clump-finder-params-numbers}. For $N_\mathrm{min}=2000$ the velocity dispersion increases from $5$ to $10\,\mathrm{km\,s}^{-1}$ over the simulated time with negligible difference between the magnetic and non-magnetic runs. The inter-cloud velocity dispersion is slightly larger with values from $\sim5-15\,\mathrm{km\,s}^{-1}$ over time. In particular the values for $N_\mathrm{min}=200$ are significantly larger than the ones for $N_\mathrm{min}=2000$. The numbers are higher compared to the values that we find in gravitationally identified clumps, which is likely due to the more complex and more elongated structures of the \textsc{Clumpfind} clouds that therefore probe gas motions of a larger spatial region as well as the fact that also low-density gas fluctuations with high temperatures and larger velocity dispersions are selected as clumps. In particular many low-density fast-moving clumps for $N_\mathrm{min}=200$ increase $\sigma_\mathrm{inter-cloud}$. If we restrict the clouds to dense structures above $\langle\rho\rangle_\mathrm{min}=10^{-24}\,\mathrm{g\,cm}^{-3}$ both the internal as well as the inter-cloud velocity dispersion are smaller and comparable or in better agreement with the values derived for the gravitationally selected clouds.  This is consistent with the tendency that we find for different radii, see Section~\ref{sec:clump-radius-variation}.

Overall, the selected clumps by \textsc{Clumpfind} allow for cloud shapes that better fit the real clouds. However, the derived properties are strong functions of the parameters, which we have to set somewhat arbitrarily. Choosing clouds based on the gravitational potential thus seems a more natural choice.

\section{Varying cloud radius}
\label{sec:clump-radius-variation}

\begin{figure*}
\begin{minipage}{\textwidth}
\includegraphics[width=0.33\textwidth]{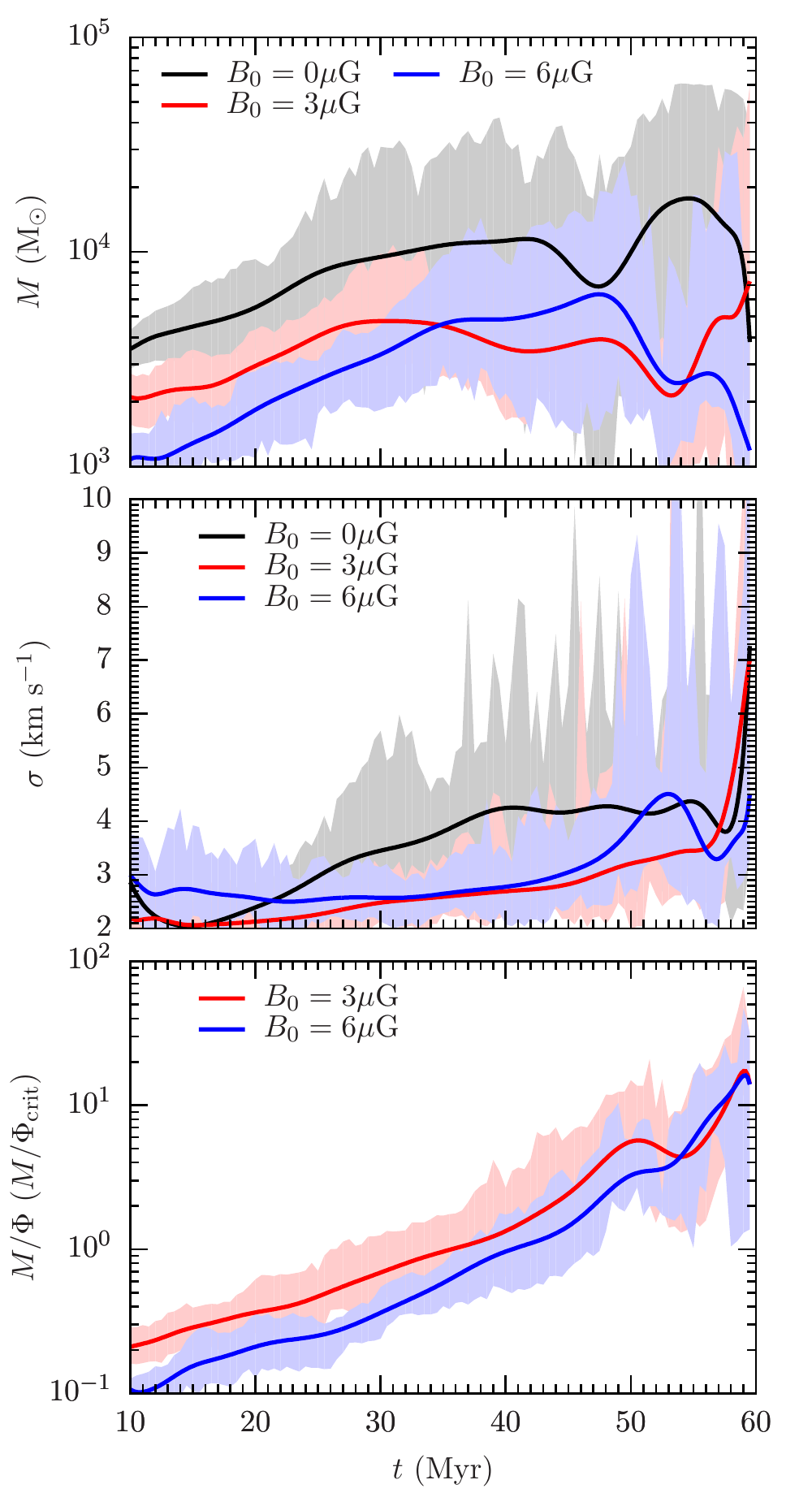}
\includegraphics[width=0.33\textwidth]{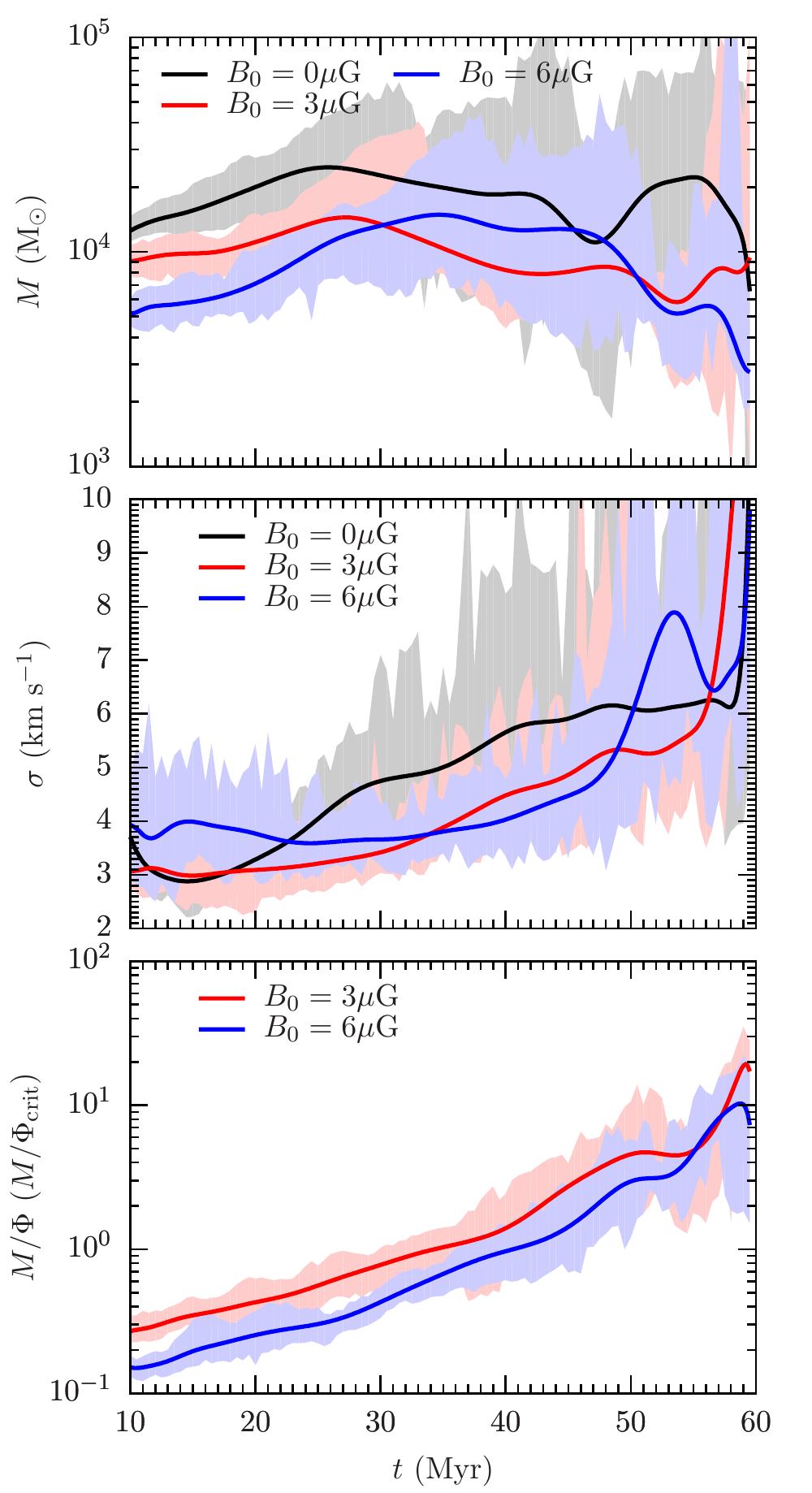}
\includegraphics[width=0.33\textwidth]{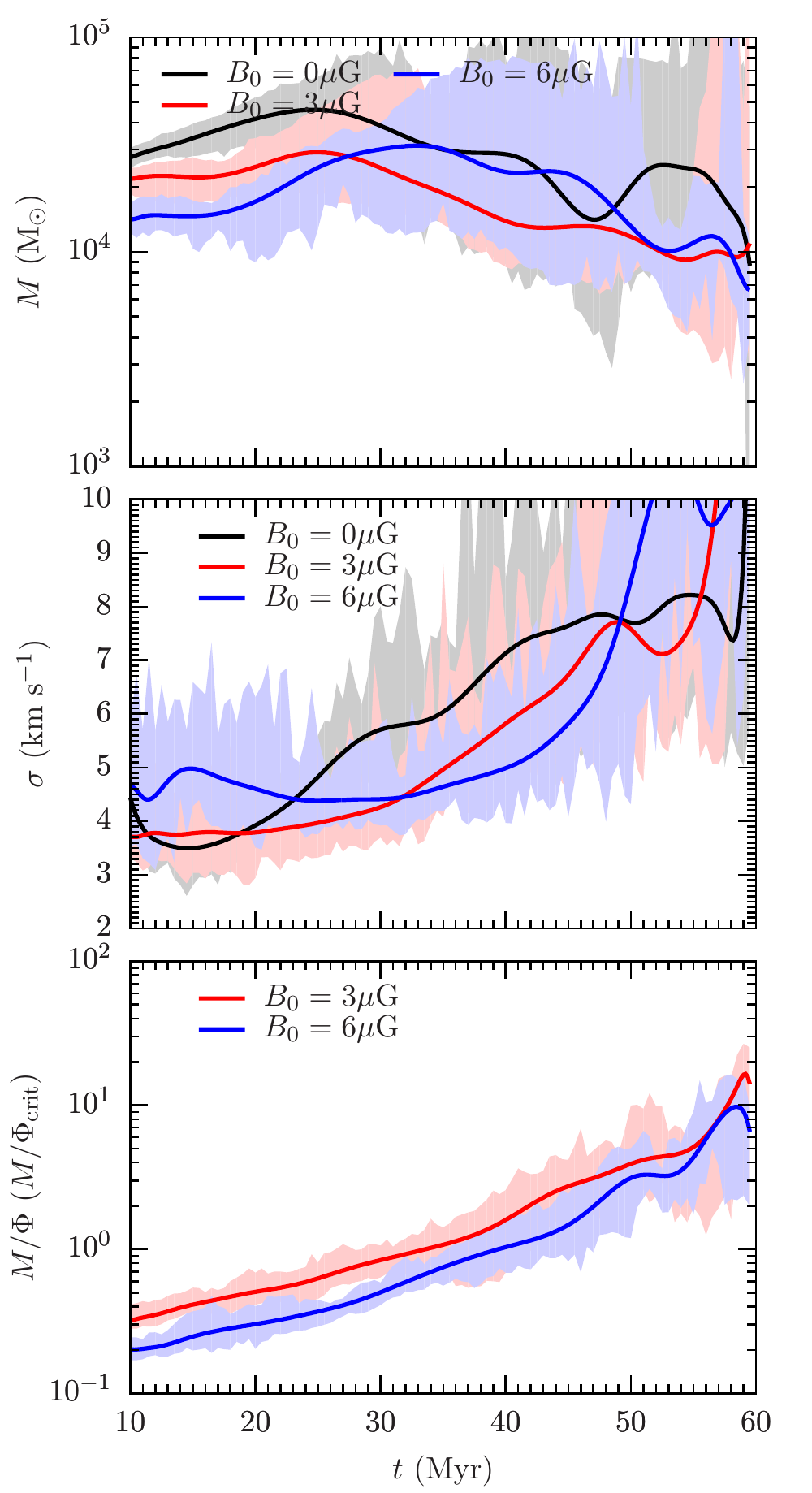}
\end{minipage}
\caption{Time evolution of the clouds for different analysis radii around the local minima of the gravitational potential. Shown are the mass (top), the internal velocity dispersion (centre) as well as the mass-to-flux ratio (bottom). From left to right the analysis radii are $10$, $20$, and $30\,\mathrm{pc}$. The total masses increase for larger analysis radii by a factor of a few, which is expected because the regions of larger radii cover more mass. Similarly the velocity dispersion increases for increasing $r_\mathrm{cloud}$. However, the mass-to-flux ratio remains unchanged for different analysis radii.}
\label{fig:L7-clump-time-evol-diff-radii}
\end{figure*}
We vary the cloud radius of the analysis volume around the local minima of the gravitational potential. In Fig.~\ref{fig:L7-clump-time-evol-diff-radii} we chose radii of $10$, $20$, and $30\,\mathrm{pc}$ from left to right. All lines show the median values, the shaded area is bounded by the 25 and 75 percentile of the distribution along the ordinate. The top panel shows the total mass, which indicates that a three times larger radius increases the total mass by approximately an order of magnitude. The internal velocity dispersions (centre) increase less strongly from left ($2-5\,\mathrm{km\,s}^{-1}$) to right ($\sim4-9\,\mathrm{km\,s}^{-1}$). The mass-to-flux ratio (bottom) does not show a dependence on the chosen radius of the analysis sphere. This strengthens our conclusions about the formation and accretion scenario of molecular clouds.

\bibliographystyle{mnras}
\bibliography{astro.bib}

\bsp	
\label{lastpage}
\end{NoHyper}
\end{document}